\documentclass[12pt,preprint]{aastex}

\begin{document}
\title{A New Approach in Understanding Growth and Decay of the Sunspots}

\author{K. M. Hiremath \altaffilmark{1} and Lovely, M. R\altaffilmark{1,2}}
\affil{Indian Institute of astrophysics, Bangalore-560034, India}
\email{hiremath@iiap.res.in}
\altaffiltext{1}{Indian Institute of Astrophysics, Bangalore-560034, India}
\altaffiltext{2}{ Sree Krishna College, Guruvayur, Kerala-680102, India;
Leave on the Faculty Improvement Programme}
\begin{abstract}
From the previous study (Hiremath 2009b; Hiremath 2010), on the genesis of solar cycle and activity
phenomena, it is understood that sunspots are formed at different
depths by superposition of Alfven wave perturbations of a strong
toroidal field structure in the convective envelope and after attaining a critical
strength, due to buoyancy,  raise toward the surface along
the rotational isocontours that have positive (0.7-0.935 $R_{\odot}$)
and negative (0.935-1.0 $R_{\odot}$) rotational gradients.
Owing to physical conditions in these two rotational gradients,
from the equation of magnetic induction, sunspot's
area growth and decay problem is solved separately.
It is found that rate of growth of sunspot's area during
its evolution at different depths is function of steady and
fluctuating parts of Lorentzian force of the ambient medium,
fluctuations in meridional flow velocity, radial variation of
rotational gradient and $cot(\vartheta)$
(where $\vartheta$ is co-latitude). While rate of decay of sunspot's area
at different depths during its evolution mainly depends upon
magnetic  diffusivity, rotational gradient and $sin^{2}(\vartheta)$.
Gist of this study is that growth and decay of area of the sunspot
mainly depends upon whether sunspot is originated in the
region of either positive or negative rotational gradient.

On the surface, as fluctuating Lorentz forces and meridional
flow velocity during sunspots' evolution are considerably negligible
compared to steady parts, analytical solution for growth
of sunspot area $A$ is $A(t)=A_{0}e^{(U_{0}cot\vartheta)t/2}$
(where $A_{0}$ is area of the sunspot during
its' initial appearance and $U_{0}$ is steady part of meridional
flow velocity on the surface, $\vartheta$ is co-latitude and  $t$ is a time variable). 
Similarly analytical solution for decay
of sunspot's area on the surface follows the relation
$A(t)=C_{1}e^{-({\Omega_{0}^{2}R_{\odot}^{2}sin^{2}\theta\over{\eta}})t}+C_{2}$
(where $C_{1}$ and $C_{2}$ are the integrational constants,
$R_{\odot}$ is radius of the sun, $\Omega_{0}$ steady
part of angular velocity and $\eta$ is the
magnetic diffusivity). For different latitudes and
life spans of the sunspots on the surface during their evolutionary
history, both the analytically derived theoretical area growth
and decay curves match reasonably well with the observed
area growth and decay curves.
\end{abstract}
\vfill\eject
\section{Introduction}
Since discovery of the sunspots by Galileo, genesis of their
22 year cyclic activity in general and, their formation and decay
during their evolutionary stages in particular still remain a mystery.
The study of sunspots' origin, formation and decay is important owing
to the observed fact that variation of sunspot occurrence activity is related
with the solar irradiance that in turn affects the earth's
environment and the climate (Prabhakaran Nayar {\em et. al.} 2002;
Hiremath and Mandi 2004; Soon 2005; Badruddin, Singh and Singh 2006; 
Perry 2007; Feymann 2007; Tiwari and Ramesh 2007 and
references there in; Scafetta and West 2008
Komitov 2009, Hiremath 2009b and references there in).

 Present general consensus is that the sunspots
originate below the solar surface due to an unknown
dynamo mechanism. Due to very high conductivity of the solar
 plasma and assuming that raising flux tube does not acquires extra flux
from the ambient medium, sunspots isorotate with the internal plasma and due to buoyancy
raise toward the surface along the path of rotational isocontours.
 This implies that  sunspots are very good tracers
of the internal dynamics and structure of the solar interior.
Hence if the sunspots that have first and second days appearance
on the surface, and if one computes their initial rotation rates,
then one can infer rotation rate of the internal solar plasma
where the sunspots' foot points are anchored.
Recent studies (Javaraiah and Gokhale 1997; Javaraiah 2001; 
Hiremath 2002; Sivaraman {\em et. al.} 2003) show that
variation of initial rotation rates obtained from the
daily motion of sunspot groups with respect to their
life spans on the surface is almost similar to the radial variation
of the internal rotation profile of the solar plasma.

\begin{figure}[h]
\begin{center}
 \includegraphics[width=20.5pc,height=18pc]{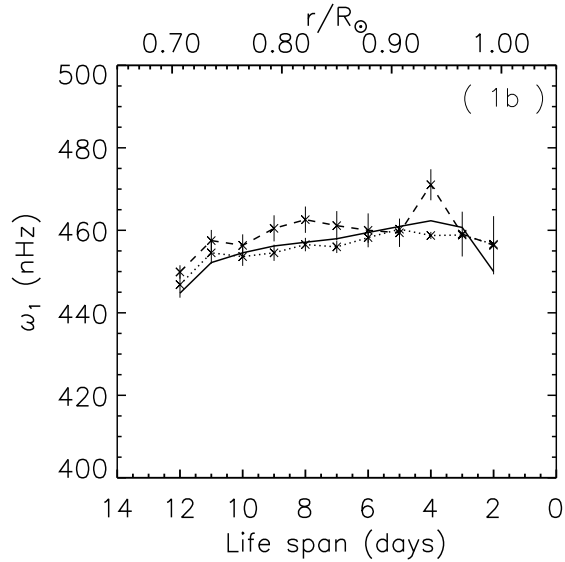}
\caption{The dashed and the dotted curves are the
variation of initial rotation rates of the sunspot groups
with respect to their life spans (Hiremath 2002). The continuous curve is the radial
variation of the internal rotation as inferred from the 
helioseismologyi (Antia, Basu and Chitre 1998).
 }
\end{center}
\label{sunspot_area_age}
\end{figure}

 From Hiremath's (2002) paper, results are reproduced in Fig 1
that illustrates a comparison between the variation of
initial rotation rates of the sunspot groups for different
life spans and radial variation of internal rotation profile
as inferred (Antia, Basu and Chitre 1998) from the helioseismology.
Note the striking similarity
between these two profiles. In order to reach closer to the
reality of the physics of convection zone and dynamics of the flux
 tubes, in the same study, the rate of change
of initial rotation rates of the sunspot groups (that represent
the acceleration or deceleration of the flux tubes in the
ambient plasma) are compared (the Fig 5(b) of Hiremath (2002) with the radial profile of
gradient of rotation that is computed from the radial variation of rotation
of the plasma inferred from the helioseismology). Again we get
a striking similarity between these profiles.
To conclude from that study, for different life spans,
initial sunspot dynamics over the surface
represents the internal dynamics in different layers of the
convection zone.
For example initial anchoring of a flux tube whose life span is 10 days is near
base of the convection zone and initial anchoring of a flux tube
whose life span is 5 days is in the middle of the convective envelope.

Observations show that there are three important stages in the sunspot's evolutionary
history : (i) a well developed sunspot (that consists of umbra and
penumbra) is formed due to coalescing of the emerging flux regions, (ii) once
stabilized sunspot is formed, its area increases and reach the
maximum value and, (iii) decay of the sunspot from it's maximum area
to minimum area and ultimately disintegrating into smaller
active regions and diffusion of the flux on the surface.

 As for the first and last stages,
there are many studies that explain the formation and
decay parts of the sunspot's evolutionary history.
The first stage is supposed to be due to convective
collapse, a kind of instability that has been invoked to
explain the kilo gauss fields on the surface (Parker 1978; Spruit 1979; Hasan 1985).
Once flux element is formed, different adjacent flux
elements coalesce and sunspot is formed. Owing to their strong
magnetic field structure, sunspots inhibit the ambient convection
resulting in reduction of temperature and density. Ultimately
lower density of the flux tube results in raising (due to
buoyancy) from the convection zone to the surface.
Contrary to this conventional view, Parker (1992) has
proposed that sunspots are basically formed due to
coalescence of magnetic elements by the vortices.
According to him, flux tubes are surrounded by vortex
flows that attract other vortices leading to coalescence
of different flux elements. On the other hand Meyer {\em et. al.} (1974),
have different view on the formation of the flux tubes.
According to them a strong converging flow is necessary
to form the sunspots. That means sunspots might be
formed at the boundary of the convective cells, with
an outflow at the surface and an inflow in the deeper
layers. Where as Hiremath (2009b; 2010), by updating Alfven's (1943)
seminal idea of sunspot formation, came to the conclusion
that sunspots are formed due to superposition of Alfven wave perturbations of
the underlying steady part of large scale toroidal magnetic field
structure and travel along isorotational contours in order to reach 
at the proper activity belt on the surface.

 There are many studies on the decaying phase of the sunspot.
Cowling (1946) was the first person to investigate the decay
part of the sunspot area. Bumba (1963) obtained a linear
decay law for the recurrent spot groups and exponential
decay law for the non-recurrent spot groups. Where as
some of the previous studies (Petrovay and Moreno-Insertis 1997;
Petrovay and van Driel-Gesztelyi 1997)
 indicate the quadratic decay ({\em i.e.}, sunspot area
as quadratic function of time) and other studies 
(Solanki 2003 and references there in) 
indicate the linear decay law.
Moreno-Insertis and Vazquez (1988) and Martinez Pillet,  Moreno-Insertis
and Vazquez (1993)
 conclude that the present sunspot
data do not allow any distinction between either linear
or quadratic decay law. To add to these decay laws,
log-normal distribution (Martinez Pillet, Moreno-Insertis
and Vazquez 1993) also fit the decay of umbrae.

  There are following theoretical studies to understand the sunspot
decay. First theoretical study in supporting the results of
linear decay laws is by Zwan and Gokhale (1972). Such a linear
decay law suggests that flux loss takes place everywhere
within the spot irrespective of their different sizes.
Zwan and Gokhale (1972) assumed a current sheet around the
sunspot and turbulent diffusion inside the tube. In this
case Ohmic diffusion dictates the decay of the current sheet
and hence as spot decays to smaller area, thickness of the
current sheet reduces. In fact such current sheets around the
sunspots have been observed by Solanki, Rueedi and Livingston (1992).
In contrast, Simon and Leighton (1964) and Schmidt (1968)
propose that the sunspots are decayed by the erosion of the
sunspot boundary which implies that $dA/dt$ is proportional
to $A^{1/2}$, where A is area of spot. Supporting the erosion model, 
Petrovay and Moreno-Insertis (1997) proposed that turbulent diffusivity
depends strongly on the field strength. Their model predicts
the quadratic decay and spontaneous current sheet around the
sunspot.

 Though there are many studies on the first and last phases
of the sunspot evolution, the second stage of a sunspot, viz., 
physics of a growth phase, during it's life time is not understood. Moreover, it is not
clear whether all the three phases in a sunspot's life time
remain same or different over the whole solar cycle. That means: 
is there any year to year variations in the area gradients (rate of
change of area $dA/dt$ with respect to time, where A(t) is
time dependent area of the sunspots and $t$ is time variable) of the
sunspots during it's increasing (second phase) and
decaying (last phase)? Is there any connection between
the evolutionary history of the sunspots and underlying
deeper dynamics or this phenomenon is simply due
to surface convection. Some of these important issues are
addressed in this study.

As for year to year changes in gradient of sunspots' area,
for the year 1955-1965 and for different life spans, Hiremath 
(2005, with summer student Mr. Subba Rao), computed both growth
$dA_{1}/dt$ and the decay $dA_{2}/dt$ rates of the sunspots
and came to the following conclusions. 
For the same life span, to reach their maximum areas sunspots 
take different times as cycle progresses. That is, in the beginning
of the cycle, area-age curves are nearly gaussian and as
cycle progresses area-age curve follow the simple linear
decay law. Further they conclude: (i) during the beginning of the 
solar cycle, sunspots' {\em rate of growth} and {\em rate of decay} are 
larger compared at end of the solar cycle, (ii) in the beginning
of the solar cycle, in order to reach their maximum areas,
sunspots  increase their area at a rate of $\sim$ 100 mh (millionths 
hemisphere)/day where as at the end of solar cycle sunspots increase
their area at the rate of $\sim$ 50 mh/day and,
(iii) sunspots decay faster ( $\sim$ 75 mh/day) in
the beginning of the solar cycle compared to the
end of the solar cycle ($\sim$ 25 mh/day). The last
conclusion is similar to the conclusion
from the recent study (Hathaway and Choudhary 2008) on
decay of area of the sunspot groups.
Active regions are centers of solar activity ranging from flares to CMEs. 
They are believed to be locations where magnetic flux bundles erupt from deep interior 
in the convection zone to emerge at solar surface in the form of sunspots 
due to magnetic buoyancy. Further the complexity of sunspot groups plays 
an important role in determining the active regions (Zirin 1988). 
The difference in the total energy output between a solar minimum and maximum,
indirectly associated with the sunspots, is about
0.1\%. Even this small energy changes in the sun's output over 11-year
solar cycle can intensify wind and rainfall activities and therefore has 
a major impact on global weather and climatic patterns 
in the Earth's climatic parameters. Therefore it is useful to investigate 
how the sunspot groups themselves eventually grow and decay. The growth and decay of sunspot groups also 
play an important role in the day to day irradiance variations (Wilson 1981).

    If decay of sunspots were purely by ohmic dissipation, sunspots 
would have lifetimes of about 300 years by considering their 
size and photospheric conductivity (Cowling 1946). However, the
sunspots have shorter life span of $\sim$ weeks for non-recurrent
spot groups and $\sim$ months for recurrent spot groups. 
How to reconcile these observed phenomena, viz., three phases
of formation, growth and decay of the sunspots. 
In the recent study (Hiremath 2009b; Hiremath 2010), it is proposed that sunspots are formed by the superposition 
of many Alfven wave perturbations of the embedded toroidal magnetic field structure. 
Once sunspots are formed, due to buoyancy, at different depths
in the convective envelope raise along isorotational
contours and reach the surface at different latitudes.
One can notice from Fig 1 that the internal rotational profile (continuous curve),
as inferred from helioseismology (Antia, Basu and Chitre 1998), has two
rotational gradients, {\em viz.,} a positive rotational gradient
from base of the convective envelope to $0.935 R_{\odot}$ and
a negative rotational gradient from $0.935 R_{\odot}$ to $1.0 R_{\odot}$.  

From the magnetic induction equation, it is proposed in this study that growth and decay of either sunspots' area or
magnetic flux is due to interplay of both
convective source term (that in turn depends mainly upon fluctuations in the positive rotational
gradient and convection) and sink term (that in turn depends
upon fluctuations in negative rotational gradient, magnetic eddy diffusivity
and radiation effects near the surface).
That means sunspots that are formed in the region of positive
rotational gradient, while raising toward the surface,
accumulate magnetic flux from the ambient magnetic turbulent
medium and reduction of  magnetic flux in the region of negative rotational
gradient. The net magnetic flux of the sunspot that is formed in
the region of positive rotational gradient in the convective envelope
while raising it's anchoring feet and reaching toward $0.935 R_{\odot}$, 
should increase and, magnetic flux should decrease as flux
tubes' anchoring feet lifts from $0.935 R_{\odot}$ to $1.0 R_{\odot}$.
On the other hand, the sunspots that are formed in the region of negative
rotational gradient while raising toward the surface mainly experience decay phase only.
These reasonable ideas will be clear in the following
section. In order to understand and test these ideas on growth and decay
phases of the sunspots, magnetic induction equation
is solved by considering the source and the sink terms separately.
In section 2, formulation of the equations are presented. 
With reasonable approximations, analytical solution of magnetic induction 
equation for the growth of area is presented in section 3 
and solution for decay part is presented in section 4. 
In section 5, both the analytical solutions are fitted with observed
sunspots' growth and decay phases of the sunspots and 
conclusions from these results are presented.


\section {Formulation of the equations}
It is assumed that, in the convective envelope, fluid is incompressible.
We also assume that the magnetic eddy diffusivity $\eta$ 
 with value represented by the
appropriate average. Magnetic field ${\bf B}$ and the
velocity field ${\bf V}$ vectors are expressed as
\begin{equation}
  {\bf B}  = {P}{\bf \hat{\hbox{I}}}_\vartheta + {T}{\bf \hat{\hbox{I}}}_\varphi \ ,
\end{equation}
\begin{equation}
{\bf V} = {U}{\bf \hat{\hbox{I}}}_\vartheta + {r \Omega sin\theta}{\bf \hat{\hbox{I}}}_\varphi \ ,
\end{equation}
where ${\bf \hat{\hbox{I}}}_\vartheta$ and ${\bf \hat{\hbox{I}}}_\varphi$
are the unit vectors along heliographic latitude $\vartheta$ and longitude $\varphi$ of the
sun; $P$, $T$, $U$ and $\Omega$ are scalar functions. $P$ and $T$ are
scalar functions that represent poloidal and toroidal parts of the
the magnetic field structures and $U$ and $\Omega$ are scalar functions
that represent poloidal (meridional) and toroidal (angular velocity) parts
of the velocity field structures. Equation of continuity is
\begin{equation}
{{\partial \rho}\over{\partial t}} + \rho \nabla . {\bf V} = 0 .
\end{equation}
As the life spans ($\sim$ weeks to months) of sunspots
are very much larger than the time scales ($\sim$ minutes) 
of ambient density perturbations in the convective envelope,
we have ${{\partial \rho}\over{\partial t}}=0$ and the resulting
equation is
\begin{equation}
{\Large\bf {{ \rho \nabla . {\bf V} = 0}}} 
\end{equation}
where $\rho$ is the density. Similarly as magnetic diffusivity is assumed 
to be constant, magnetic induction equation is
\begin{equation}
{{\partial {\bf B}}\over{\partial t}} = curl ({\bf V}\,\times {\bf B}) 
+ \eta \nabla^{2} {\bf B} .
\end{equation}
This magnetic induction equation determines growth and decay of the sunspot.
The first term on right hand side (RHS) is the source term that enhances
the magnetic flux of the sunspot and second term on RHS is the
sink term that attempts to destroy the generated magnetic flux.
As magnetic induction equation in turn depends upon
velocity and diffusivity $\eta$, these two source and sink terms are
important and dictate the growth and decay of the
sunspots. We solve the induction equation by
considering the source and sink terms separately for the following reasons.
In case of region of positive rotational gradient from base of convective envelope
 to $0.935R_{\odot}$, rate of increase of magnetic flux that mainly
depends upon fluctuations of increase in rotational gradient is dominant
compared to magnetic diffusivity. As for region of negative 
rotational gradient from $0.935R_{\odot}$ to $1.0R_{\odot}$, fluctuations in decreasing rotational
gradient, increasing magnetic diffusivity ( as magnetic diffusivity
$\eta \, \sim T^{-3/2}$, where $T$ is ambient
temperature) and dominant radiational effects near the surface remove and destroy the magnetic flux.

\section{Solution for growth of the sunspot}
After substituting equations (1) and (2) in equation (5)
and also by satisfying the continuity equation (4), by
considering a source (first) term of the toroidal
component of the induction equation in spherical coordinates is 

\begin{equation}
{{\partial T}\over{\partial t}} = ({{UTcot\theta}\over{r}}) + (P sin \theta{{\partial \Omega}\over{\partial \theta}} - {{U}\over{r}}{{\partial T}\over{\partial \theta}}) + (T {{\partial \Omega}\over{\partial \phi}} -  \Omega {{\partial T }\over{\partial \phi }}) \ ,
\end{equation}
where $r$, $\theta$ and $\phi$ are radial, latitudinal and longitudinal
variables in spherical coordinates. The last term in RHS of
the above equation can be simplified further as follows
\begin{equation}
\Omega = {{\phi_2 - \phi_1}\over{t_2 - t_1}} = {{\partial \phi}\over{\partial t}} \ ,
\end{equation}
where $\Omega$ is the angular velocity, $\phi_1$ and $\phi_2$ are 
changes in longitudinal displacement from time $t_1$ and $t_2$ respectively. Hence we have following equations
\begin{equation}
{\partial \phi} = \Omega {\partial t} \ ,
\end{equation}
\begin{equation}
T{{\partial \Omega}\over{\partial \phi}} = {{T}\over {\Omega}}{{\partial \Omega}\over{\partial t}} \ ,
\end{equation}

\begin{figure}
\centering
{\label{fig:linear fit}\includegraphics[width=7.5cm,height=7.5cm]
{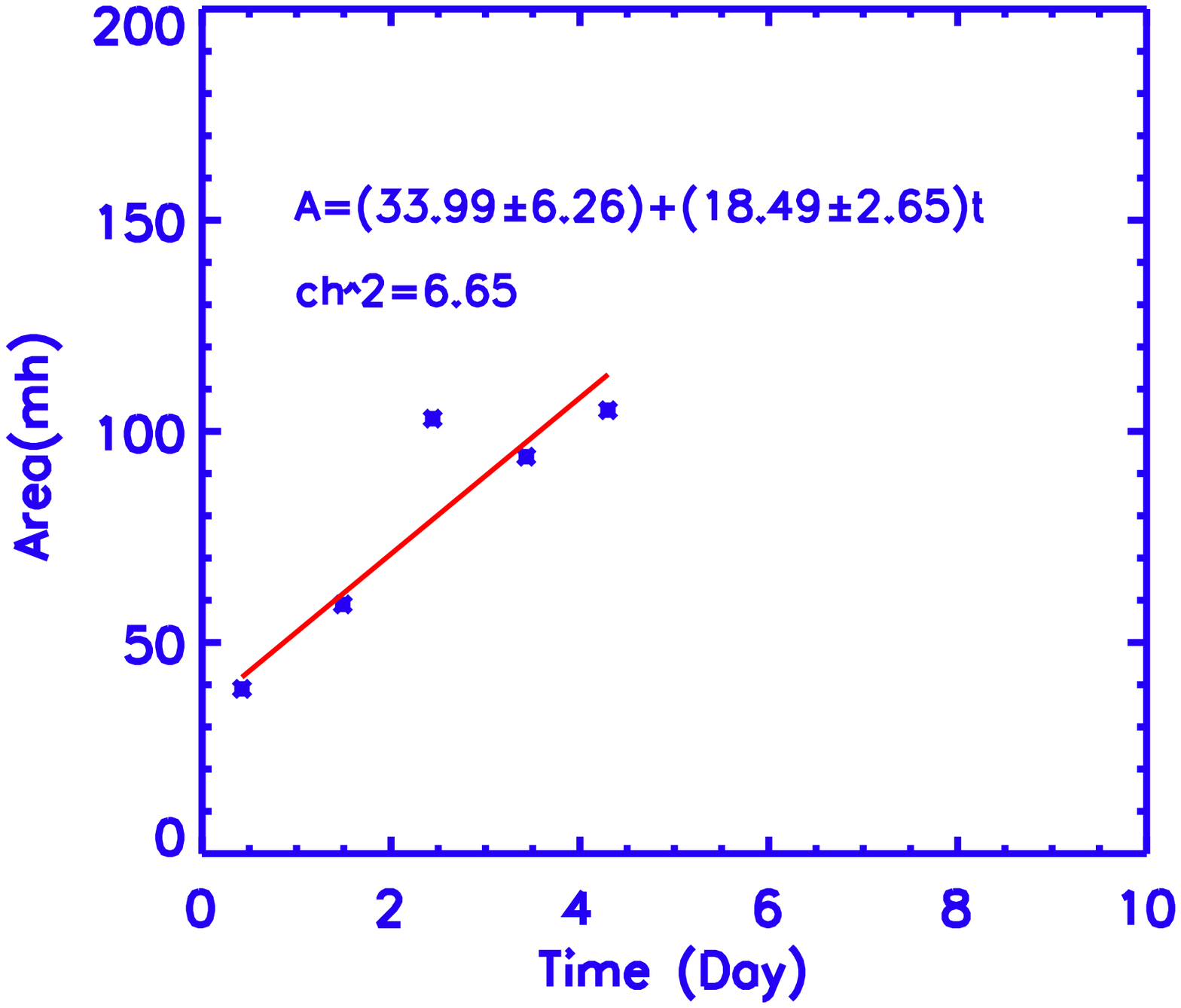}}
{\label{fig:quadratic fit}\includegraphics[width=7.5cm,height=7.5cm]
{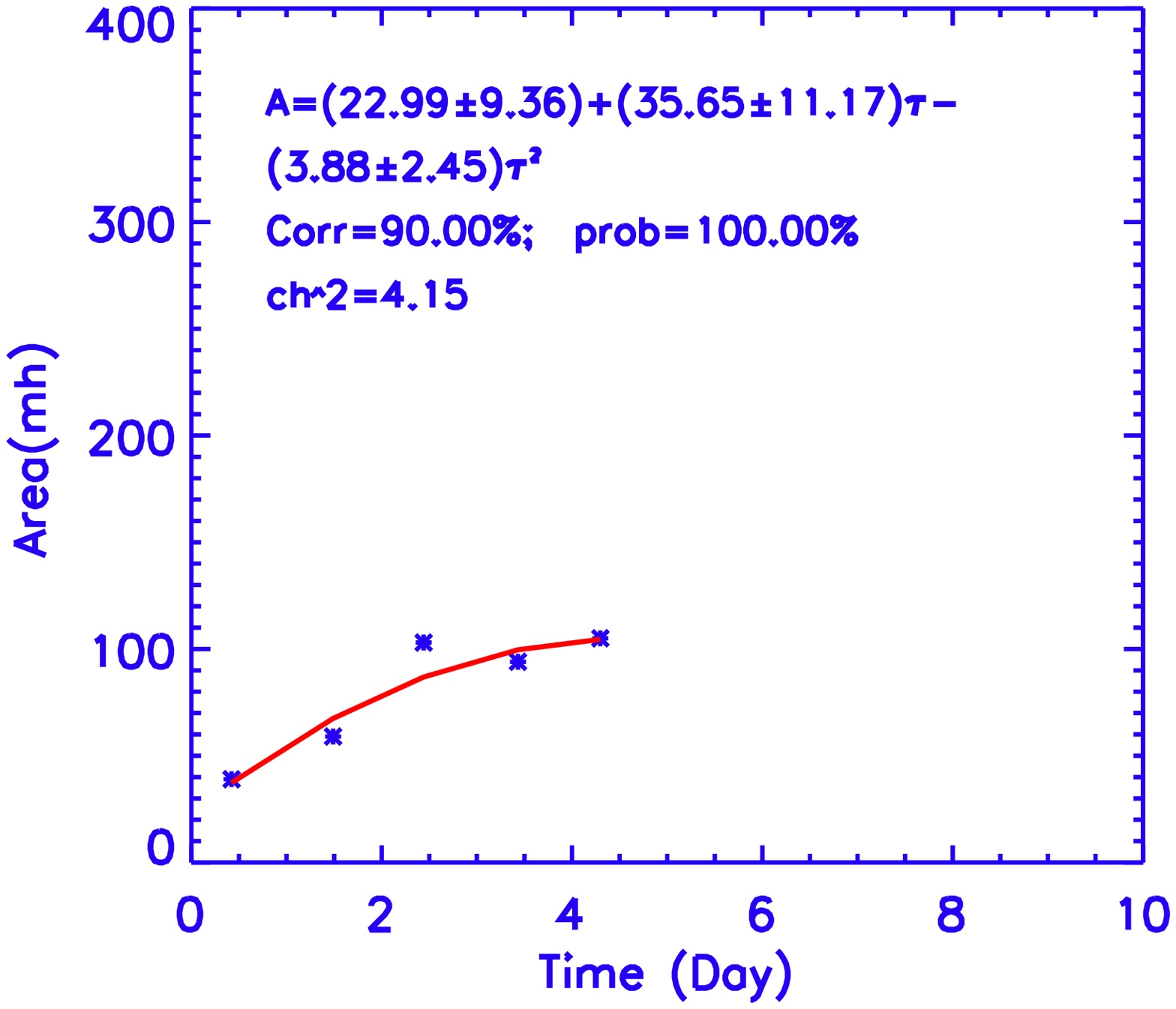}}\\
{\label{fig:exponential fit}\includegraphics[width=8.5cm,height=7.5cm]
{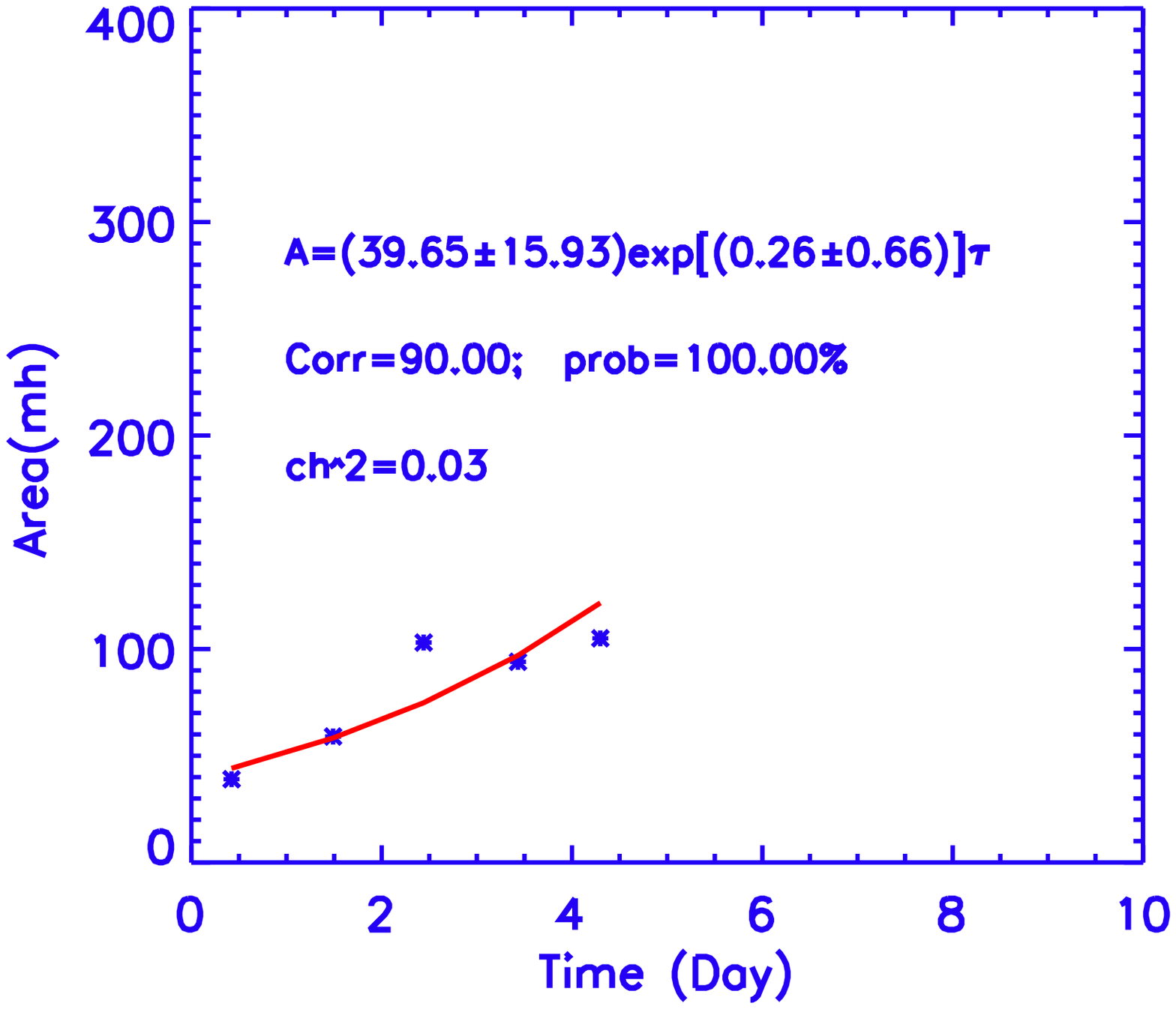}}
\caption{Evolution of growth of area A of non-recurrent sunspot group at a latitude region of 0 -10$^\circ$ that has lifespan of 8 days. Red line is theoretical area growth curve over plotted on the observed area growth curve (blue cross points).}
\end{figure}

\begin{figure}
\centering
{\label{fig:Linear fit}\includegraphics[width=7.5cm,height=7.5cm]
{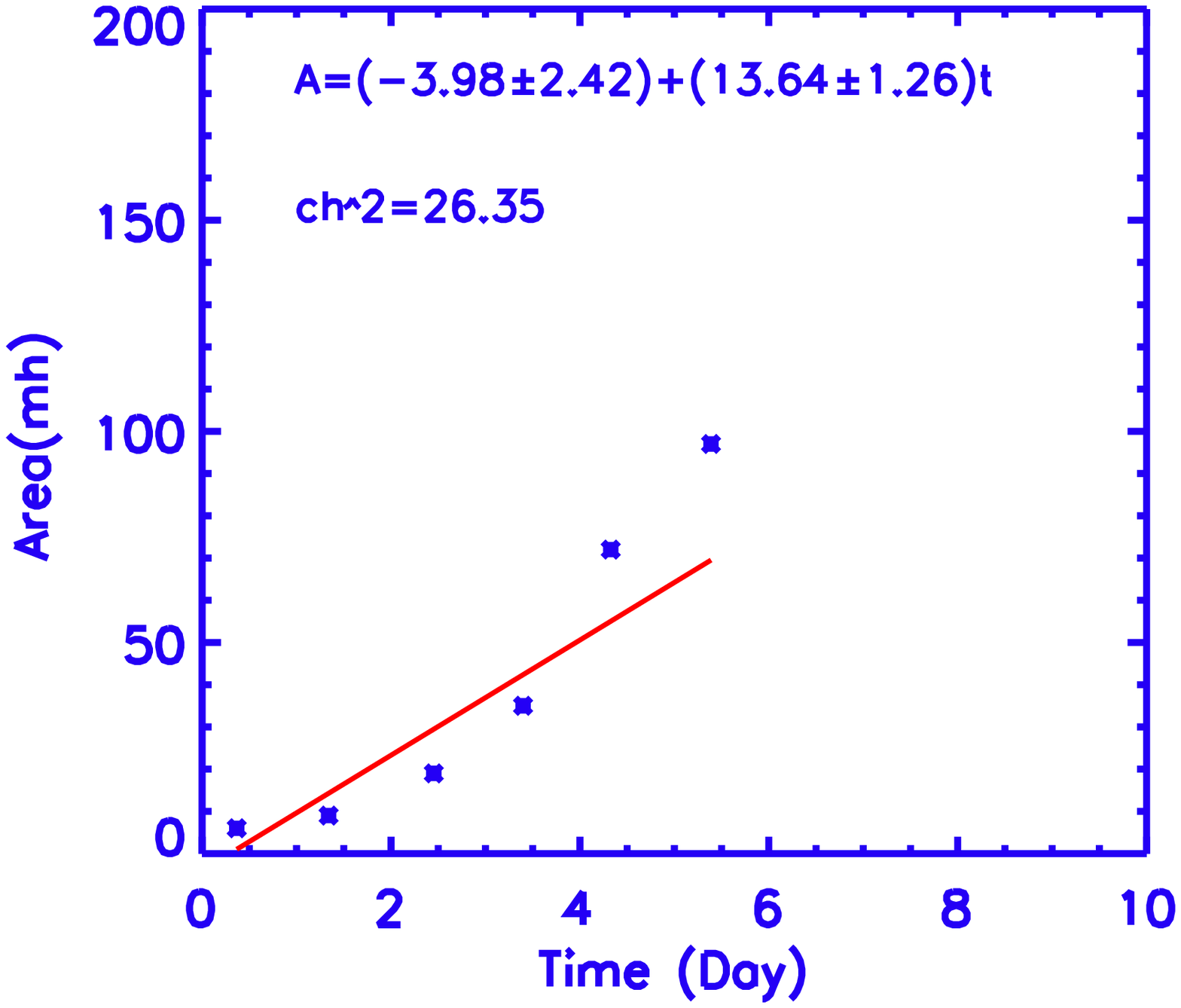}}
{\label{fig:Quadratic fit}\includegraphics[width=7.5cm,height=7.5cm]
{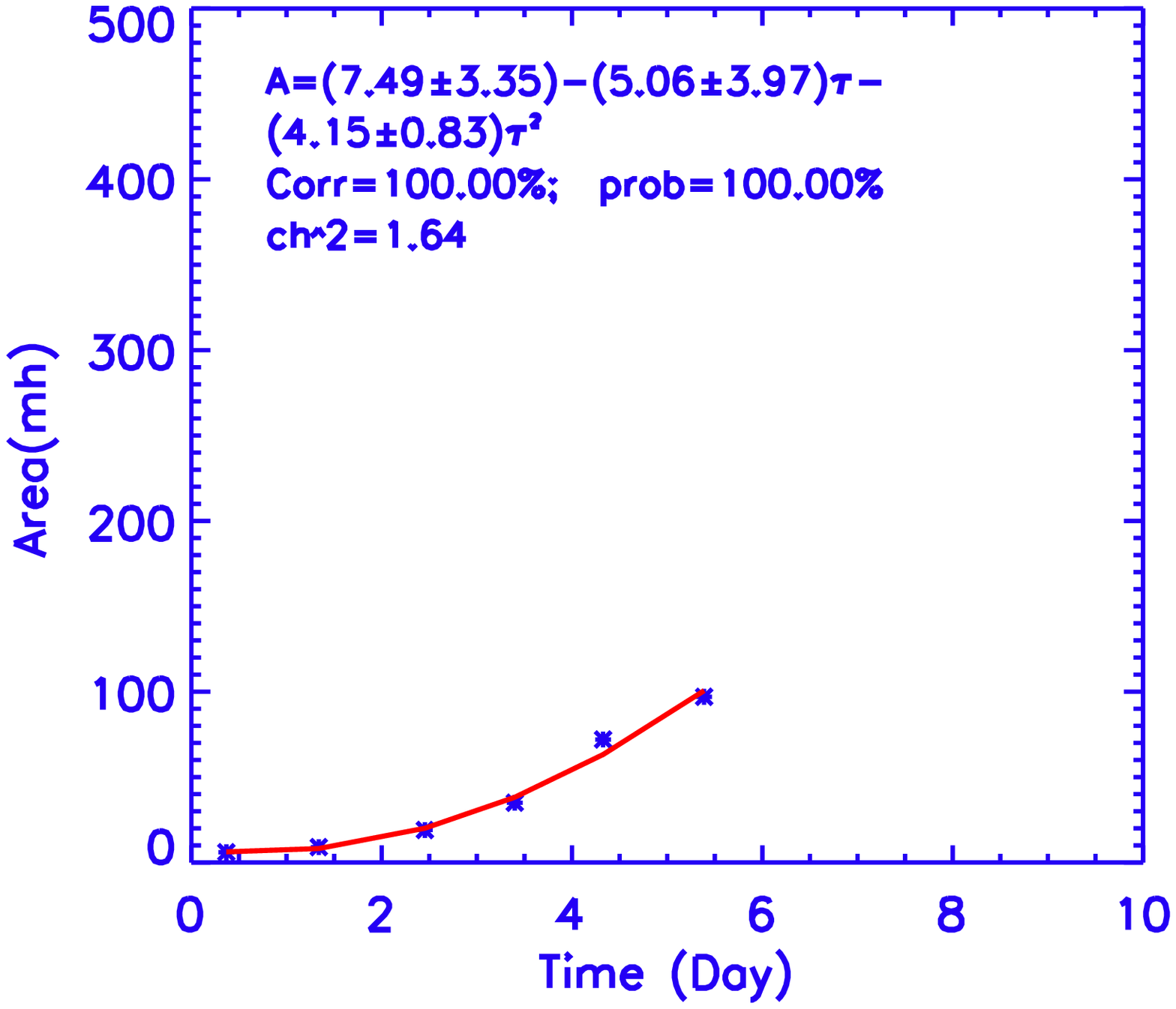}}\\
{\label{fig:Exponential fit}\includegraphics[width=8.5cm,height=7.5cm]
{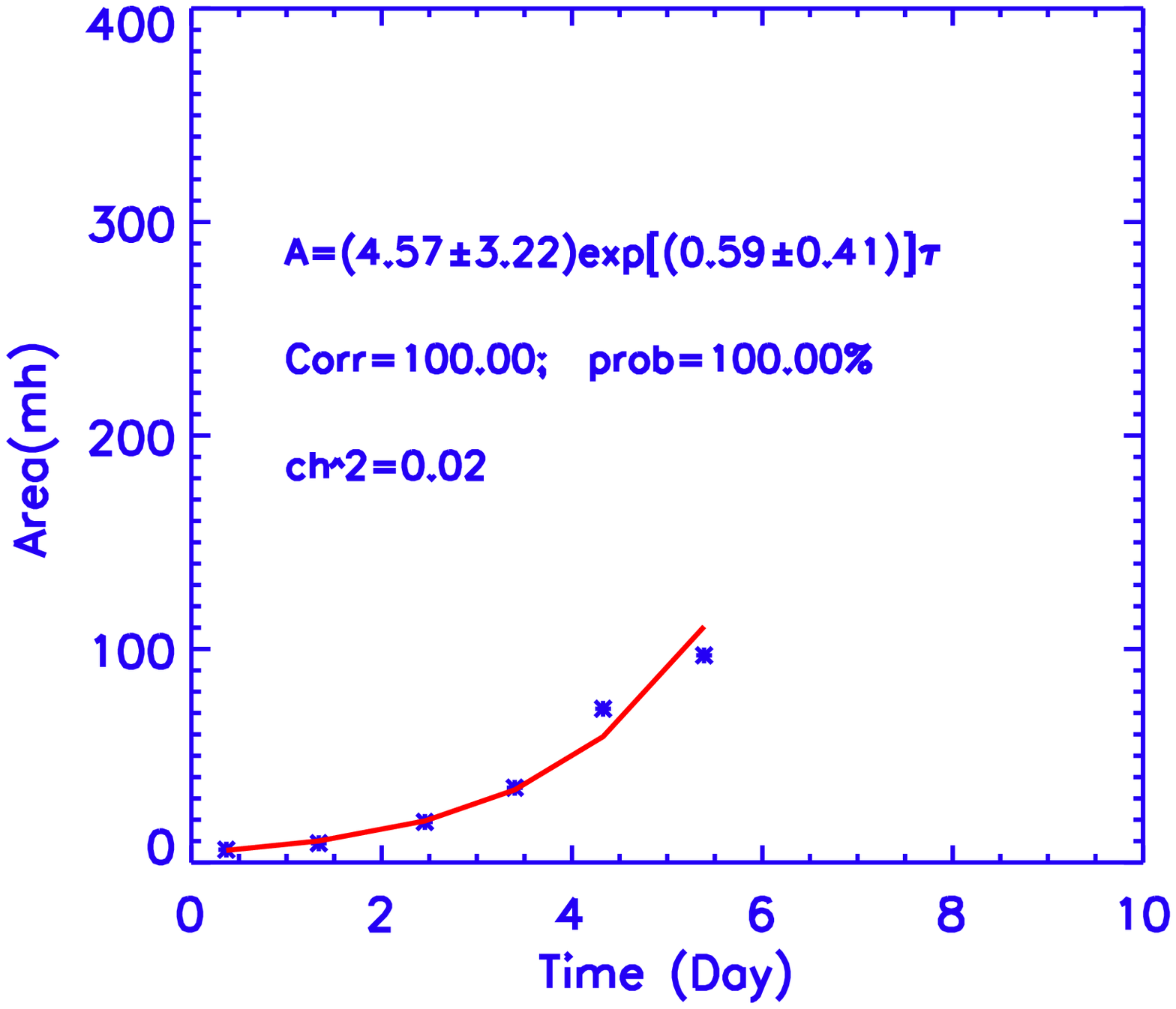}}
\caption{Evolution of growth of area A of non-recurrent sunspot group at a latitude region of 0 -10$^\circ$ that has lifespan of 8 days. Red line is theoretical area growth curve over plotted on the observed area growth curve (blue cross points).}
\end{figure}

\begin{figure}
\centering
{\label{fig:linear fit}\includegraphics[width=7.5cm,height=7.5cm]
{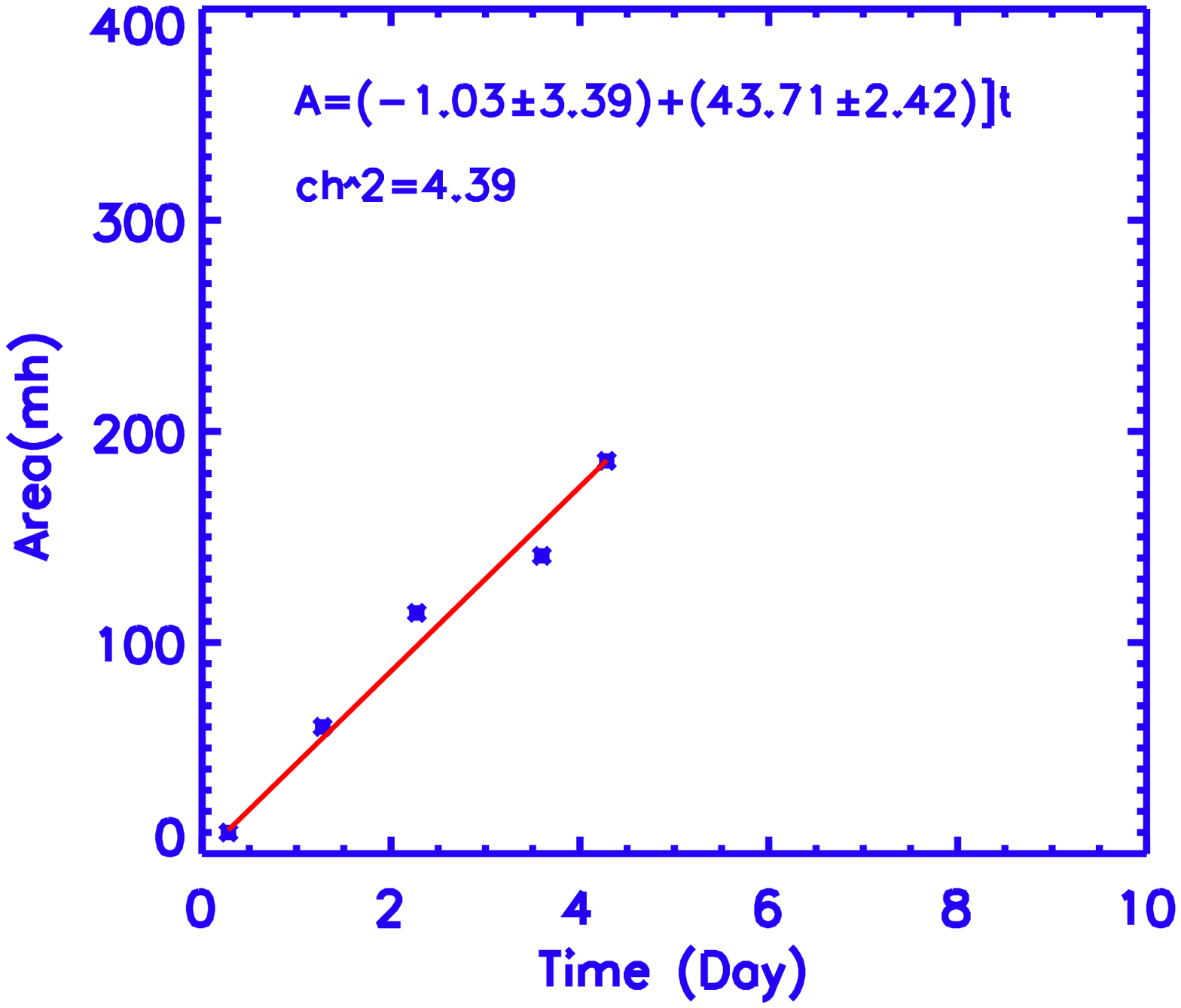}}
{\label{fig:quadratic fit}\includegraphics[width=7.5cm,height=7.5cm]
{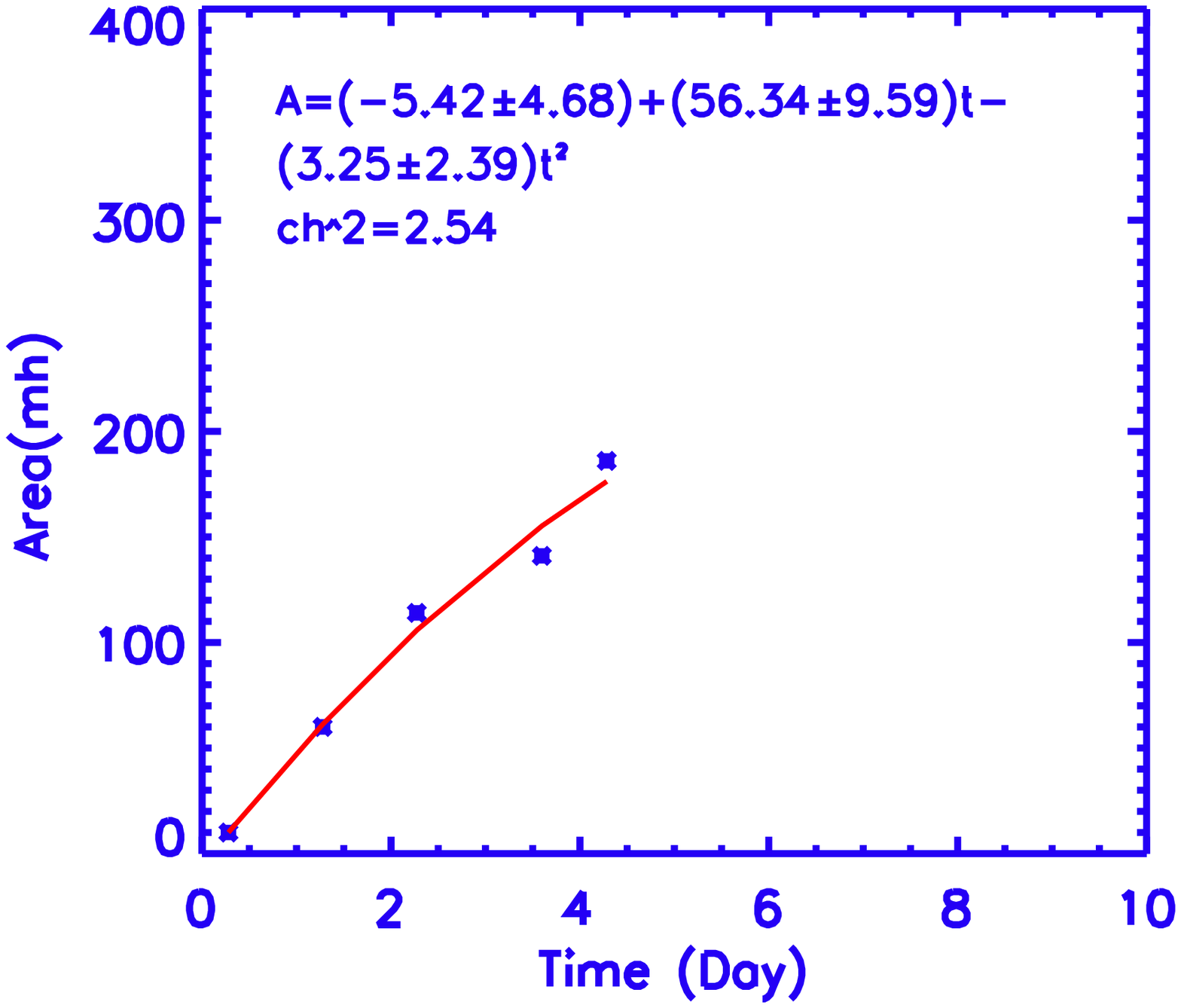}}\\
{\label{fig:exponential fit}\includegraphics[width=8.5cm,height=7.5cm]
{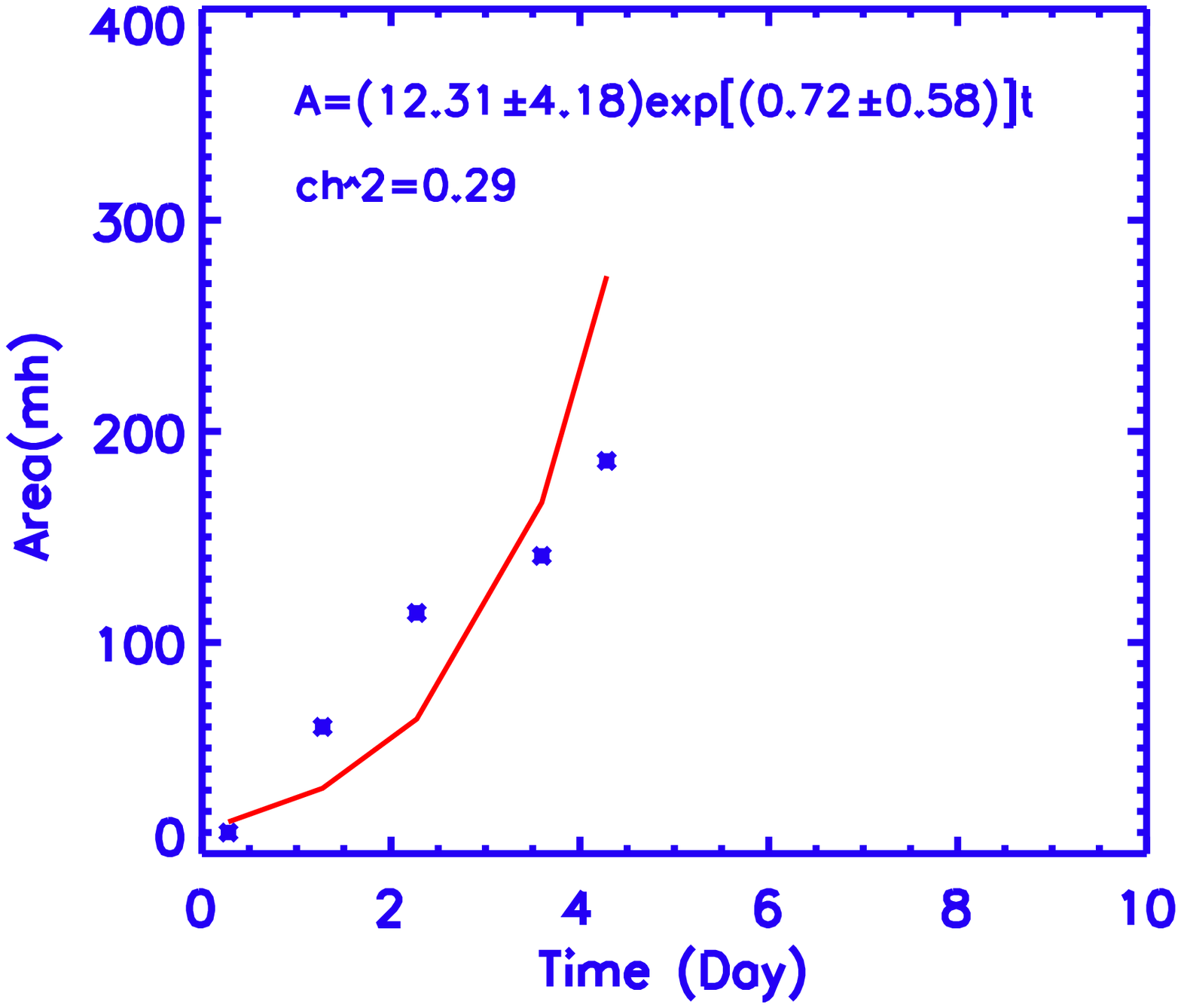}}
\caption{Evolution of growth of area A of non-recurrent sunspot group at a latitude region of 10 -20$^\circ$ that has lifespan of 8 days. Red line is theoretical area growth curve over plotted on the observed area growth curve (blue cross point).}
\end{figure}

\begin{figure}
\centering
{\label{fig:linear fit}\includegraphics[width=7.5cm,height=7.5cm]
{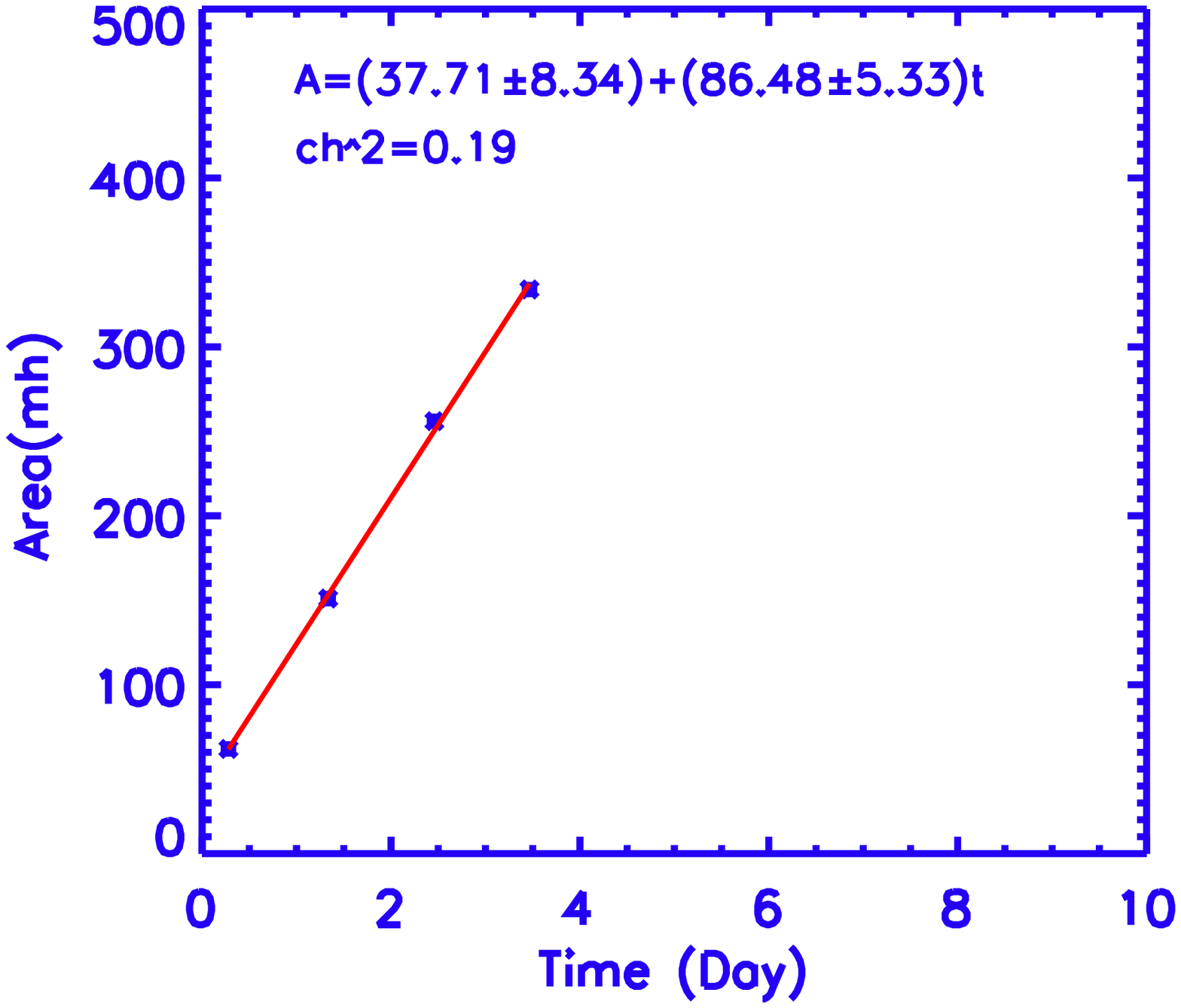}}
{\label{fig:quadratic fit}\includegraphics[width=7.5cm,height=7.5cm]
{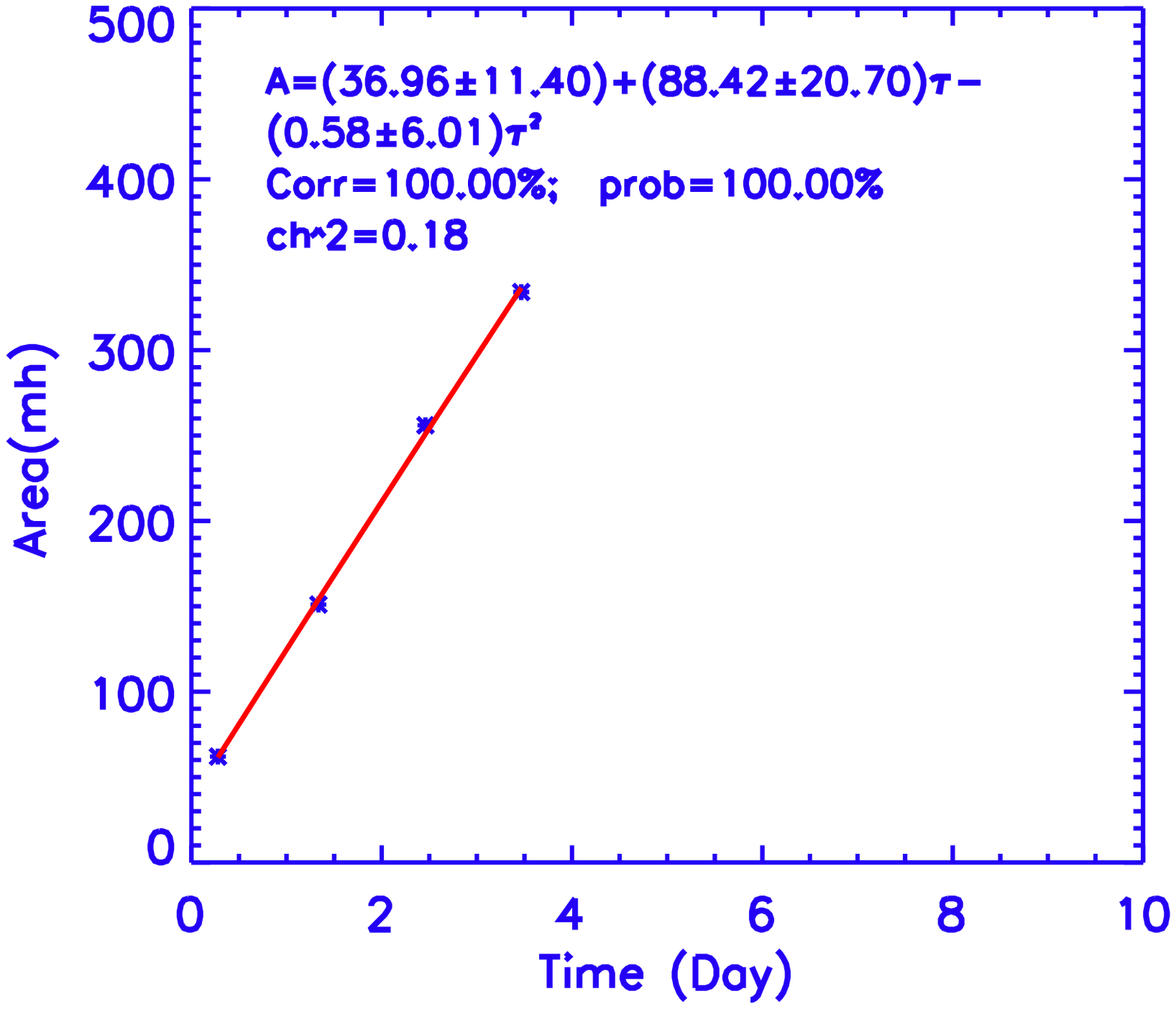}}\\
{\label{fig:exponential fit}\includegraphics[width=8.5cm,height=7.5cm]
{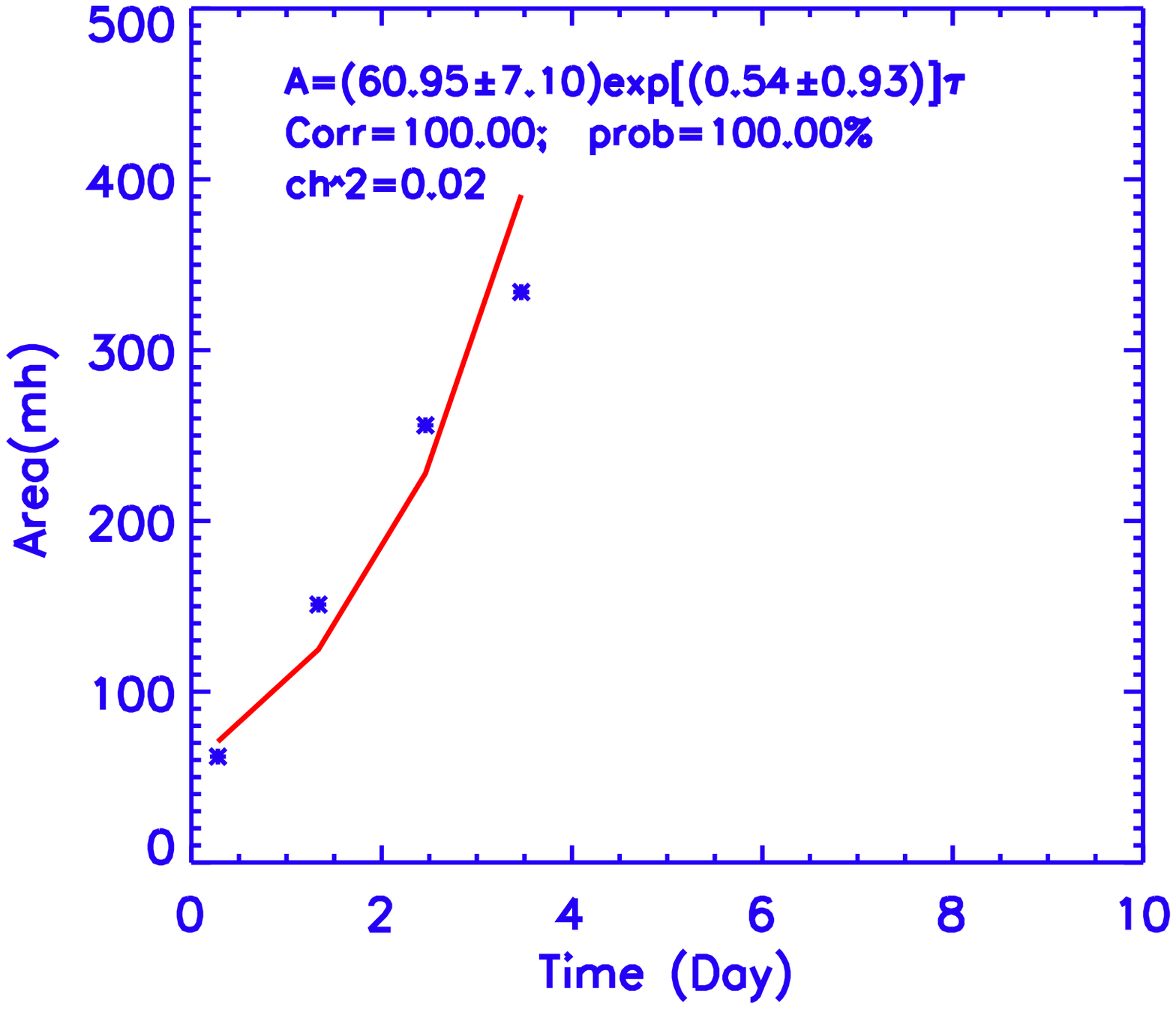}}
\caption{Evolution of growth of area A of non-recurrent sunspot group at a latitude region of 10 -20$^\circ$ that has lifespan of 10 days. Red line is theoretical area growth curve over plotted on the observed area growth curve (blue cross points).}
\end{figure}

\begin{figure}
\centering
{\label{fig:linear fit}\includegraphics[width=7.5cm,height=7.5cm]
{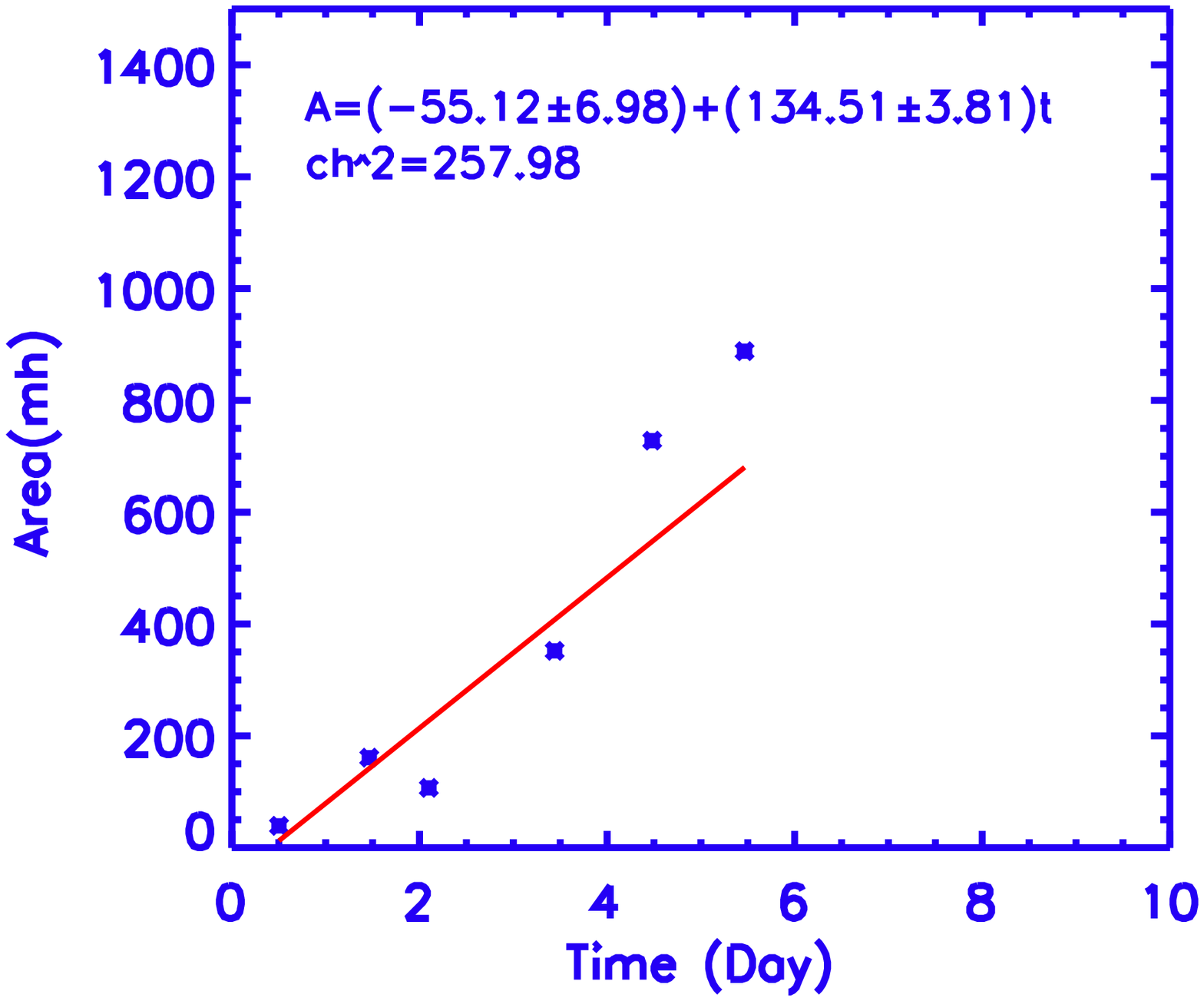}}
{\label{fig:quadratic fit}\includegraphics[width=7.5cm,height=7.5cm]
{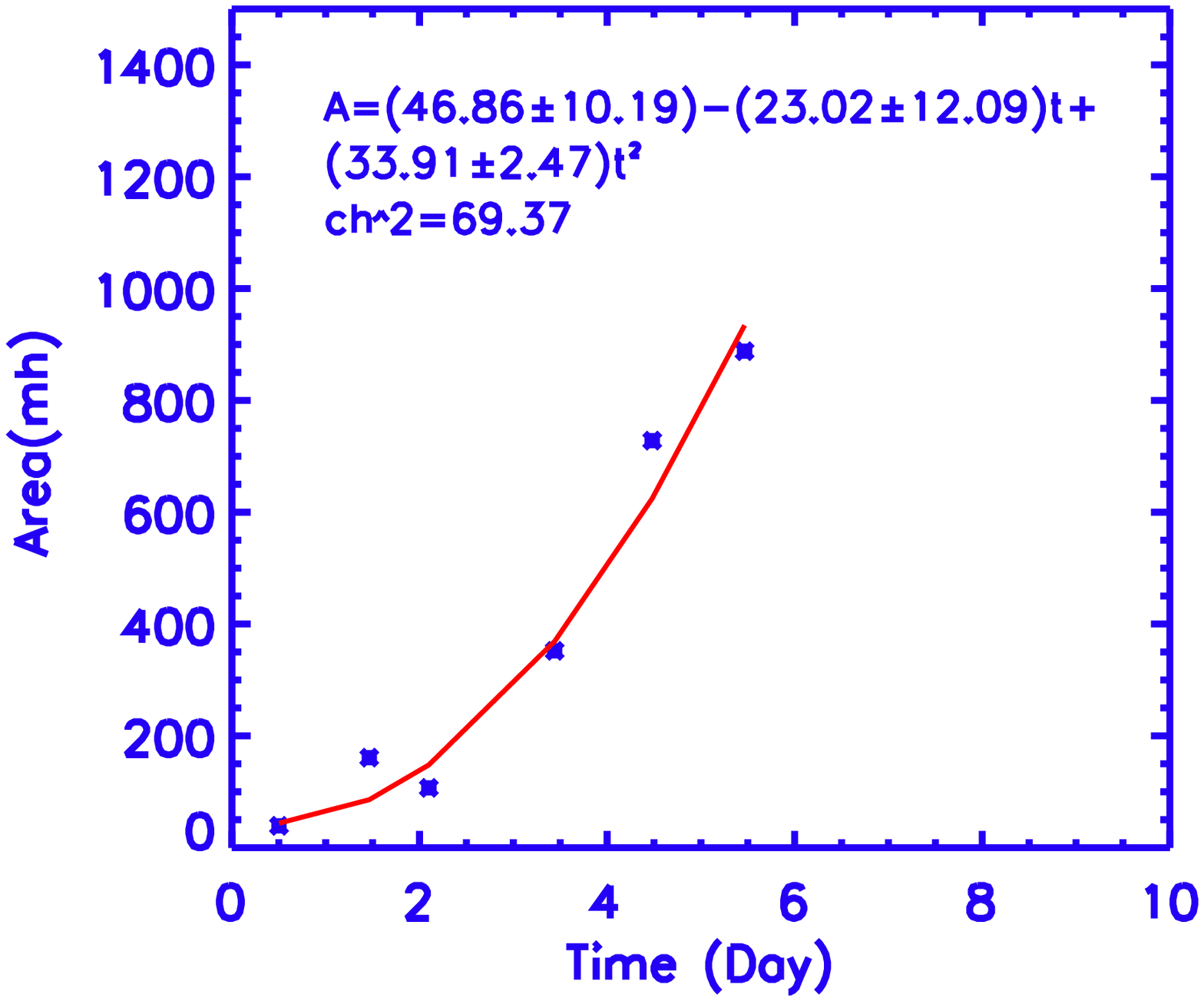}}\\
{\label{fig:exponential fit}\includegraphics[width=8.5cm,height=7.5cm]
{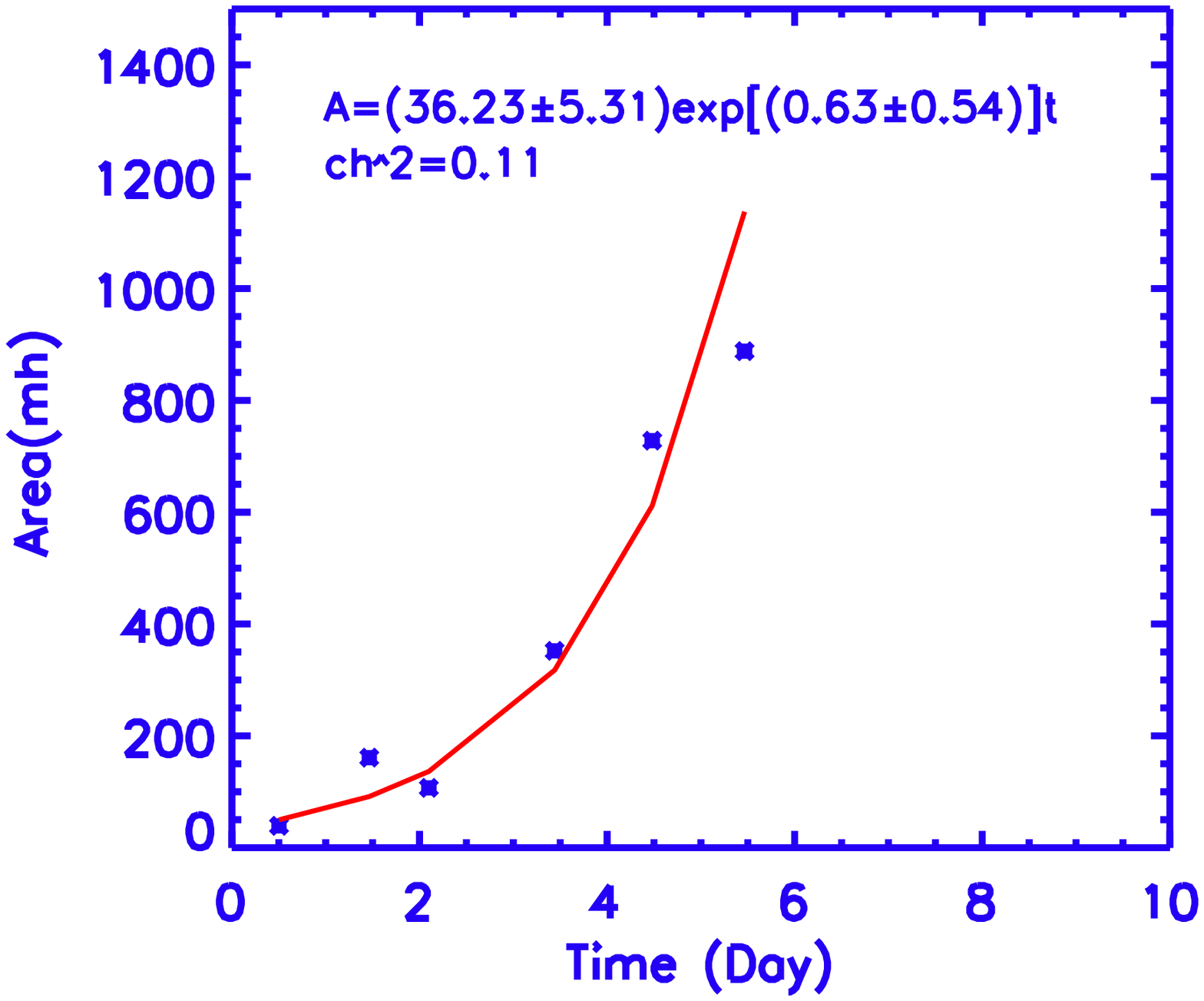}}
\caption{Evolution of growth of area A of non-recurrent sunspot group at a latitude region of 20 -30$^\circ$ that has lifespan of 10 days. Red line is theoretical area growth curve over plotted on the observed area growth curve (blue cross points).}
\end{figure}

\begin{figure}
\centering
{\label{fig:linear fit}\includegraphics[width=7.5cm,height=7.5cm]
{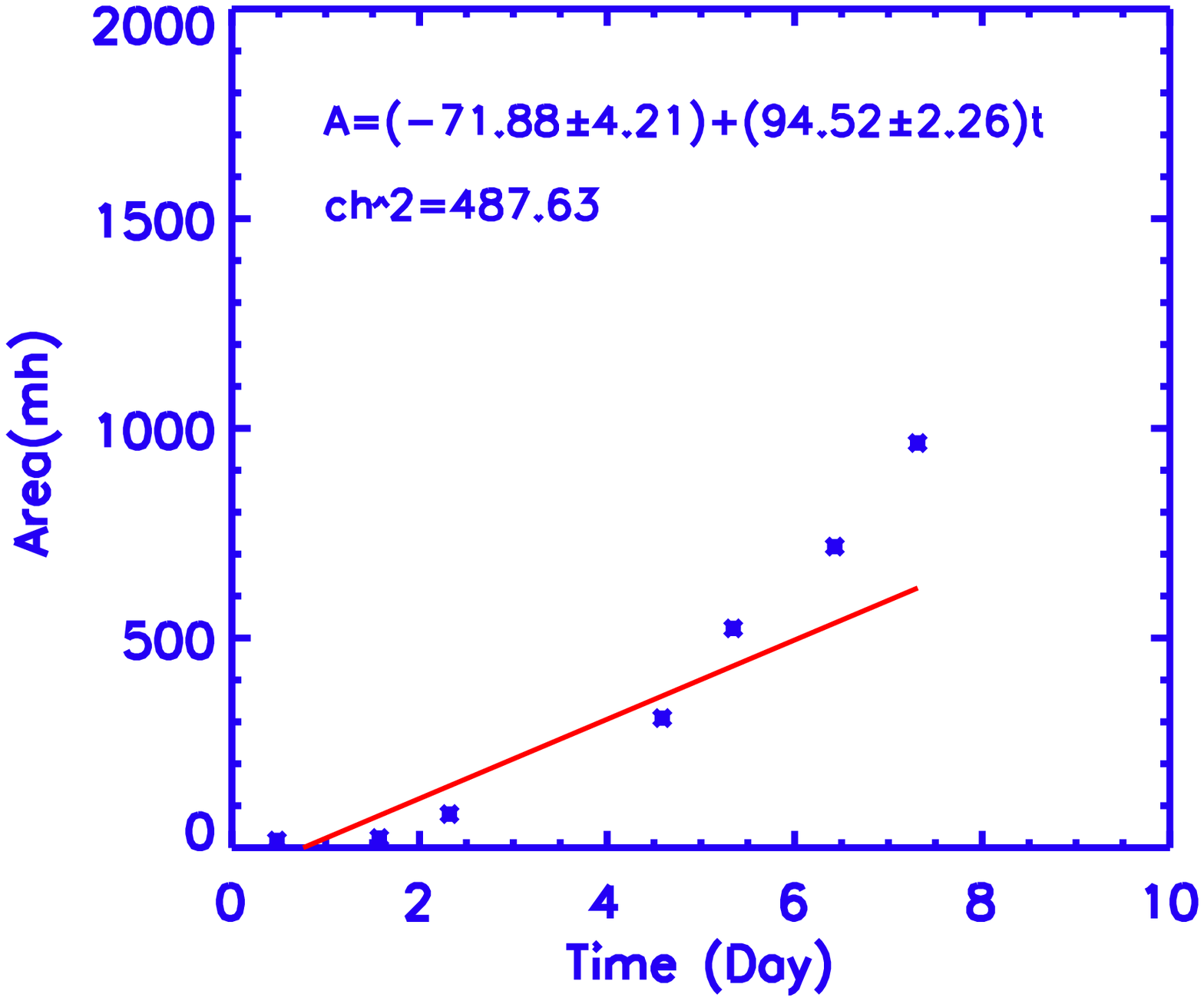}}
{\label{fig:quadratic fit}\includegraphics[width=7.5cm,height=7.5cm]
{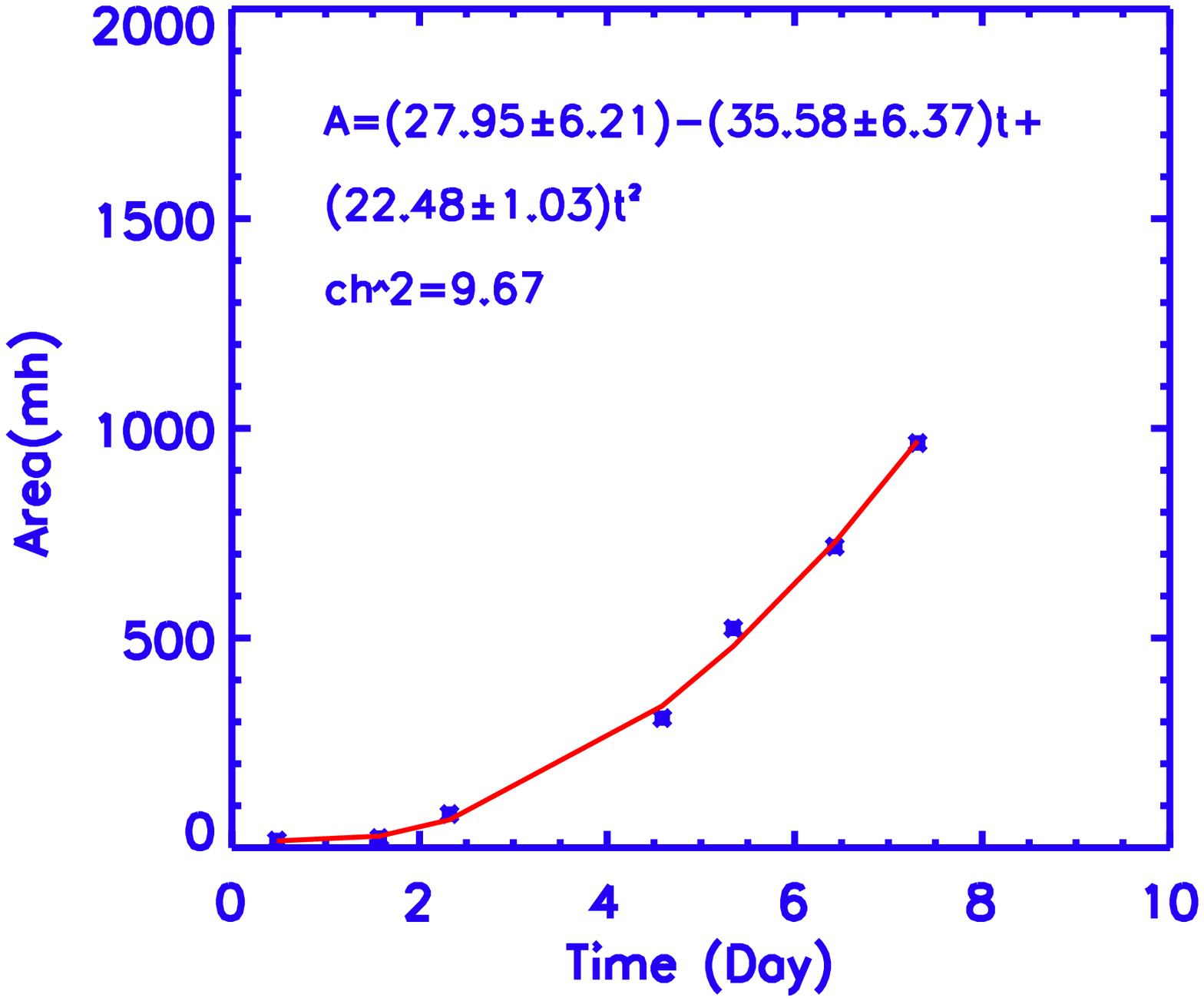}}\\
{\label{fig:exponential fit}\includegraphics[width=8.5cm,height=7.5cm]
{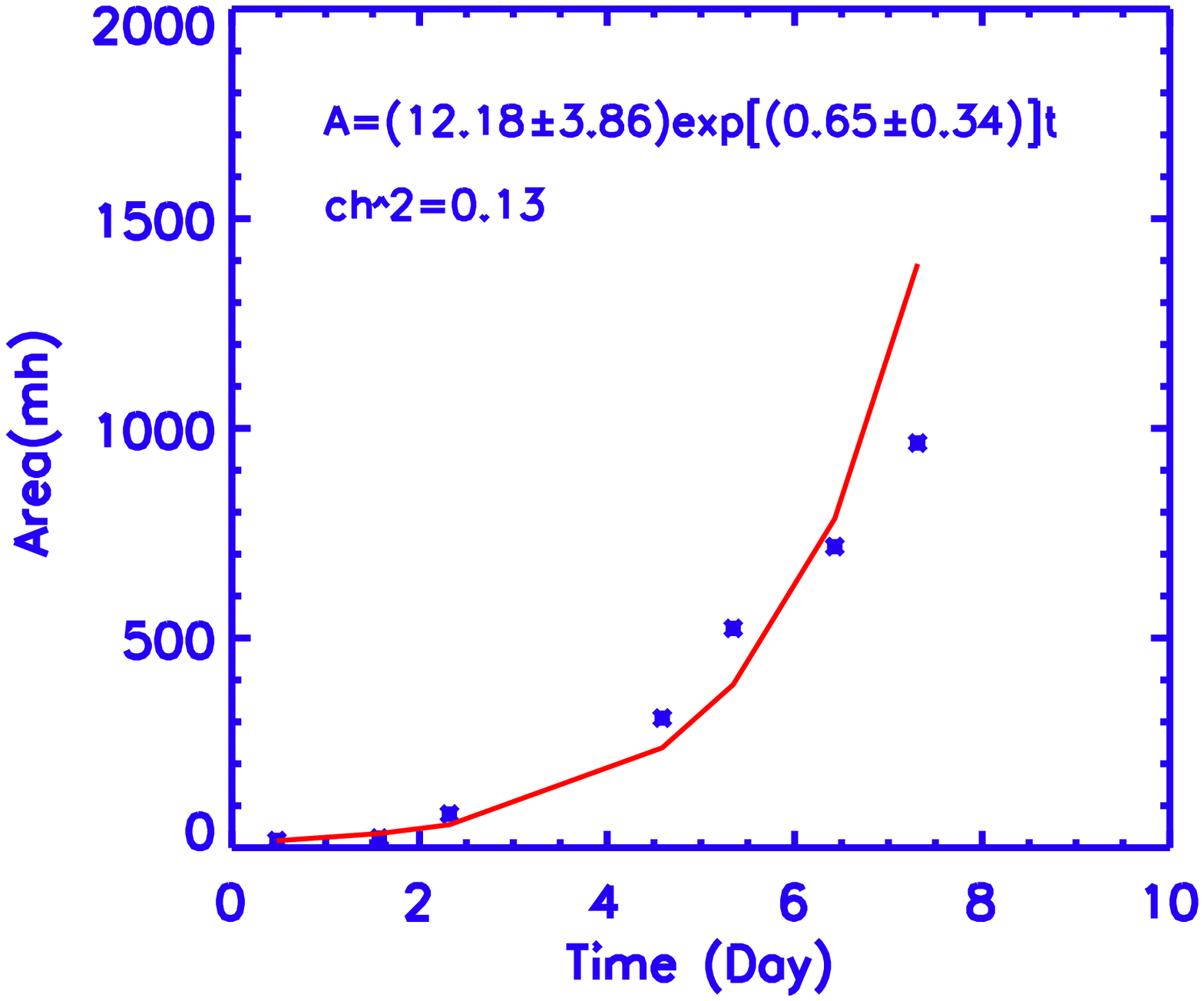}}
\caption{Evolution of growth of area A of non-recurrent sunspot group at a latitude region of 20 -30$^\circ$ that has lifespan of 10 days. Red line is theoretical area growth curve over plotted on the observed area growth curve (blue cross points).}
\end{figure}

\begin{figure}
\centering
{\label{fig:linear fit}\includegraphics[width=7.5cm,height=7.5cm]
{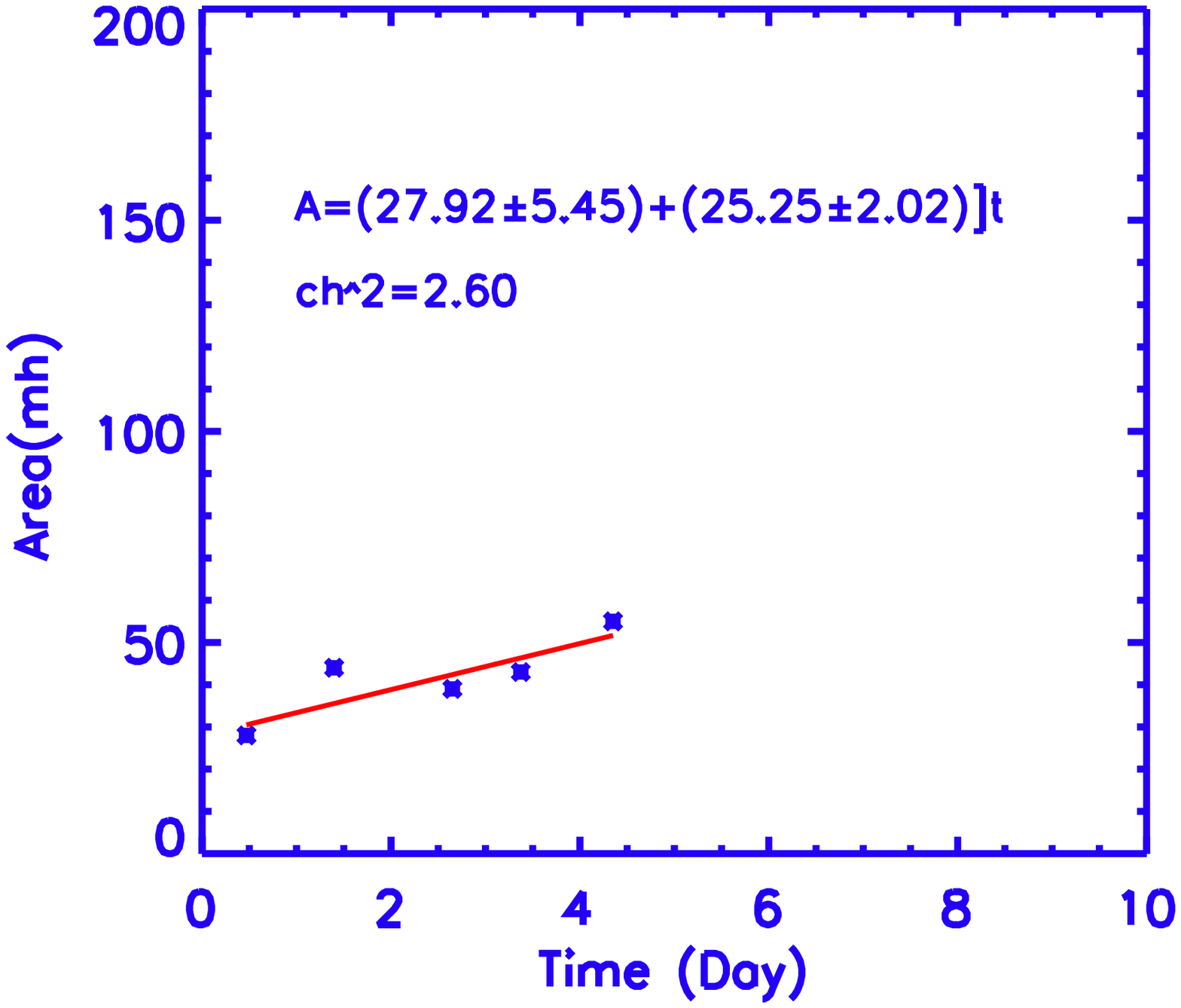}}
{\label{fig:quadratic fit}\includegraphics[width=7.5cm,height=7.5cm]
{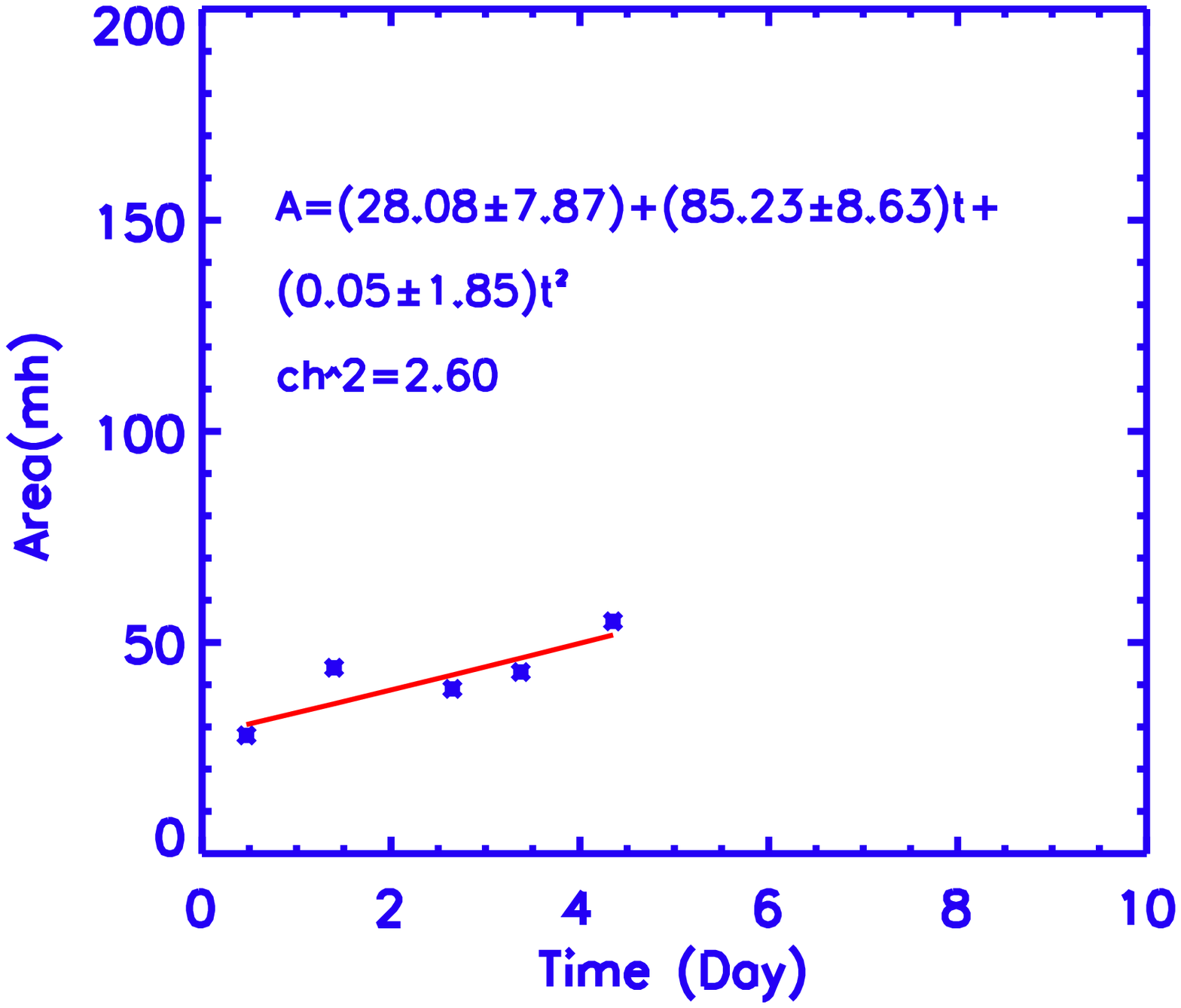}}\\
{\label{fig:exponential fit}\includegraphics[width=8.5cm,height=7.5cm]
{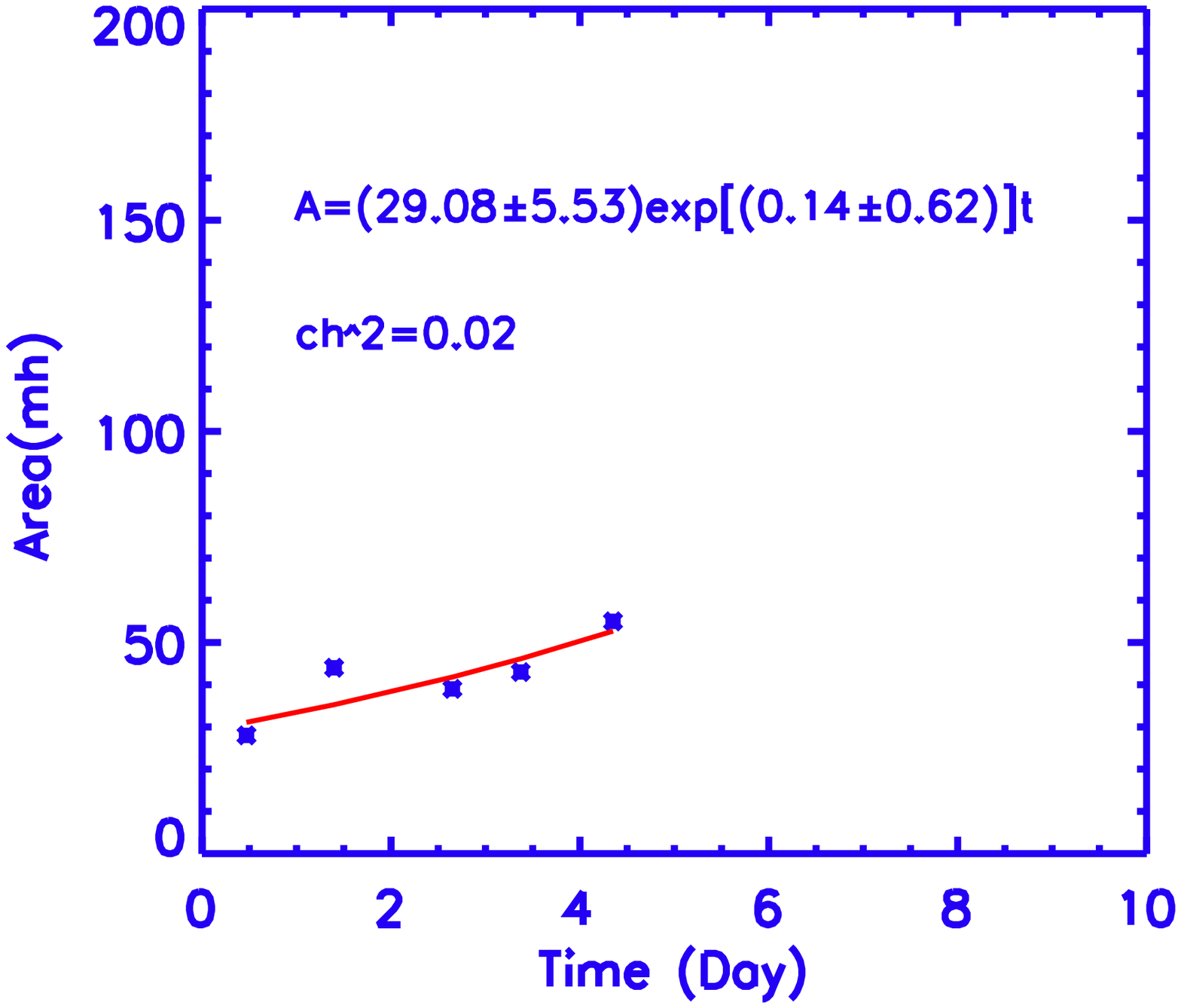}}
\caption{Evolution of growth of area A of non-recurrent sunspot group at a latitude region of 30 -40$^\circ$ that has lifespan of 8 days. Red line is theoretical area growth curve over plotted on the observed area growth curve (blue cross points).}
\end{figure}

\begin{figure}
\centering
{\label{fig:linear fit}\includegraphics[width=7.5cm,height=7.5cm]
{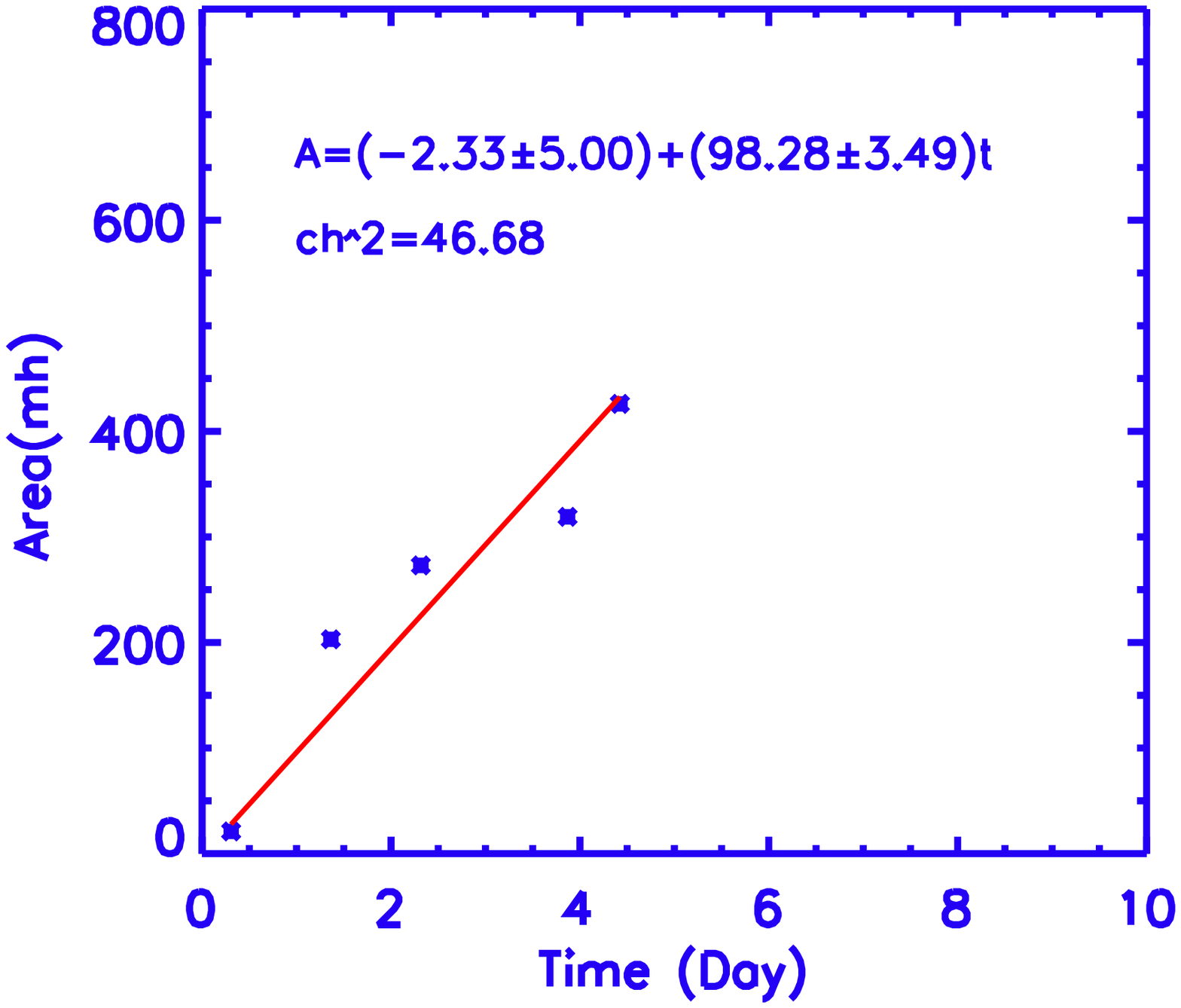}}
{\label{fig:quadratic fit}\includegraphics[width=7.5cm,height=7.5cm]
{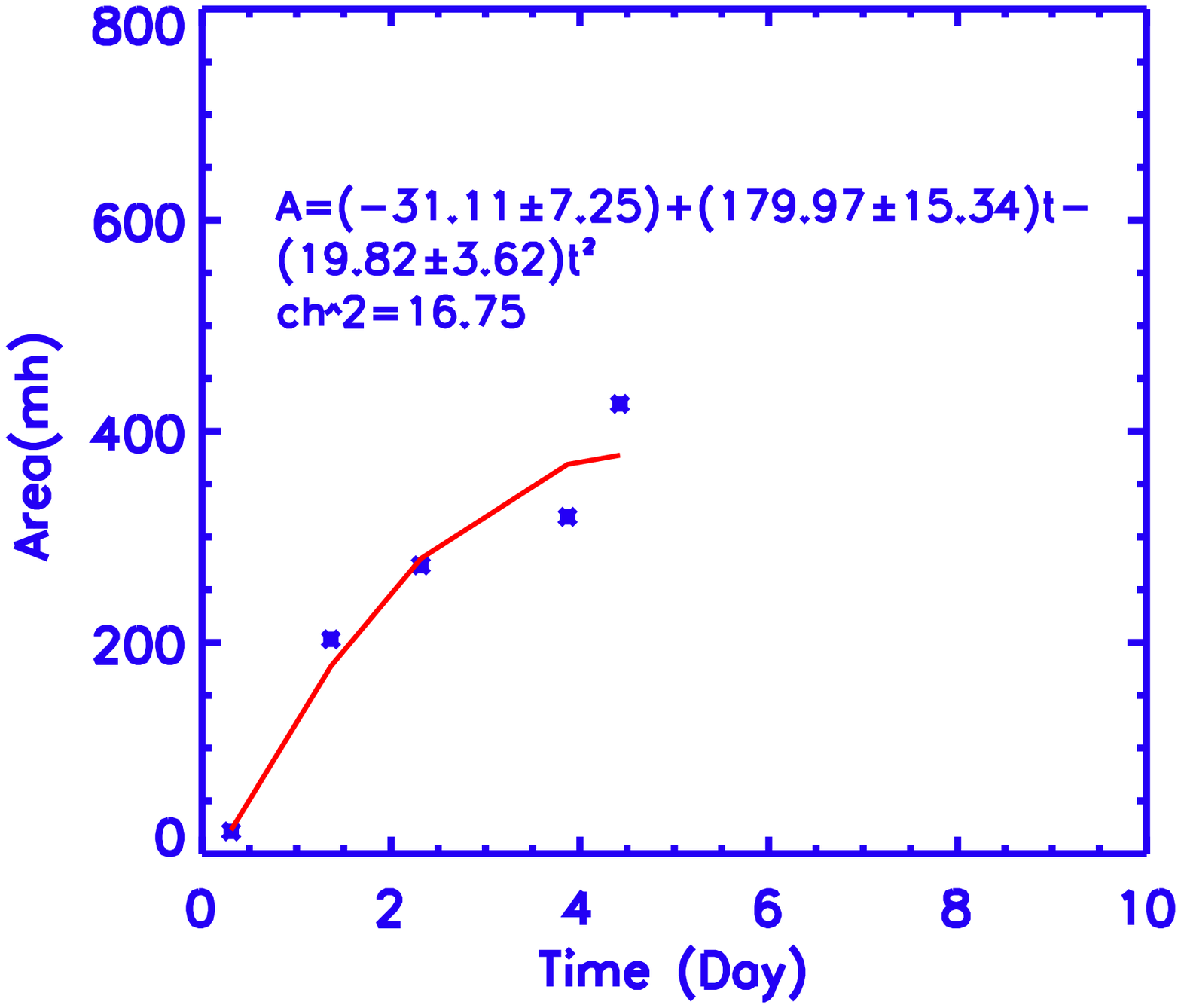}}\\
{\label{fig:exponential fit}\includegraphics[width=8.5cm,height=7.5cm]
{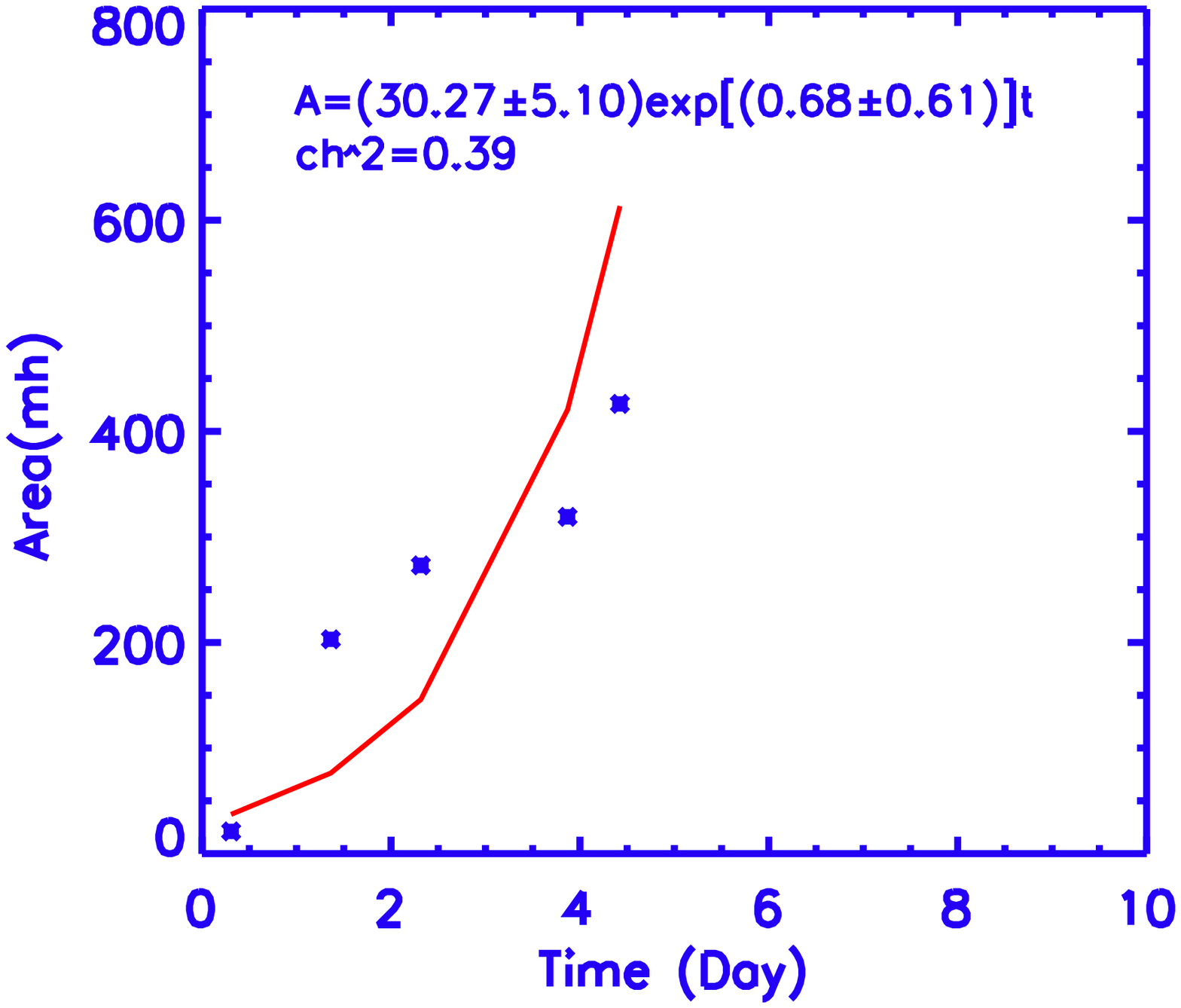}}
\caption{Evolution of growth of area A of non-recurrent sunspot group at a latitude region of 30 -40$^\circ$ that has lifespan of 9 days. Red line is theoretical area growth curve over plotted on the observed area growth curve (blue cross point).}
\end{figure}
and 
\begin{equation}
{{\partial T}\over{\partial \phi}} = 
{{\partial T}\over{\partial t}} {{\partial t}\over{\partial \phi}} =
{{\partial T}\over{\partial t}}{1 \over {\Omega}}.
\end{equation}
Hence,
\begin{equation}
(T{{\partial \Omega}\over{\partial \phi}} - \Omega{{\partial T}\over{\partial \phi}}) = ({{T}\over {\Omega}}{{\partial \Omega}\over{\partial t}} - {{\partial T}\over{\partial t}}).
\end{equation}
 With these equations, equation (6) can be written as
\begin{equation}
2 {{\partial T}\over{\partial t}} = ({{UTcot\theta}\over{r}}) + (P sin \theta{{\partial \Omega}\over{\partial \theta}} - {{U}\over{r}}{{\partial T}\over{\partial \theta}})
+({{T}\over {\Omega}}{{\partial \Omega}\over{\partial t}}).
\end{equation}
Perturb this induction equation by
taking the variables $\Omega = \Omega_0 + \Omega^{'}$, $T = T_0 + T^{'}$ and $=U_{0}+U^{'}$
such that ${{\partial \Omega_{0}}\over{\partial t}}={{\partial T_{0} }\over{\partial t}} =
{{\partial U_{0} }\over{\partial t}}=0$
and magnitudes of fluctuating $\Omega^{'}$, $T^{'}$ and $^{'}$ components are
assumed to be very small compared to steady parts $\Omega_0$ and $T_0$.
This condition also implies that magnitudes of products of the fluctuating
components are nearly zero. Further it is assumed that poloidal
component of the magnetic field structure $P$ is constant
and it's magnitude is very small compared to magnitude of toroidal
magnetic field structure. This reasonable assumption
is consistent with the observed strength of solar magnetic field structure
that during 11 years period strength of poloidal field
structure ( $\sim$ 1 G) is $<<$ strength of toroidal
magnetic field structure ( $\sim 10^{3}$ G). That means the fluctuating
term $(P sin \theta{{\partial \Omega^{'}}\over{\partial \theta}})$
is neglected. Hence resulting time dependent part of 
toroidal component of global magnetic field structure
for the Alfven wave perturbations along the direction of rotation is given as follows
\begin{equation}
2 {{\partial T^{'}}\over{\partial t}} = ({{U_{0}T^{'}+U^{'}T_{0}}\over{r}})cot\theta 
 - ({{U_{0}}\over{r}}{{\partial T^{'}}\over{\partial \theta}}
 + {{U^{'}}\over{r}}{{\partial T_{0}}\over{\partial \theta}})
+({{T_{0}}\over {\Omega_{0}}}{{\partial \Omega^{'}}\over{\partial t}}) .
\end{equation}
Derivative ${{\partial T^{'}}\over{\partial \theta}}$ can be
modified as follows
\begin{equation}
{{\partial T^{'}}\over{\partial \theta}} = 
{{\partial T^{'}}\over{\partial t}} {{\partial t}\over{\partial \theta}} = 
{{\partial T^{'}}\over{\partial t}} {{1}\over{U_{0}}} \ ,
\end{equation}
where $U_{0}$ is steady part of meridional circulation.
 
On both sides of the equation (13), multiply the term $ A = \pi S^{2}$ of the flux tube
( where $S$ is radius of the tube at a particular depth) and resulting equation for rate of change
of magnetic flux or area (as area of the sunspot is directly
proportional to magnetic flux) of the sunspot is given as
follows
\begin{equation}
{{\partial A}\over{\partial t}} =
{U_{0}cot\theta A \over{2r+1}} + {S^{2} U^{'} \over{2r+1}}(T_{0}cot\theta 
- {{\partial T_{0}}\over{\partial \theta}})   
+ {r S^{2} T_{0}  \over{(2r+1)\Omega_{0}}}{{\partial \Omega^{'}}\over{\partial t}} .
\end{equation}
This equation suggests that rate of change of area of the sunspot
is a function of steady parts of poloidal and toroidal
 velocity field structures, radial variations in fluctuations of the
meridional velocity and steady part of toroidal component 
of magnetic field structure respectively. Although momentum
equation is necessary (for the hydrostatic equilibrium of internal structure of the sun, as
 ${{\partial \Omega^{'}}\over{\partial t}}$
is proportional to fluctuating parts of advective terms, Lorentzian force
and variation in the second derivative of angular velocity), as 
fluctuating terms are assumed to 
be small (although in principle not to be neglected),
radial variation of last two terms in RHS is neglected. Hence, with the
initial conditions that at time $t=0$, area $A=A_{0}$ (initial area), solution 
yields the following relationship between increase of flux
tube area (while it raises in the positive rotational gradient) with
respect to time.
\begin{equation}
A(t) = A_{0}e^{(U_{0}cot\theta )t \over{2r+1}}
\end{equation}  
As for Alfven wave perturbations that are opposite to the direction
of angular velocity, solution for growth of the sunspot is
\begin{equation}
A(t) = A_{0}e^{(-U_{0}cot\theta )t \over{2r+1}}    \, .
\end{equation}
Hence, in the region of positive rotational gradient,
simultaneous growth and decay of the Alfven wave
perturbations exist yielding net exponential growth of the sunspot.
Another interesting property of solution (equation 16)
is that exponent of the growth part depends upon
magnitude of meridional velocity $U_{0}$, $cot \theta$
and the depth of the foot point of the flux tube where
it is anchored. It is not known how the meridional
velocity varies with depth and it is assumed to be constant.
Thus as time progresses, due to buoyancy, anchored feet
lifts from interior in the positive rotational gradient
(until it reaches maximum angular velocity at the depth
$0.935R_\odot$) with an exponential growth of area
of the sunspot. 

If one keeps the ratio ${U_{0}\over{2r+1}}$
constant at a particular depth (say near the surface),
exponent of solution for growth of area is directly proportional
to $cot \theta$. That means by the property of $cot \theta$ function, spots at the lower co-latitudes $\theta$
 (or higher heliographic latitudes) grow very fast compared to the spots
that grow at the lower co-latitudes ( or lower  heliographic latitudes,
i,e., near the equator). This important property of sunspot's growth
will be tested in the following sections.

Once sunspot's anchoring feet enters the
negative rotational gradient, the picture will
be different and it will be known from the next section
that area of the sunspot decays exponentially and ultimately 
disappears on the surface.    

\section{Solution for decay of the sunspot}
After substituting equations (1) and (2) in equation (5)
and also by satisfying the continuity equation (4), 
resulting toroidal
component of the induction equation with a sink term alone in spherical coordinates is
\begin{equation}
{\partial T\over\partial t} = \eta \big [{ 1 \over {r^2 }} {\partial^{2} T \over\partial \theta^{2}} 
+ { 1 \over {r^2}{ sin^2 \theta}}
{\partial^2 T\over \partial \phi^2}
+ {cot \theta \over {r^2}} {\partial T \over\partial \theta}
- {T \over {r^2 sin^{2} \theta}} \big] .
\end{equation}
Adopting a similar method in the previous section, this equation can be
transformed into following equation for steady part of toroidal 
component of the induction equation
\begin{equation}
{\partial^{2} T\over\partial t^{2}}-
(r^{2}sin^{2}\theta {\partial \Omega \over \partial t}
+{\Omega^{2} r^{2} sin^{2} \theta \over {\eta}}){\partial T\over\partial t}
+{\Omega sin^{2}\theta \over \eta} {\partial^{2} T \over \partial \theta^{2}}
+{\Omega sin^{2}\theta cot \theta \over \eta}{\partial T \over \partial \theta}
-{\eta T \over r^{2} sin^{2}\theta} = 0.
\end{equation}
Following similar perturbation method in the previous section, we get
the following time dependent component of the toroidal component of magnetic field
structure
\begin{equation}
A_{1}{\partial^{2} T^{'}\over\partial t^{2}}+A_{2}{\partial T^{'}\over\partial t}+ A_{3}T^{'} =0 \ ,
\end{equation}

\begin{figure}
\centering
{\label{fig:linear fit}\includegraphics[width=7.5cm,height=8.0cm]
{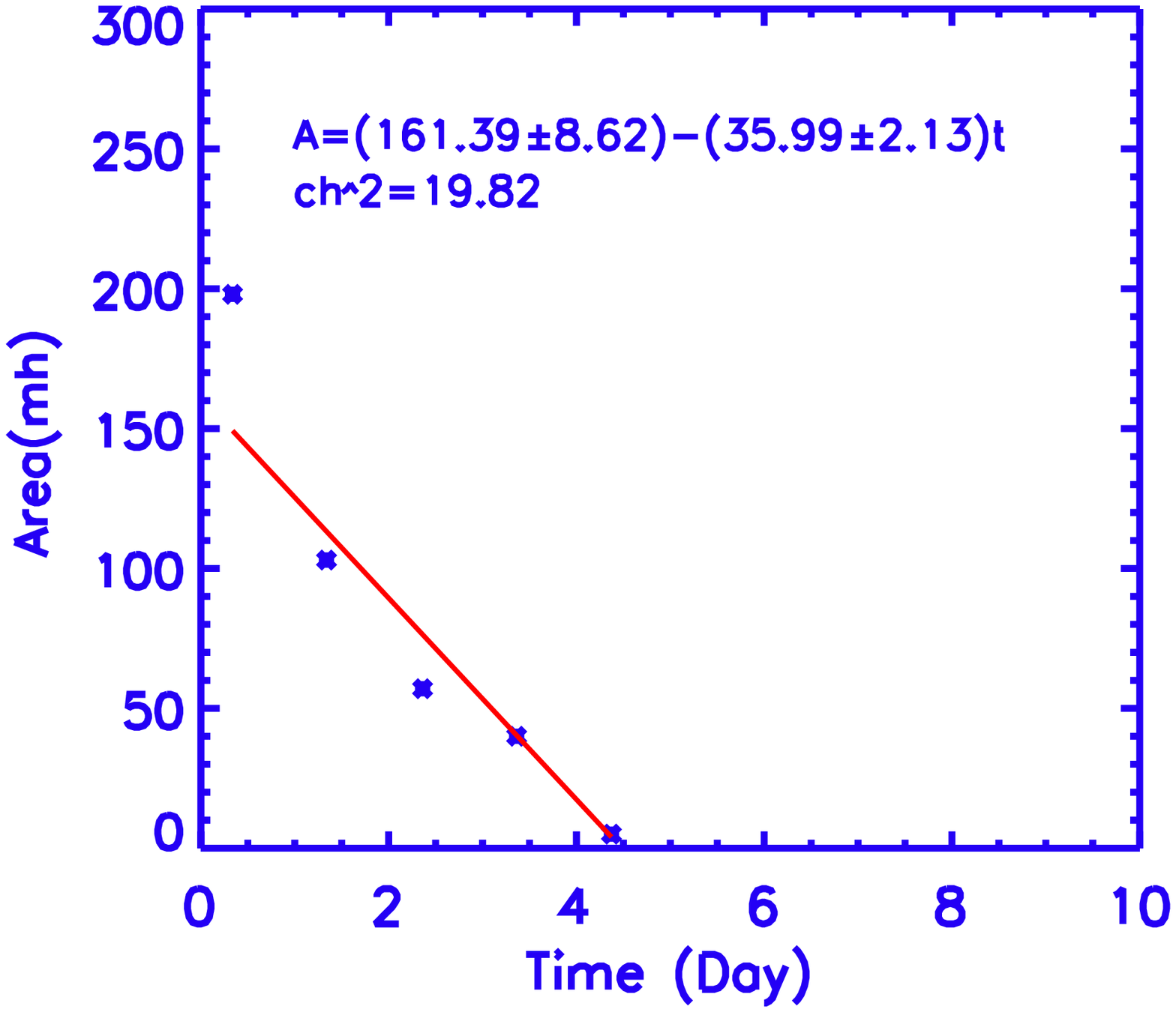}}
{\label{fig:quadratic fit}\includegraphics[width=7.5cm,height=8.0cm]
{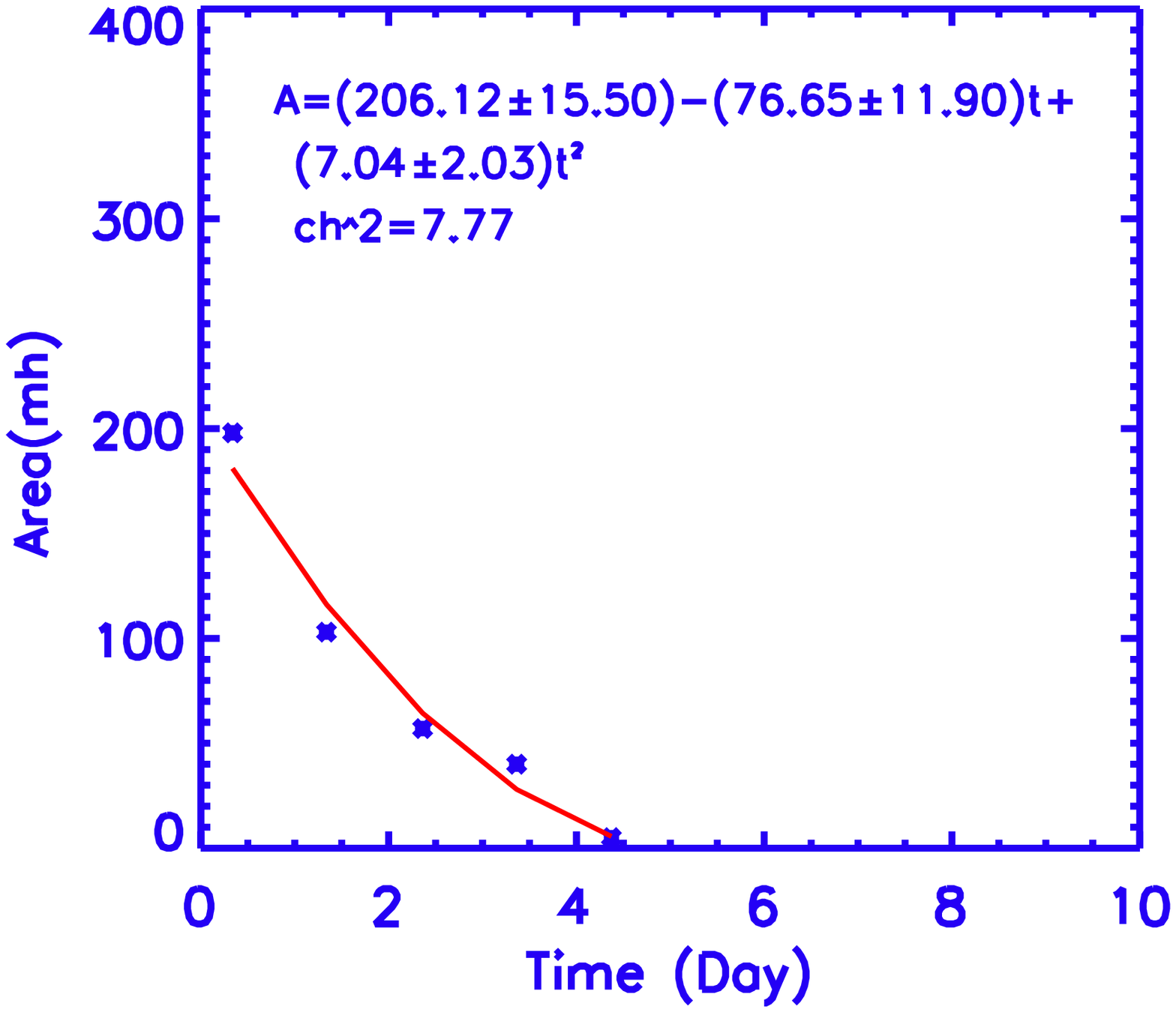}}\\
{\label{fig:log-normal fit}\includegraphics[width=7.5cm,height=8.0cm]
{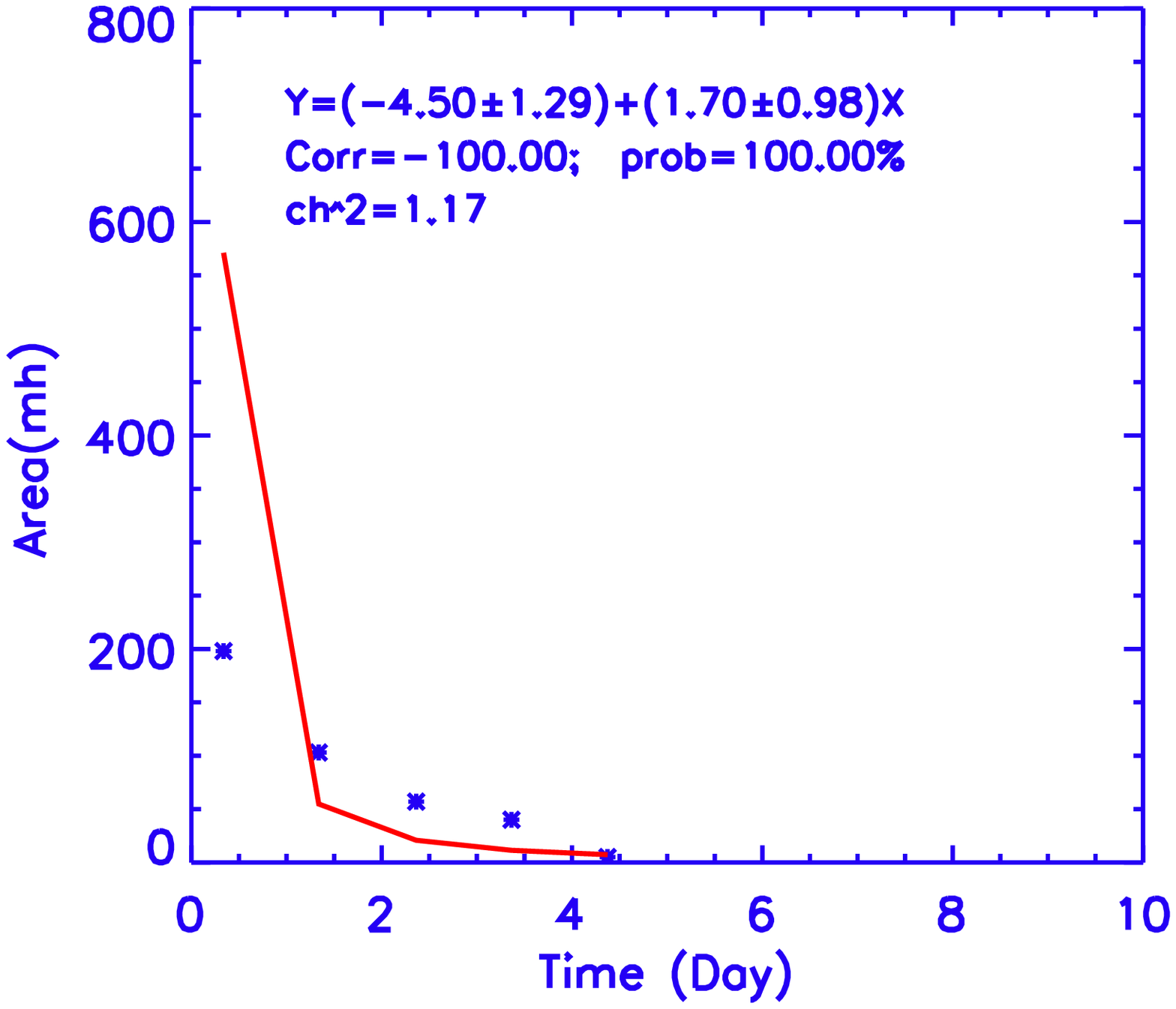}}
{\label{fig:exponential fit}\includegraphics[width=7.5cm,height=8.0cm]
{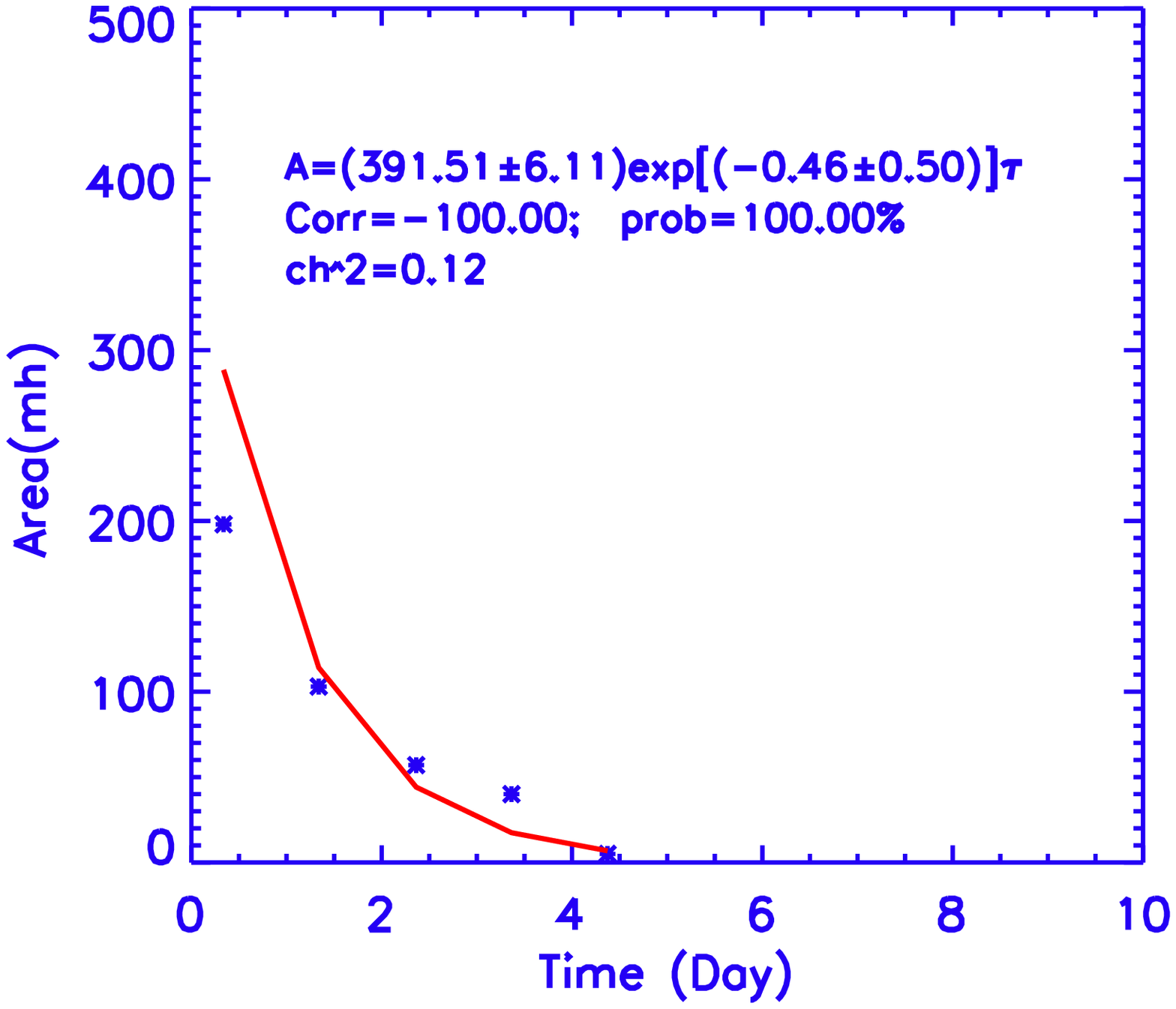}}
\caption{Evolution of decay of area A of non-recurrent sunspot group at a latitude region of 0 -10$^\circ$ that has lifespan of 9 days. In Fig (c), X=ln(Time) and Y=-ln(A). Red line is theoretical area decay curve over plotted on the observed area decay curve (blue cross points).}
\end{figure}

\begin{figure}
\centering
{\label{fig:linear fit}\includegraphics[width=7.5cm,height=8.0cm]
{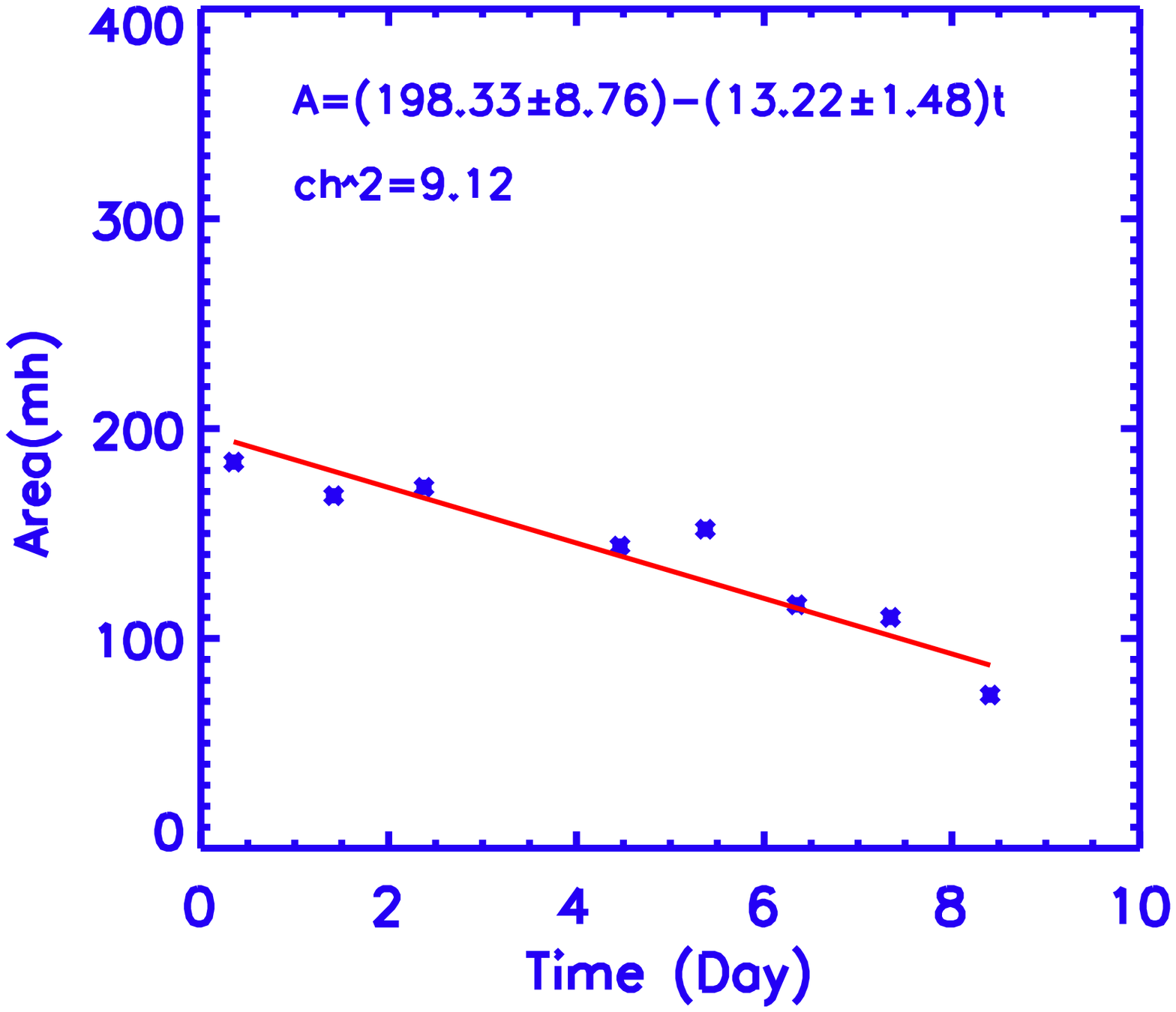}}
{\label{fig:quadratic fit}\includegraphics[width=7.5cm,height=8.0cm]
{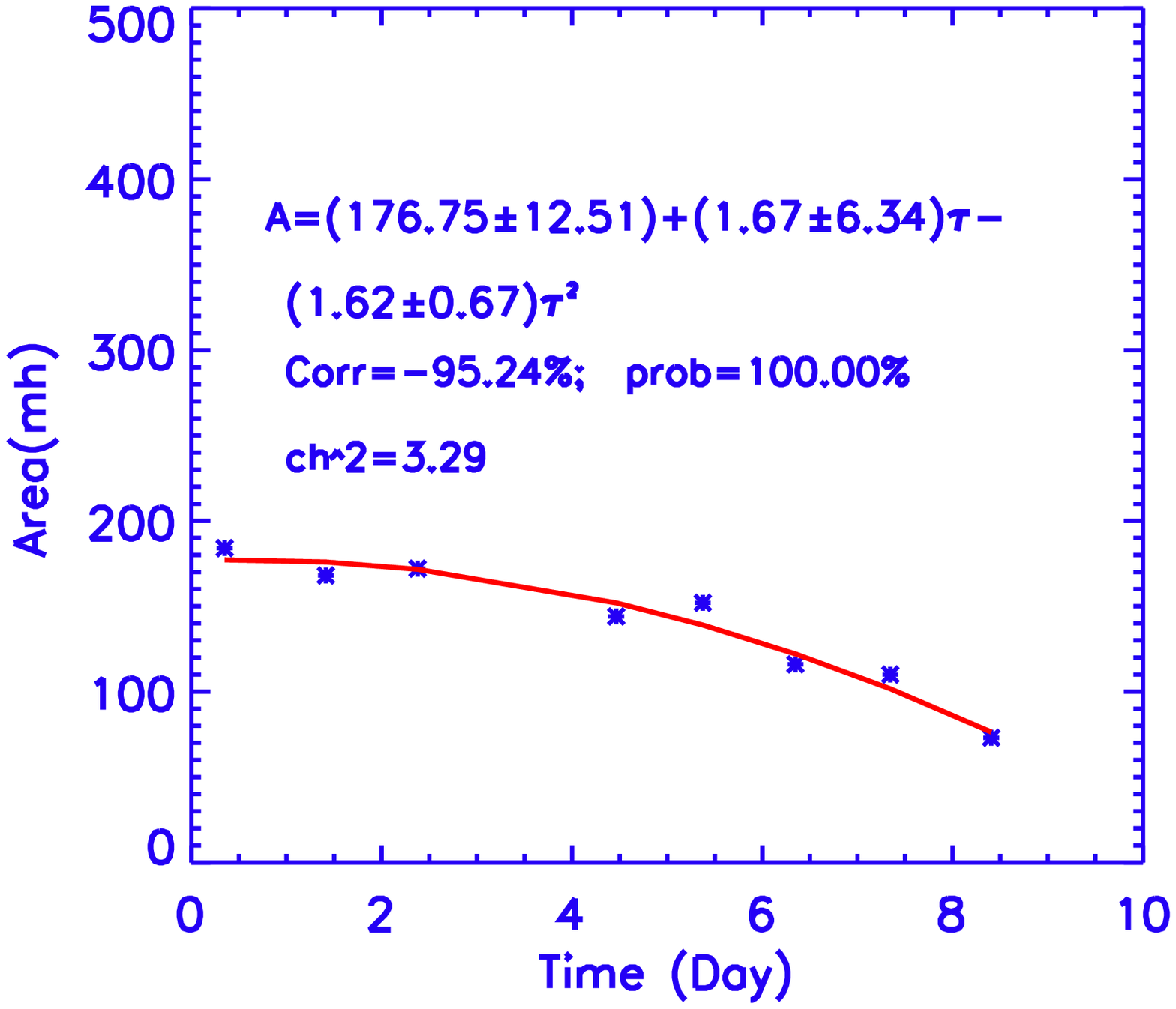}}\\
{\label{fig:log-normal fit}\includegraphics[width=7.5cm,height=8.0cm]
{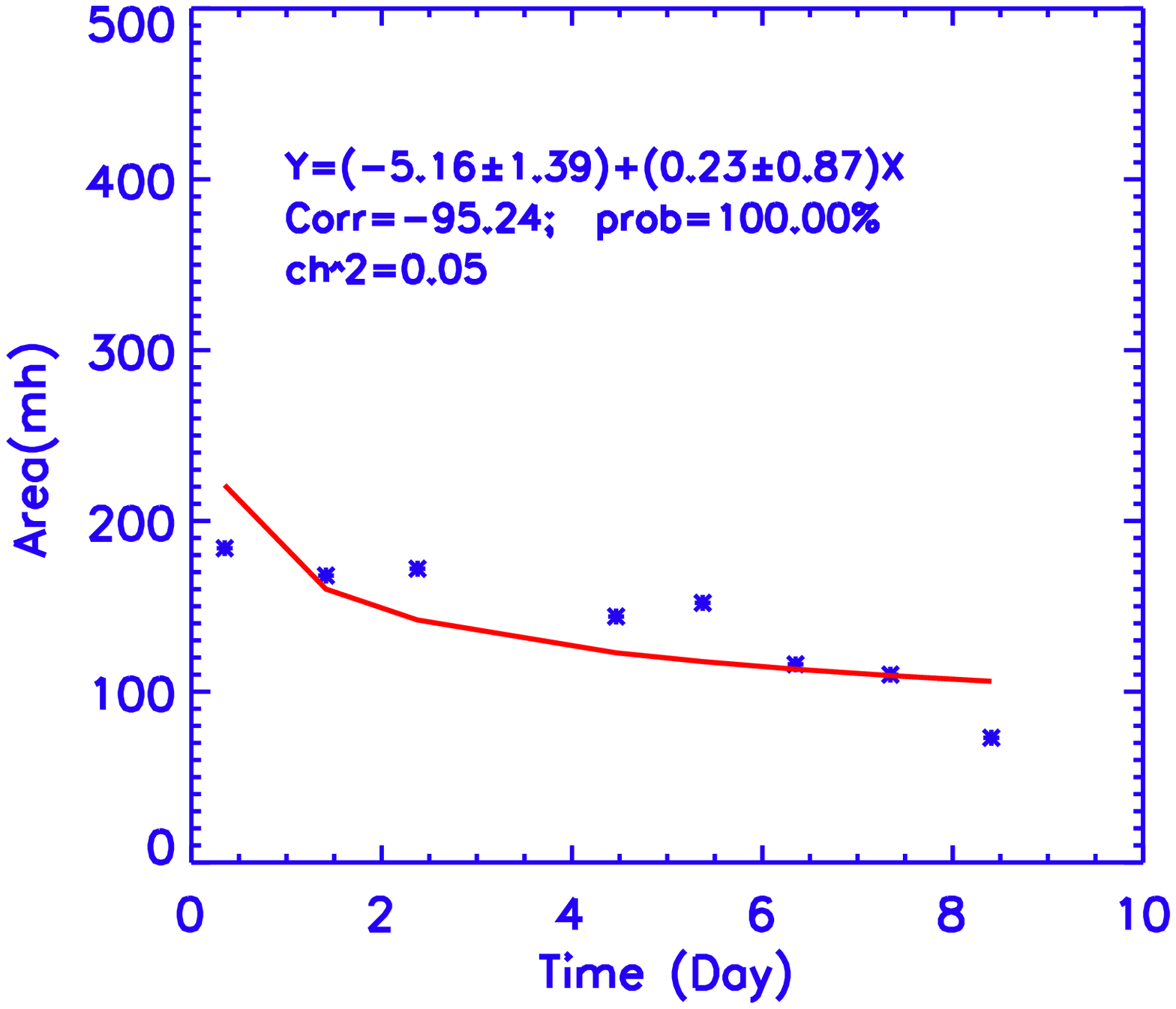}}
{\label{fig:exponential fit}\includegraphics[width=7.5cm,height=8.0cm]
{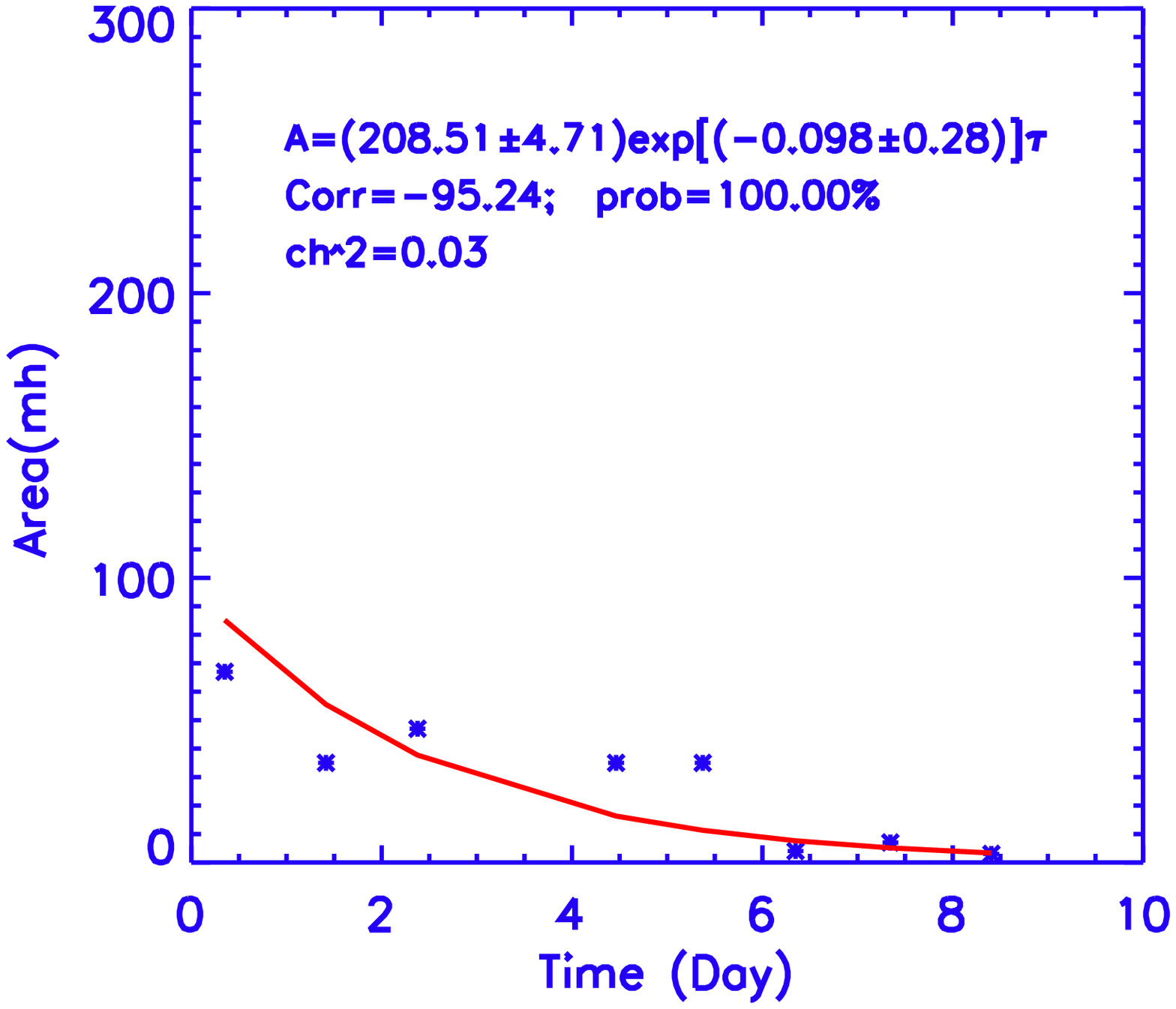}}
\caption{Evolution of decay of area A of non-recurrent sunspot group at a latitude region of 0 -10$^\circ$ that has lifespan of 9 days. In Fig (c), X=ln(Time) and Y=-ln(A). Red line is theoretical area decay curve over plotted on the observed area decay curve (blue cross points).}
\end{figure}

\begin{figure}
\centering
{\label{fig:linear fit}\includegraphics[width=7.5cm,height=7.5cm]
{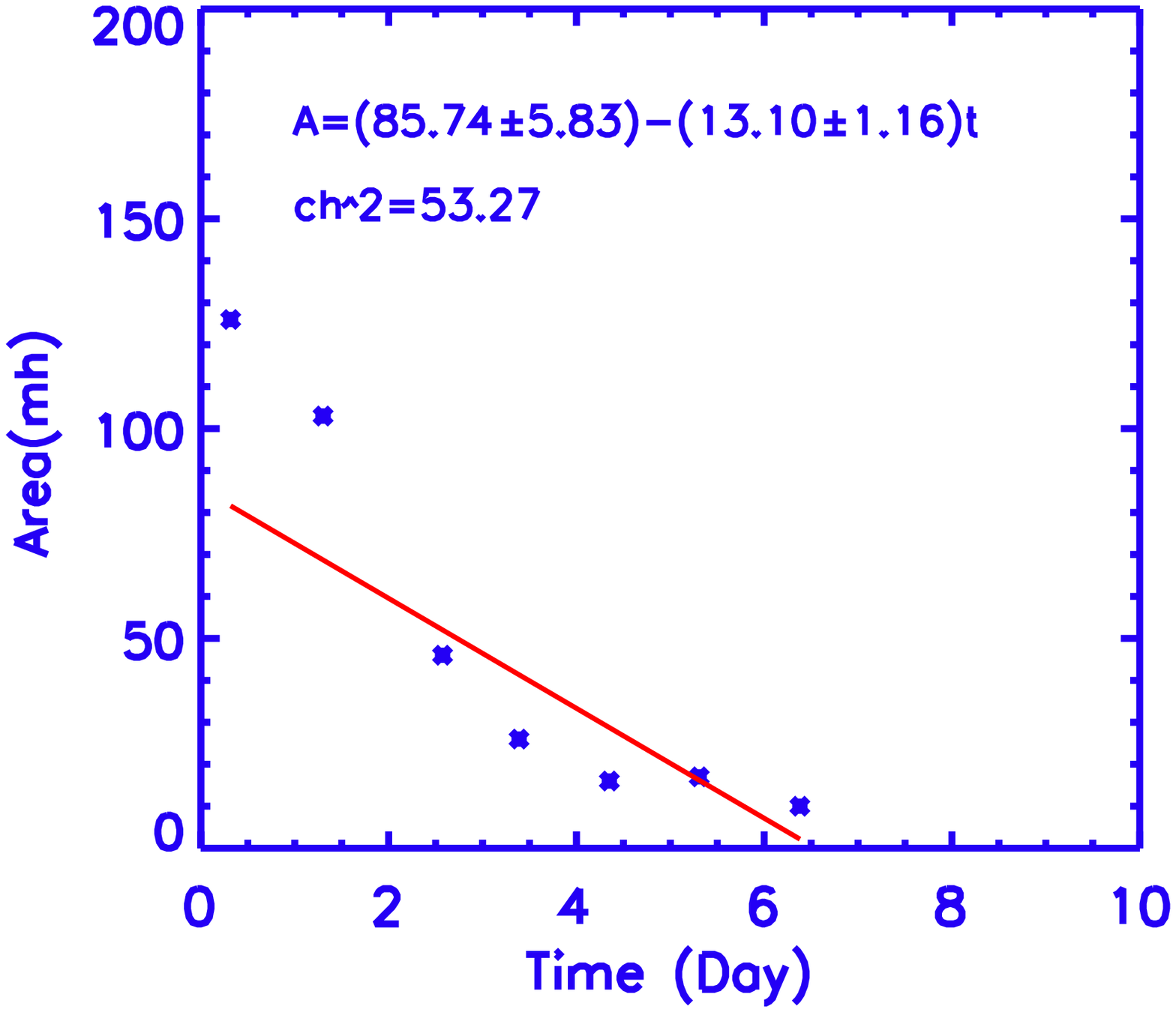}}
{\label{fig:quadratic fit}\includegraphics[width=7.5cm,height=7.5cm]
{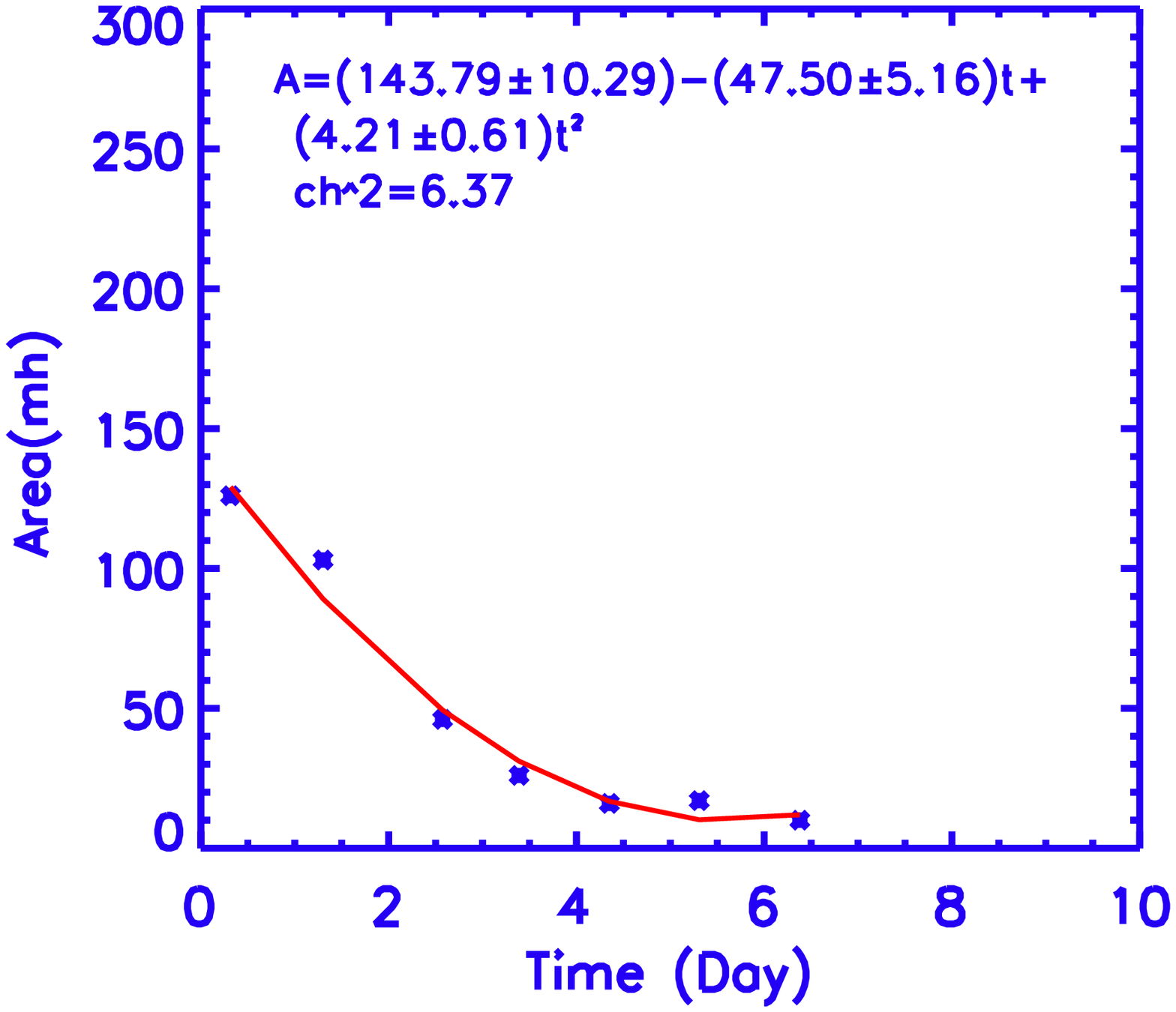}}\\
{\label{fig:log-normal fit}\includegraphics[width=7.5cm,height=7.5cm]
{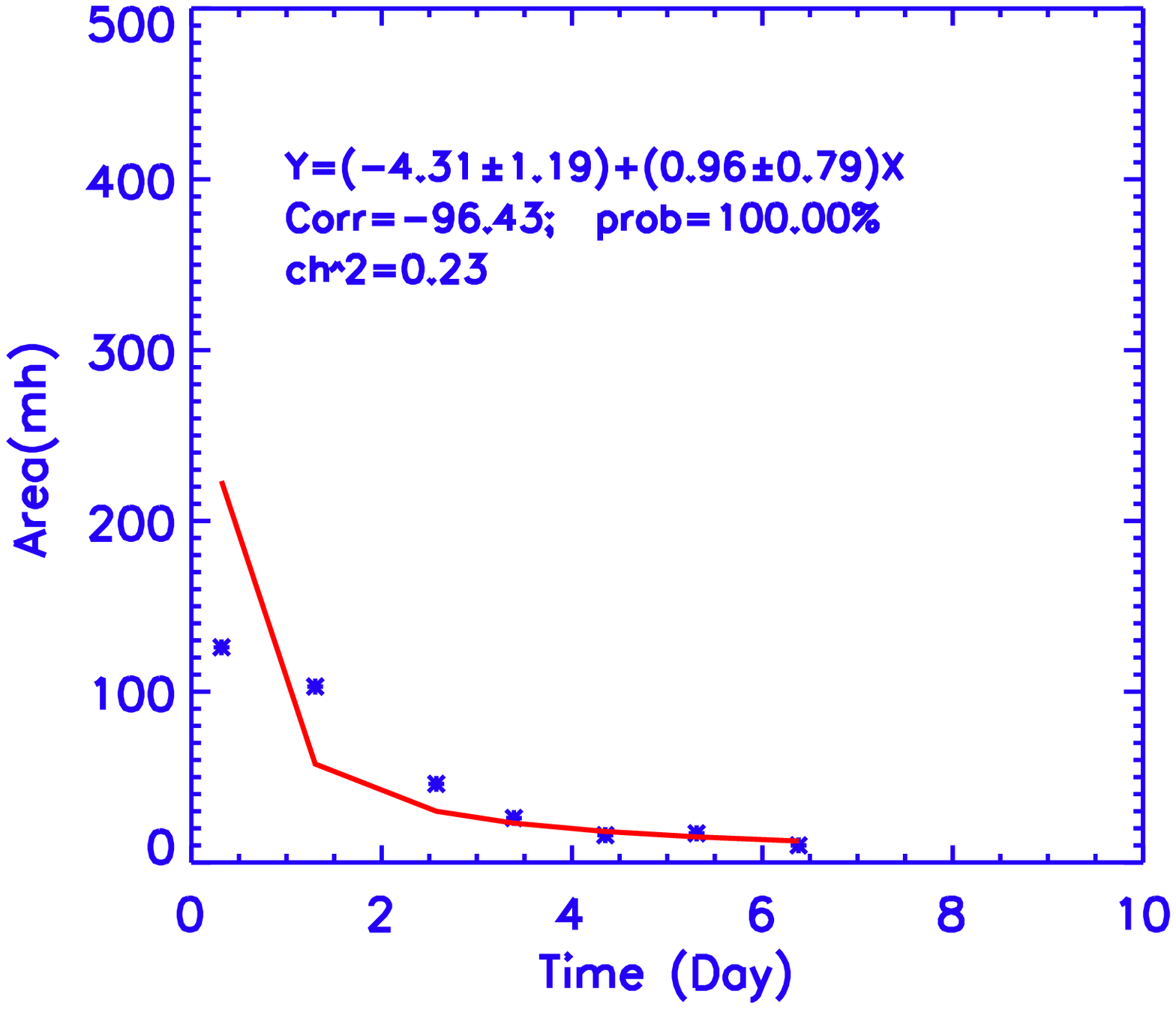}}
{\label{fig:exponential fit}\includegraphics[width=7.5cm,height=7.5cm]
{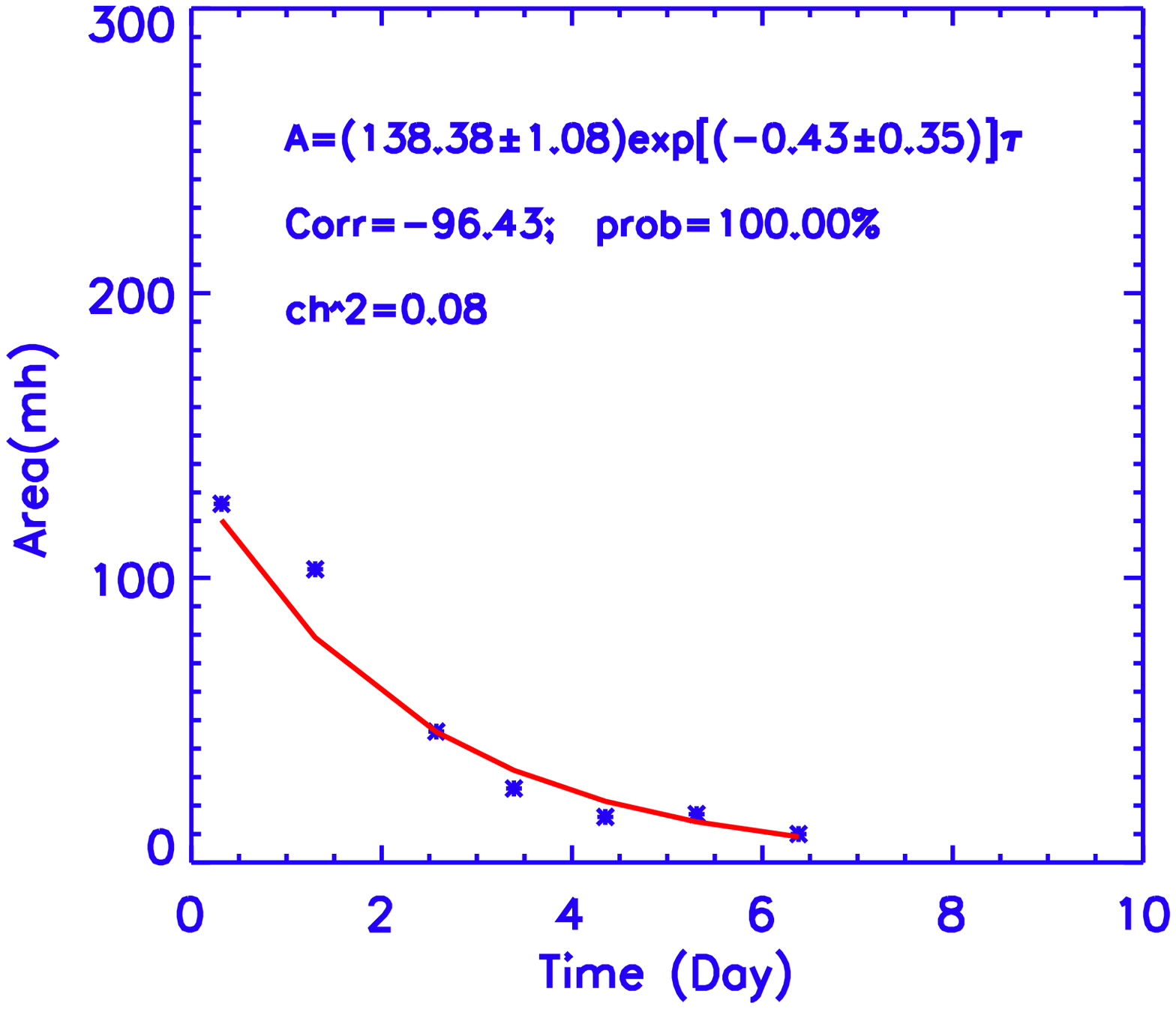}}
\caption{Evolution of decay of area A of non-recurrent sunspot group at a latitude region of 10 -20$^\circ$ that has lifespan of 10 days. In Fig (c), X=ln(Time) and Y=-ln(A). Red line is theoretical area decay curve over plotted on the observed area decay curve (blue cross points).}
\end{figure}

\begin{figure}
\centering
{\label{fig:linear fit}\includegraphics[width=7.5cm,height=7.5cm]
{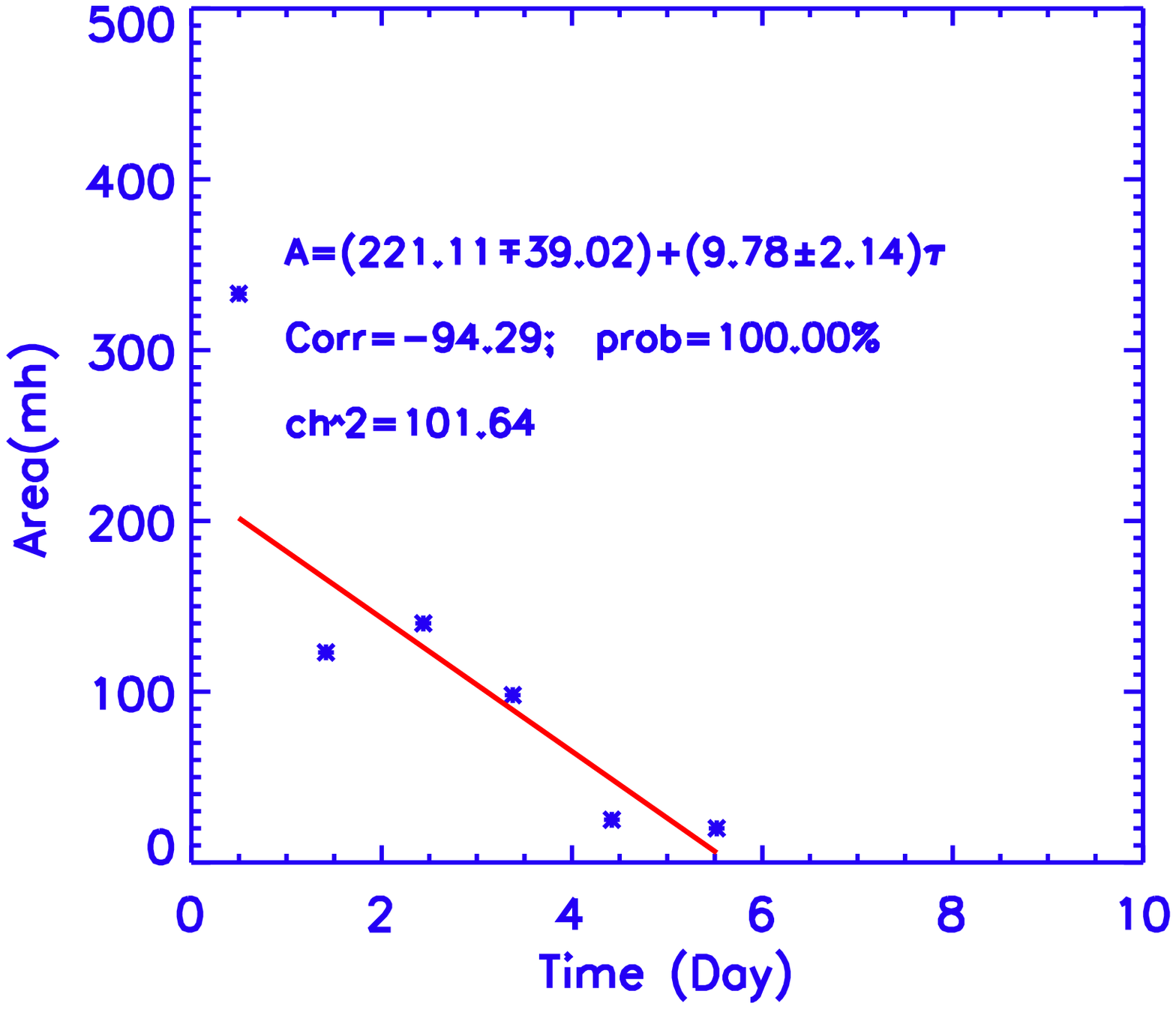}}
{\label{fig:quadratic fit}\includegraphics[width=7.5cm,height=7.5cm]
{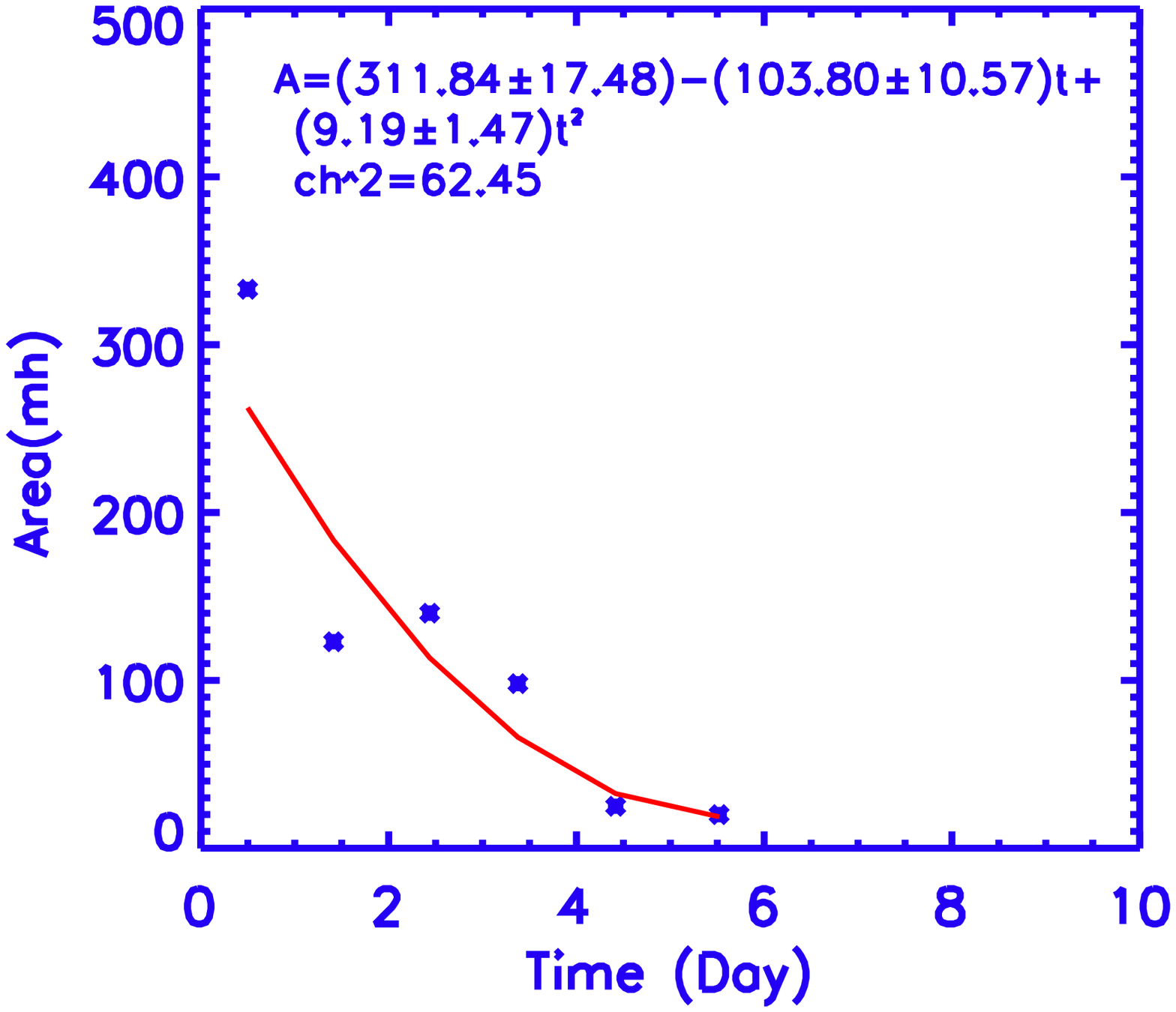}}\\
{\label{fig:log-normal fit}\includegraphics[width=7.5cm,height=7.5cm]
{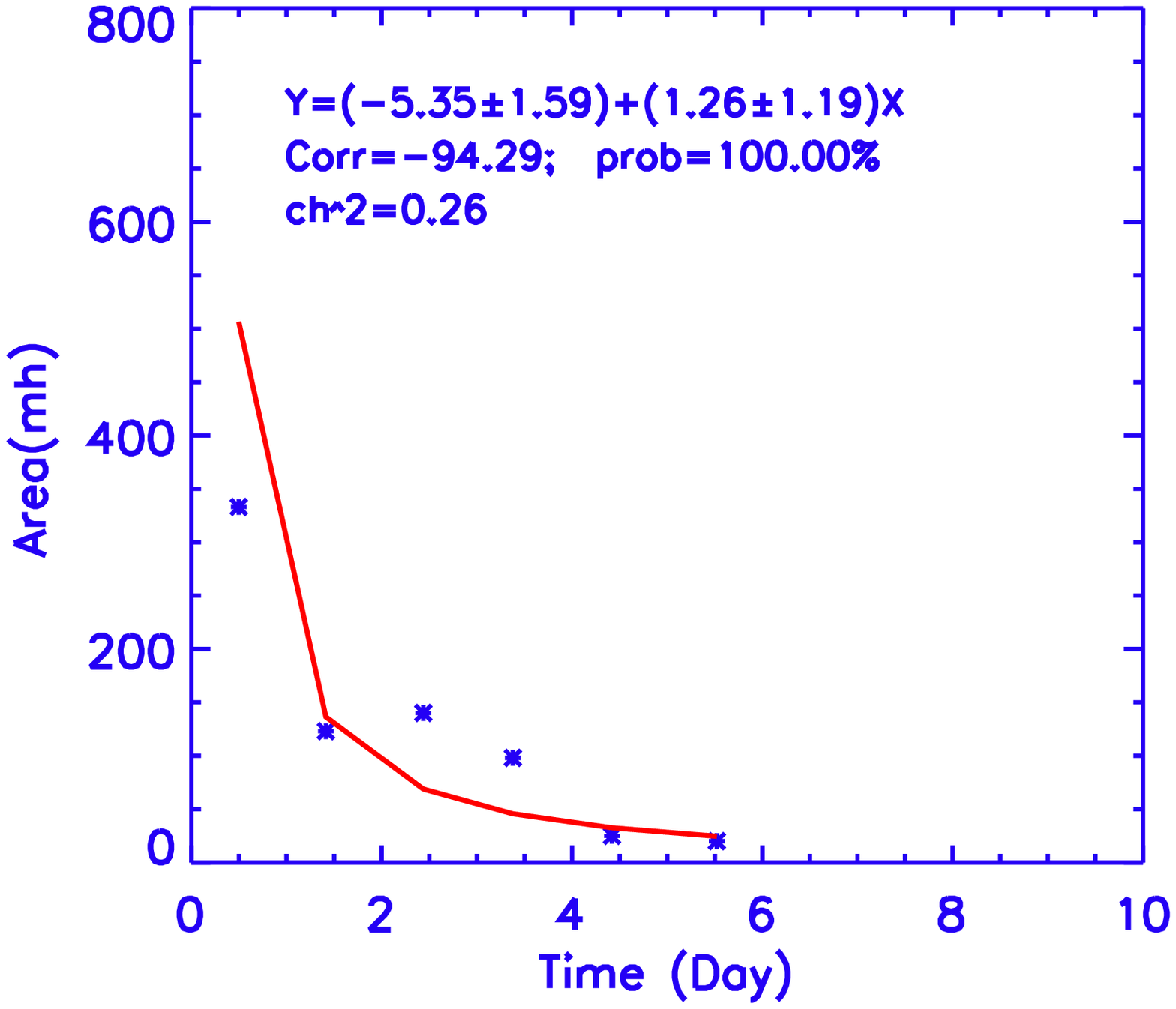}}
{\label{fig:exponential fit}\includegraphics[width=7.5cm,height=7.5cm]
{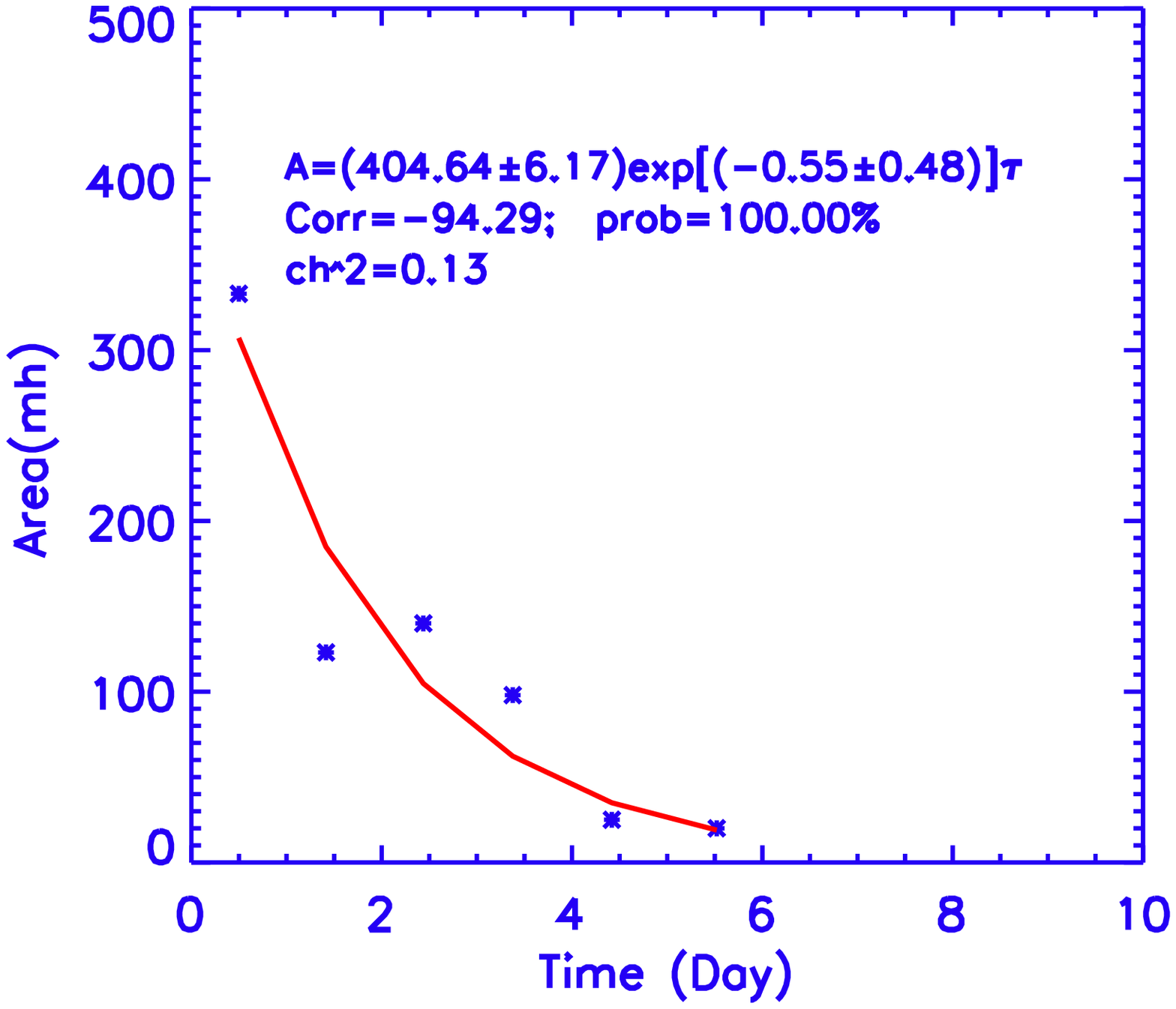}}
\caption{Evolution of decay of area A of non-recurrent sunspot group at a latitude region of 10 -20$^\circ$ that has lifespan of 9 days. In Fig (c), X=ln(Time) and Y=-ln(A). Red line is theoretical area decay curve over plotted on the observed area decay curve (blue cross points).}
\end{figure}

\begin{figure}
\centering
{\label{fig:linear fit}\includegraphics[width=7.5cm,height=7.5cm]
{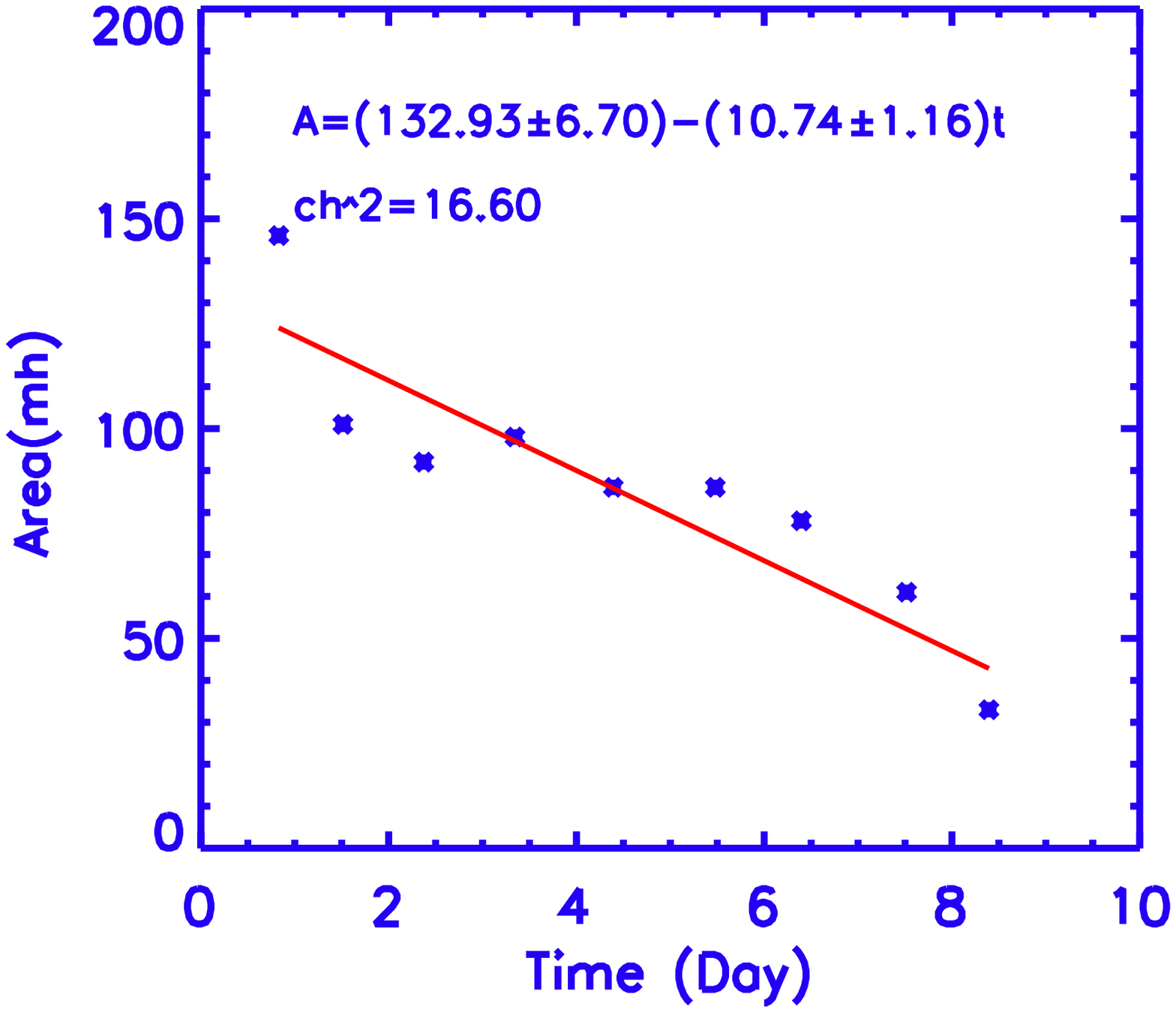}}
{\label{fig:quadratic fit}\includegraphics[width=7.5cm,height=7.5cm]
{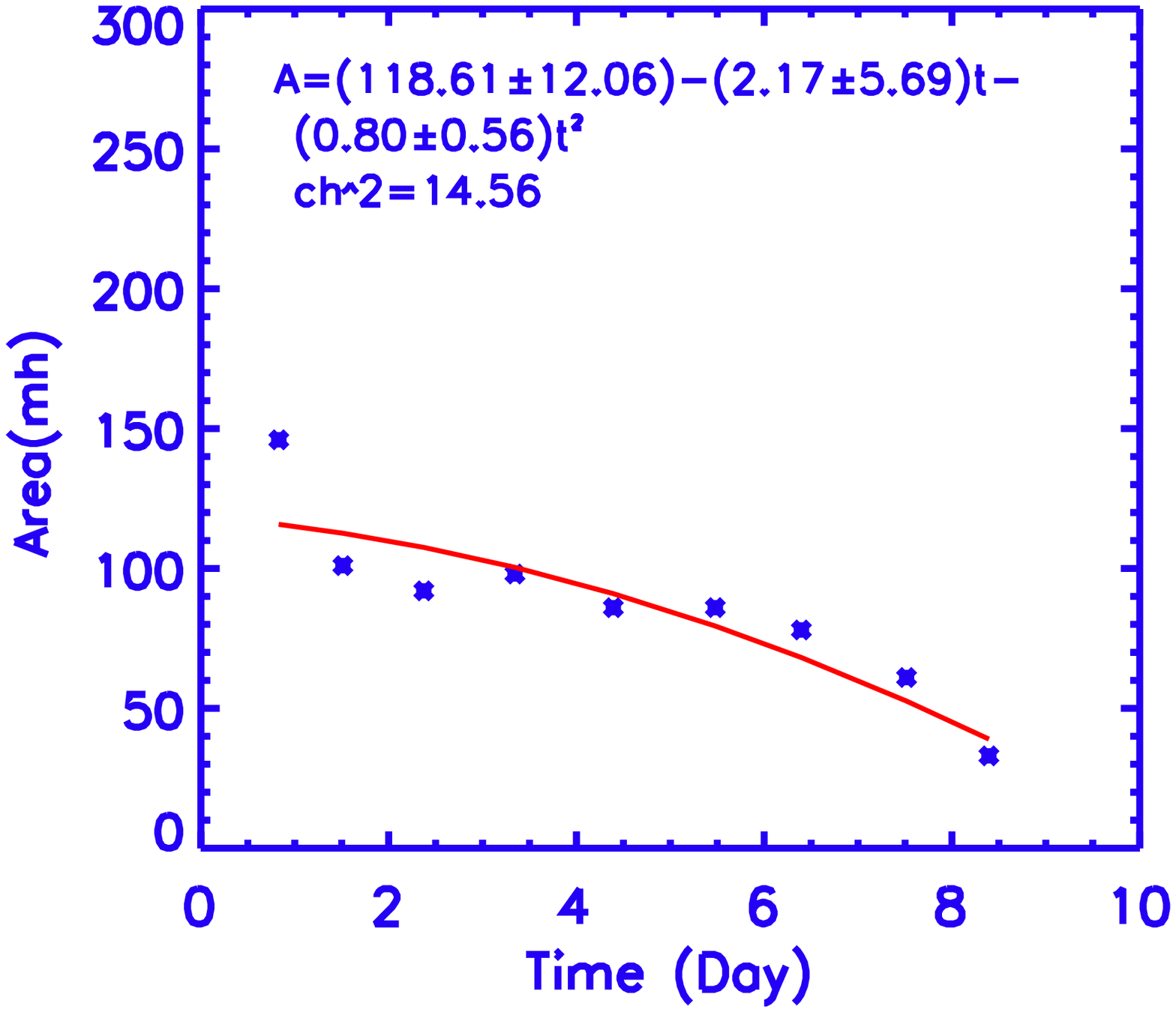}}\\
{\label{fig:log-normal fit}\includegraphics[width=7.5cm,height=7.5cm]
{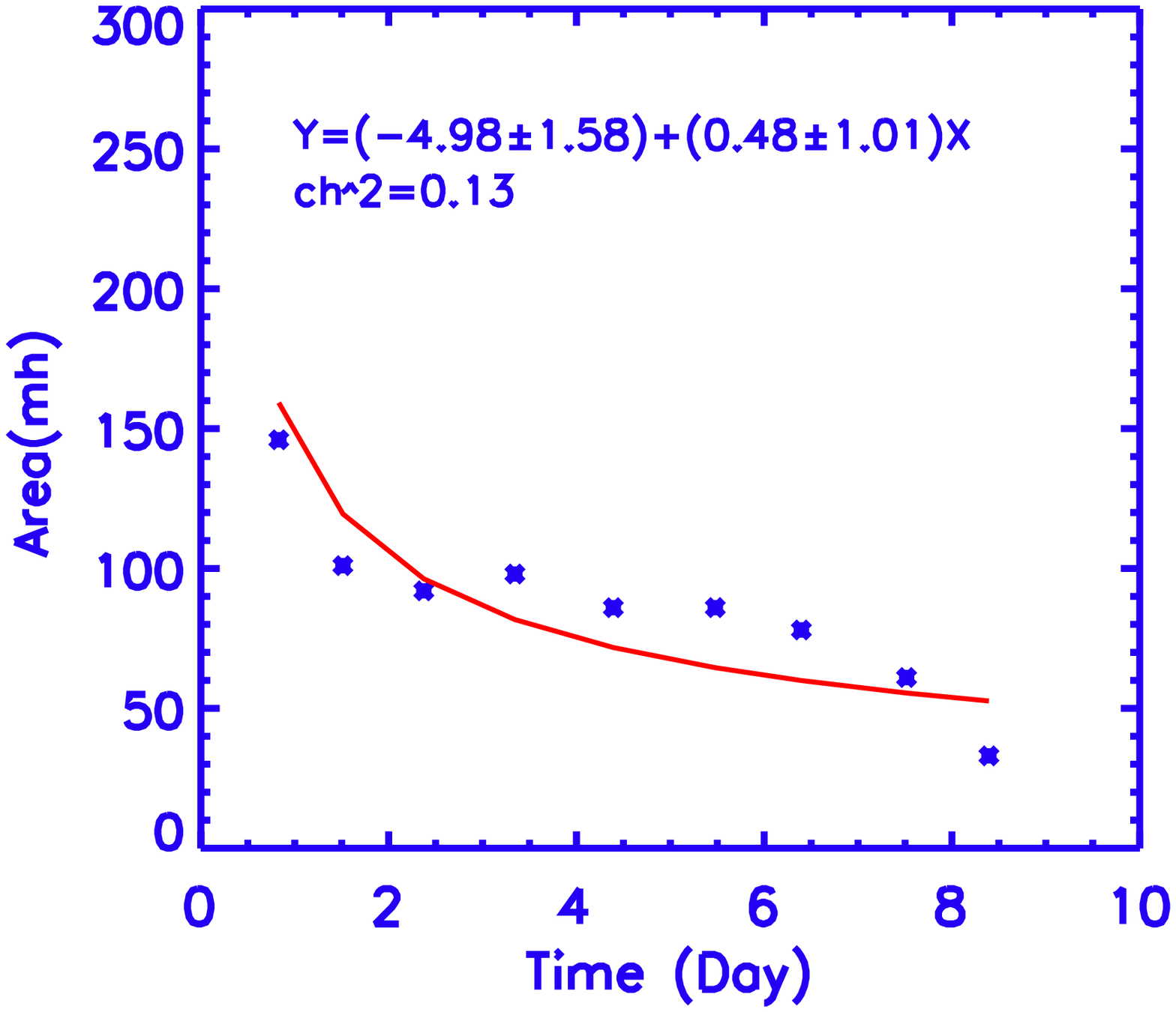}}
{\label{fig:exponential fit}\includegraphics[width=7.5cm,height=7.5cm]
{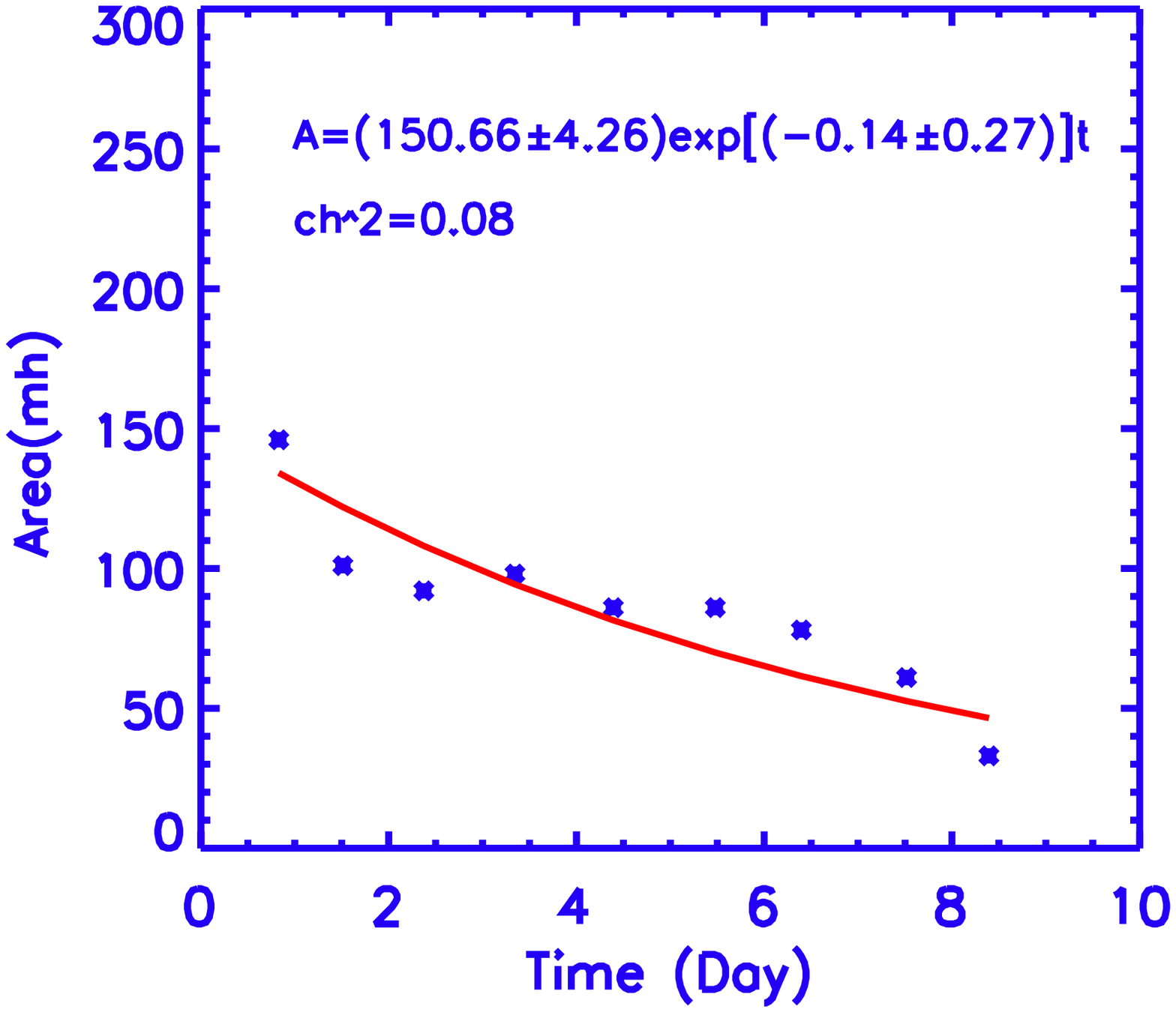}}
\caption{Evolution of decay of area A of non-recurrent sunspot group at a latitude region of 20 -30$^\circ$ that has lifespan of 10 days. In Fig (c), X=ln(Time) and Y=-ln(A). Red line is theoretical area decay curve over plotted on the observed area decay curve (blue cross points).}
\end{figure}

\begin{figure}
\centering
{\label{fig:linear fit}\includegraphics[width=7.5cm,height=7.5cm]
{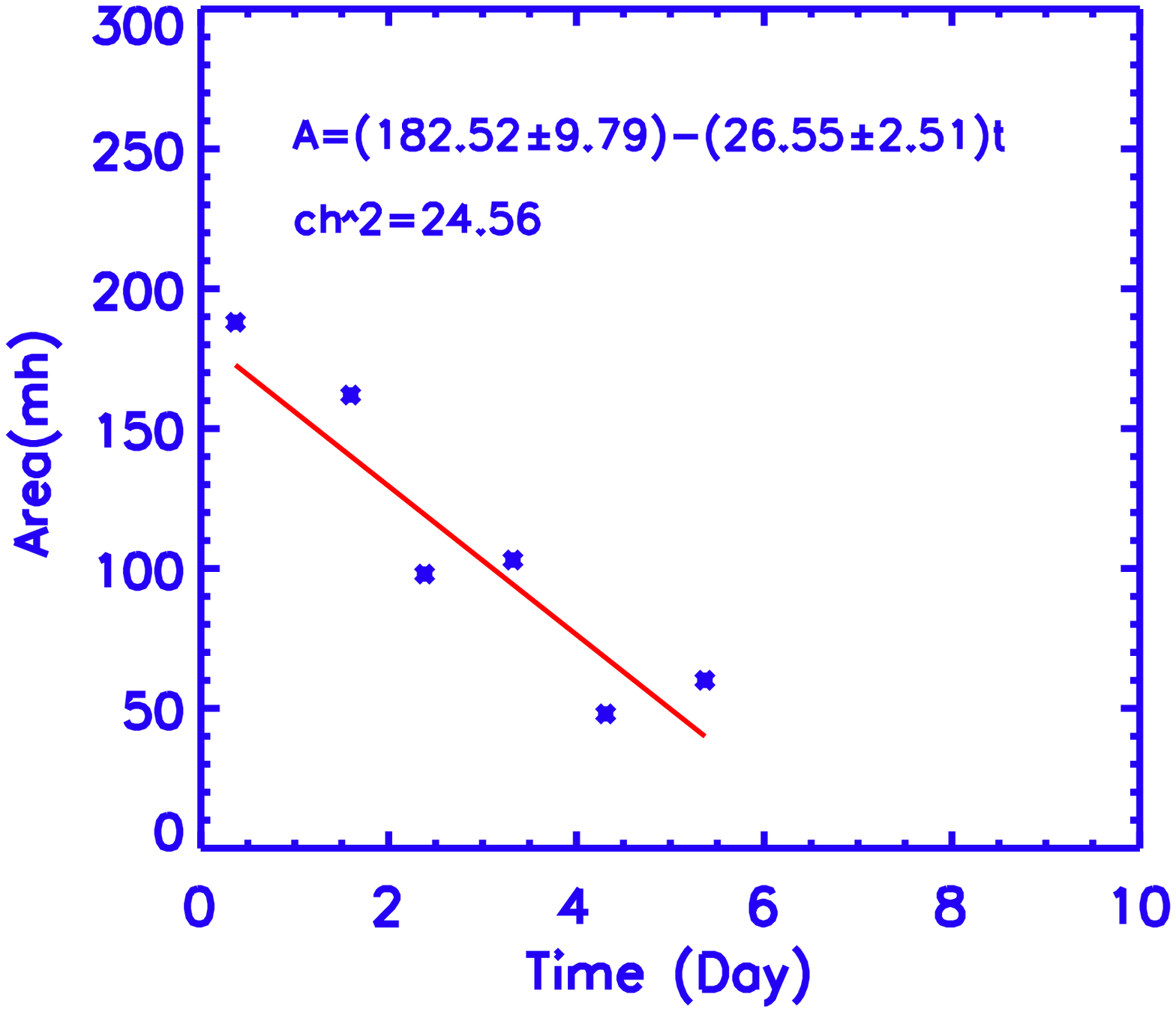}}
{\label{fig:quadratic fit}\includegraphics[width=7.5cm,height=7.5cm]
{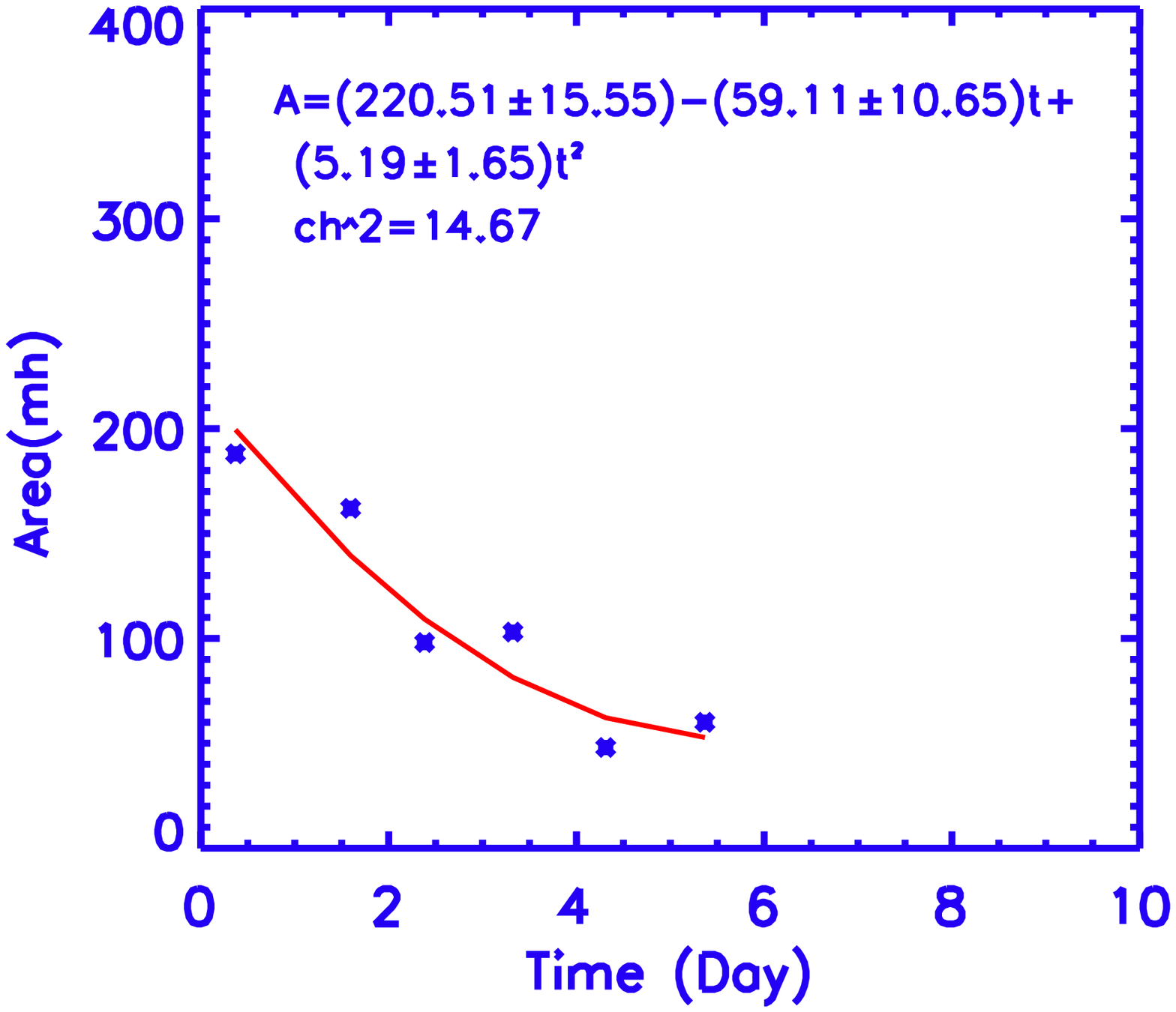}}\\
{\label{fig:log-normal fit}\includegraphics[width=7.5cm,height=7.5cm]
{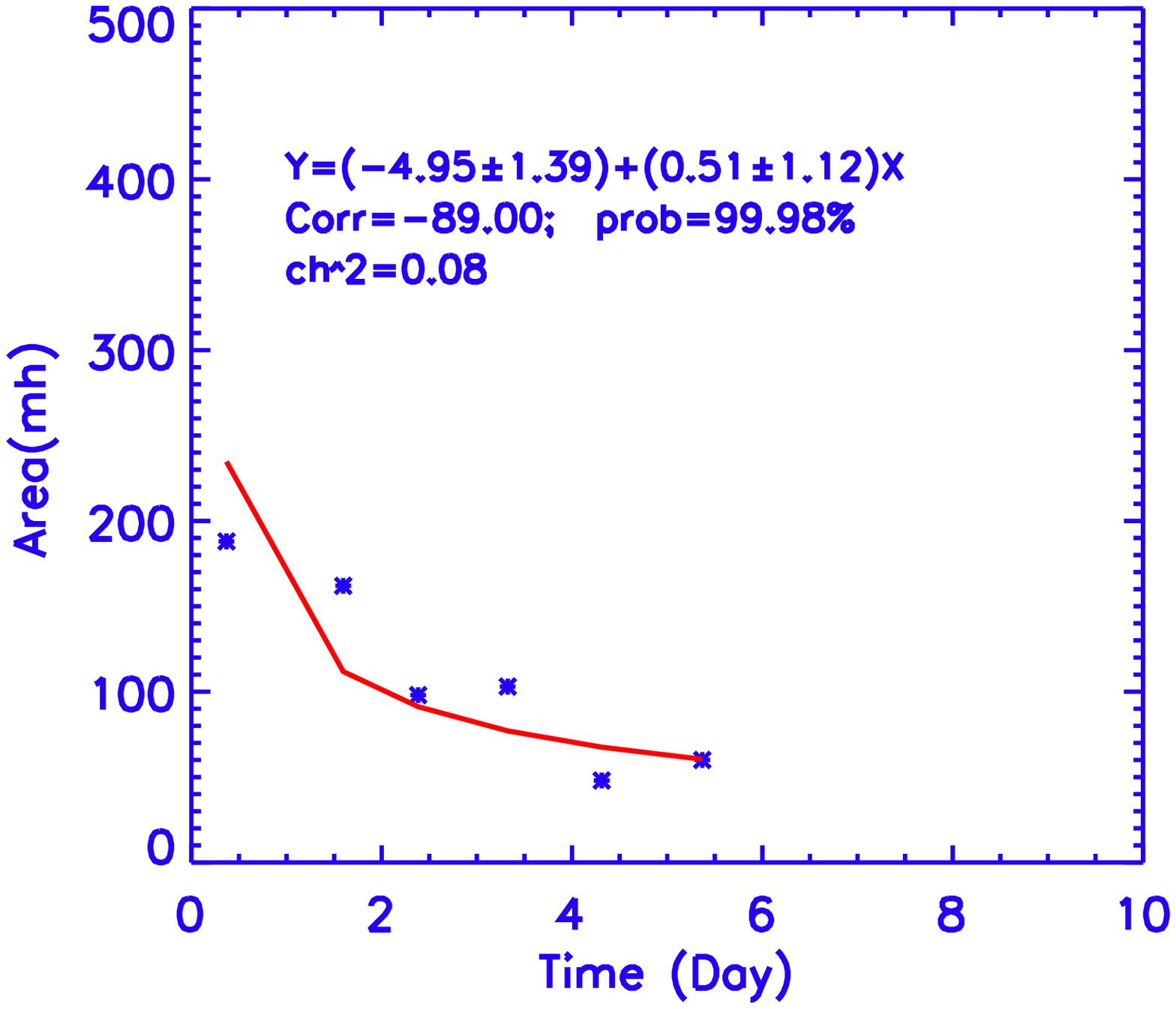}}
{\label{fig:exponential fit}\includegraphics[width=7.5cm,height=7.5cm]
{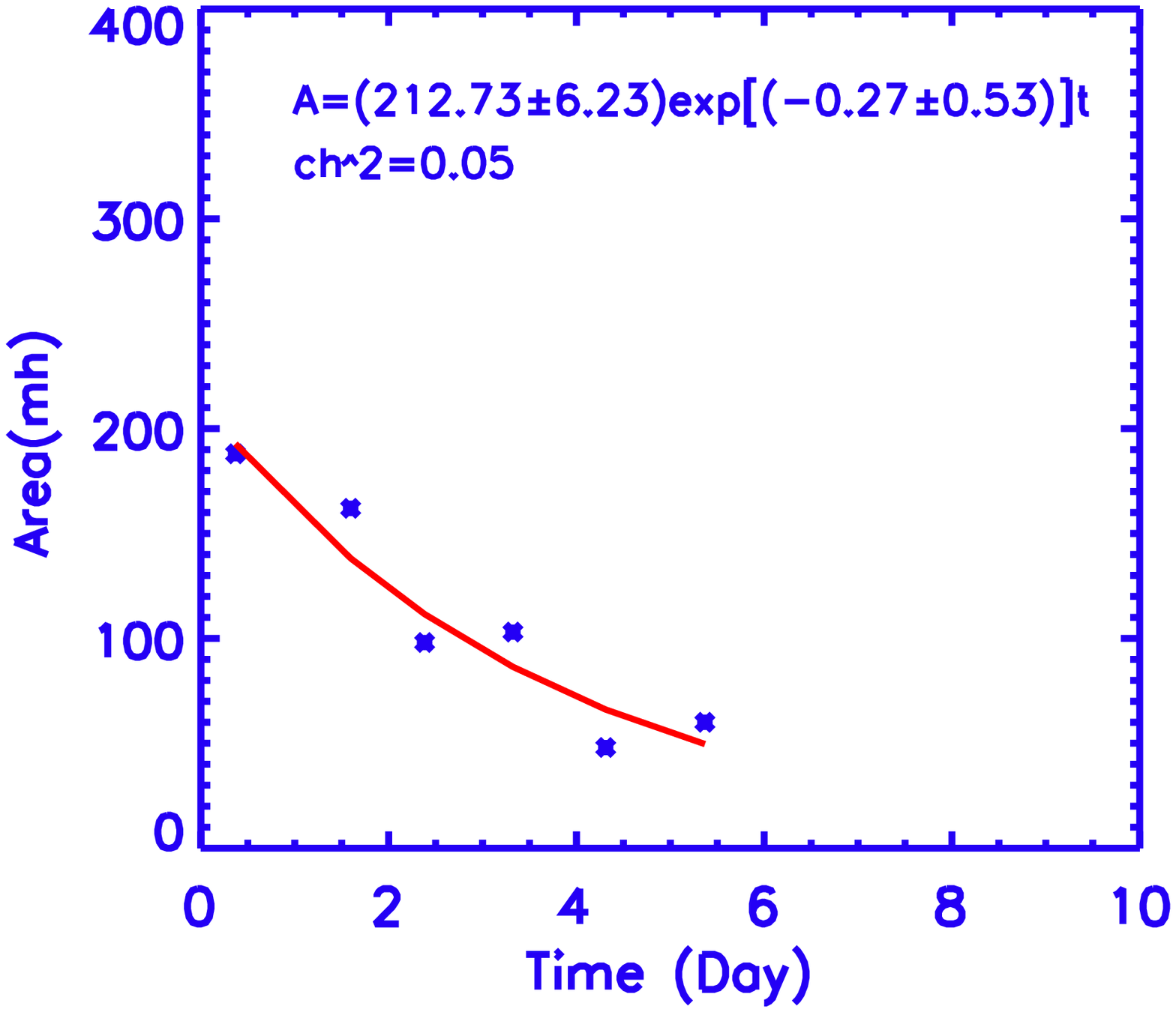}}
\caption{Evolution of decay of area A of non-recurrent sunspot group at a latitude region of 20 -30$^\circ$ that has lifespan of 10 days. In Fig (c), X=ln(Time) and Y=-ln(A). Red line is theoretical area decay curve over plotted on the observed area decay curve (blue cross points).}
\end{figure}

\begin{figure}
\centering
{\label{fig:linear fit}\includegraphics[width=7.5cm,height=7.5cm]
{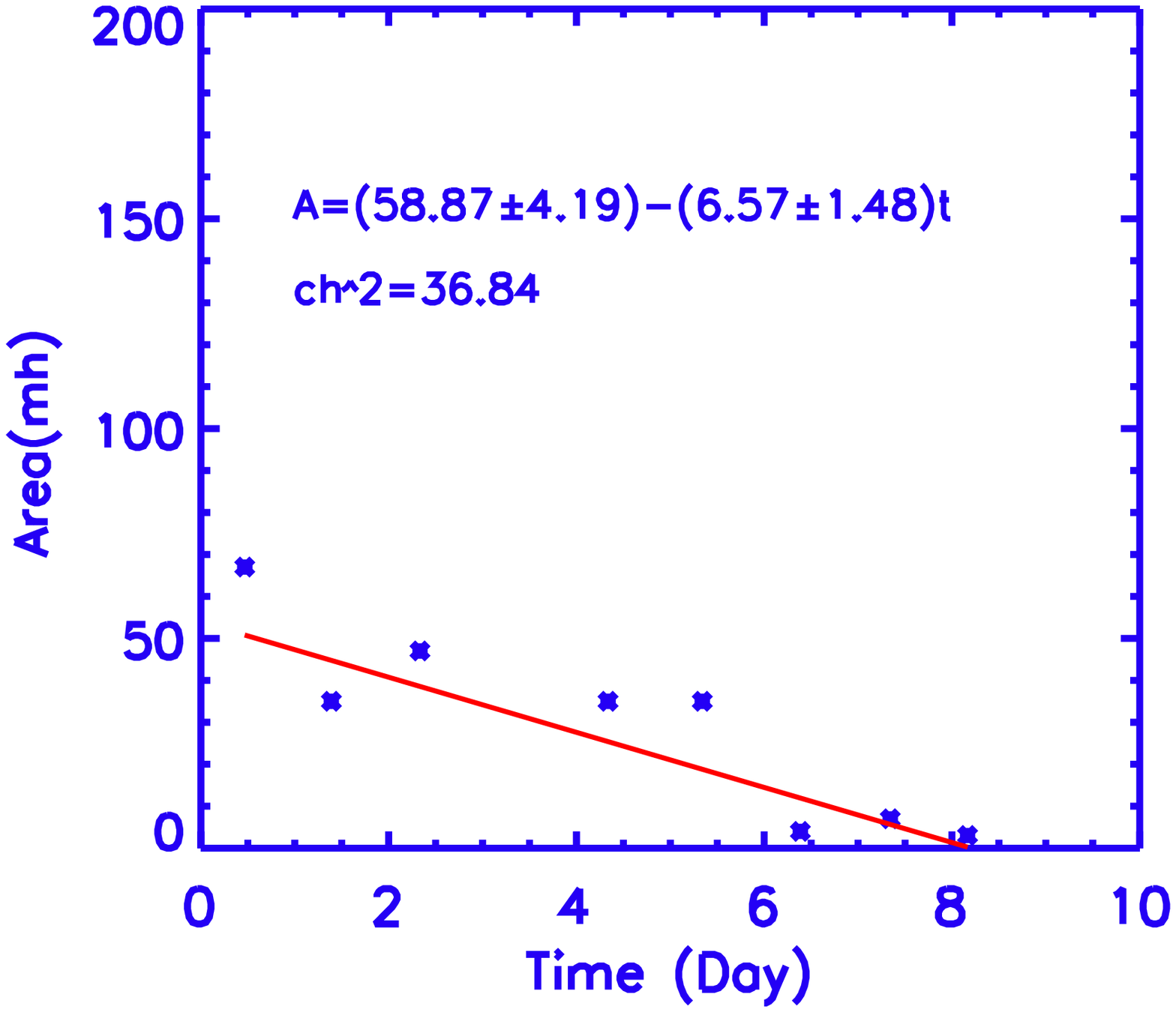}}
{\label{fig:quadratic fit}\includegraphics[width=7.5cm,height=7.5cm]
{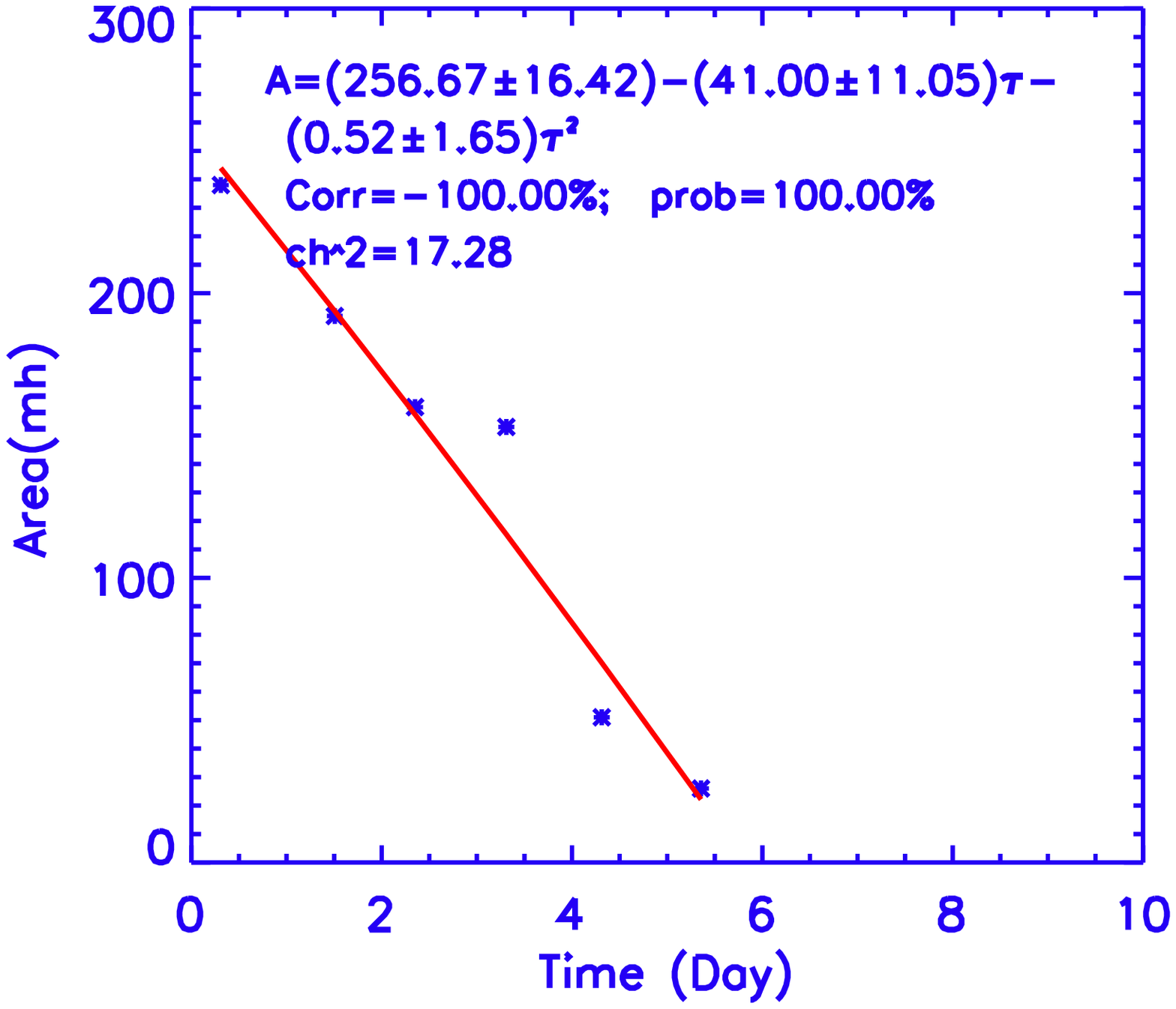}}\\
{\label{fig:log-normal fit}\includegraphics[width=7.5cm,height=7.5cm]
{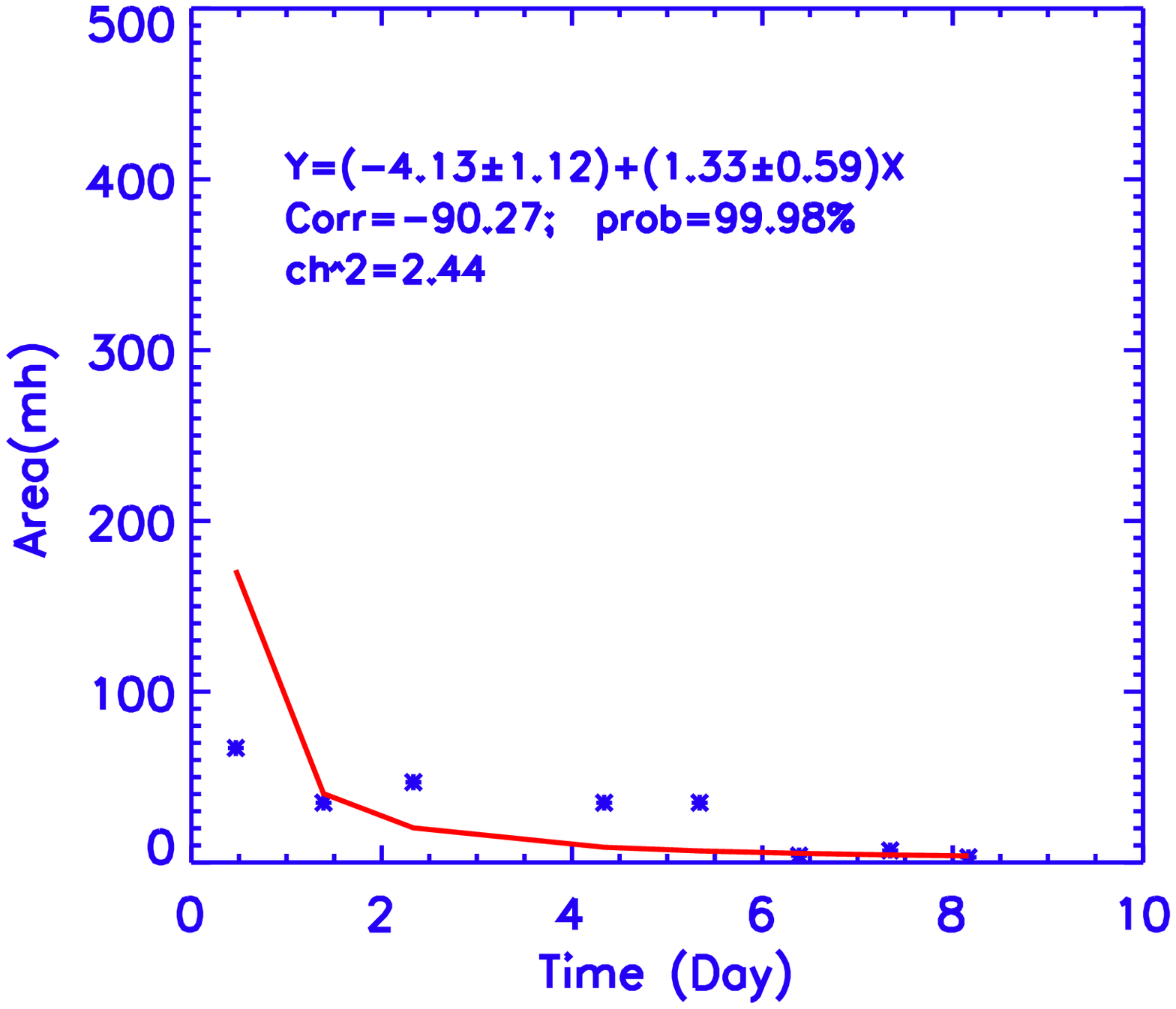}}
{\label{fig:exponential fit}\includegraphics[width=7.5cm,height=7.5cm]
{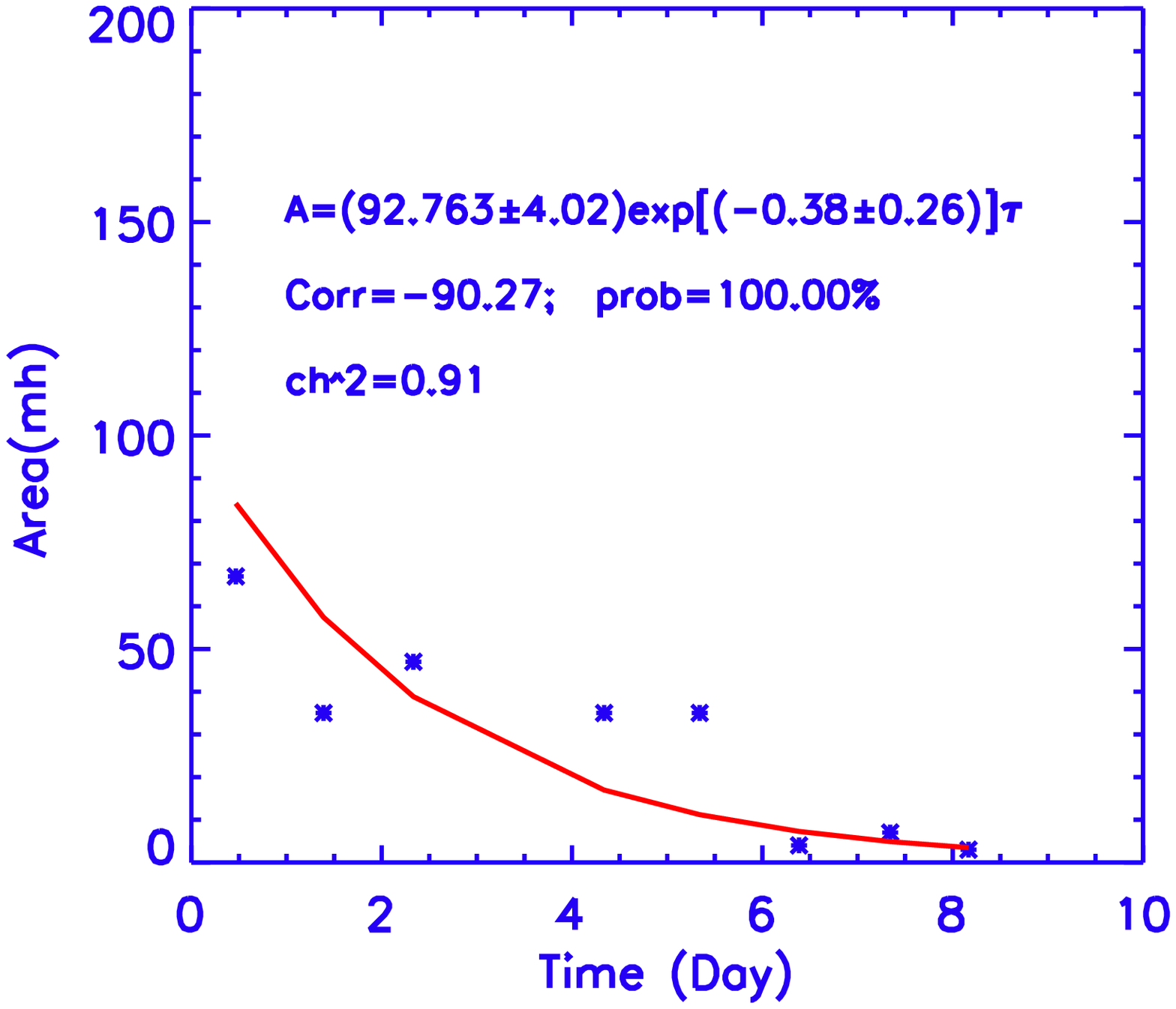}}
\caption{Evolution of decay of area A of non-recurrent sunspot group at a latitude region of 30 -40$^\circ$ that has lifespan of 8 days. In Fig (c), X=ln(Time) and Y=-ln(A). Red line is theoretical area decay curve over plotted on the observed area decay curve (blue cross points).}
\end{figure}

\begin{figure}
\centering
{\label{fig:linear fit}\includegraphics[width=7.5cm,height=7.5cm]
{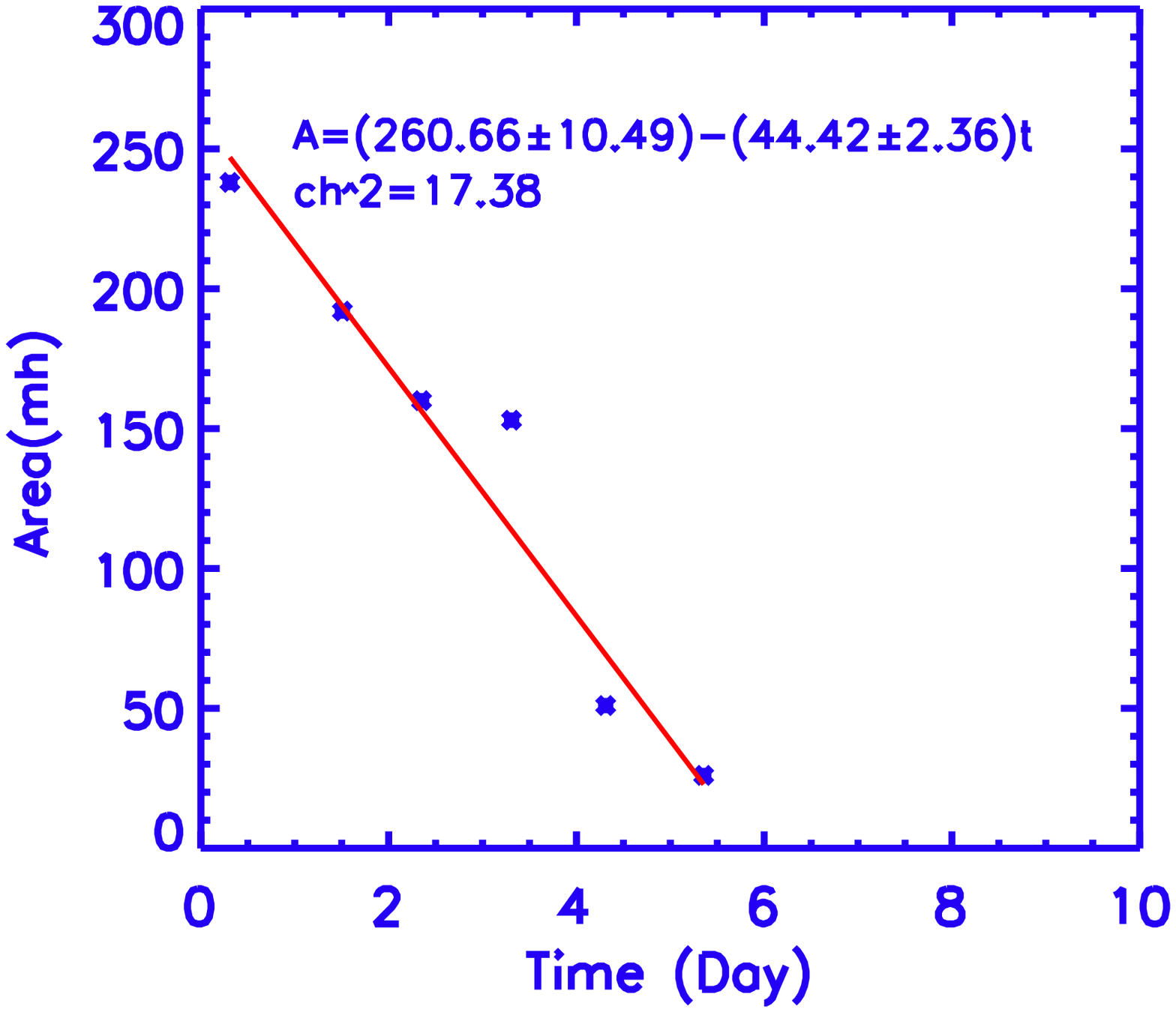}}
{\label{fig:quadratic fit}\includegraphics[width=7.5cm,height=7.5cm]
{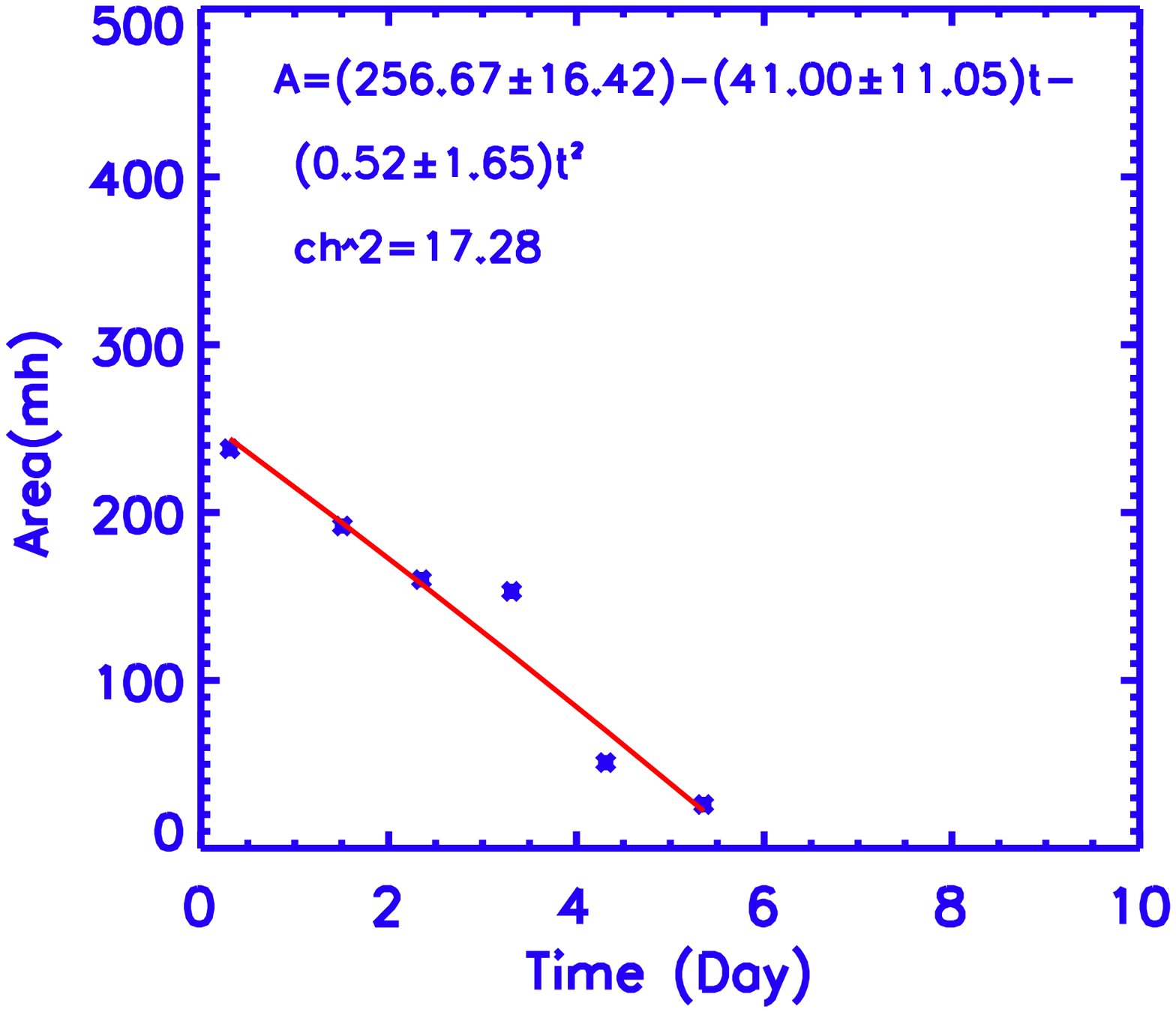}}\\
{\label{fig:log-normal fit}\includegraphics[width=7.5cm,height=7.5cm]
{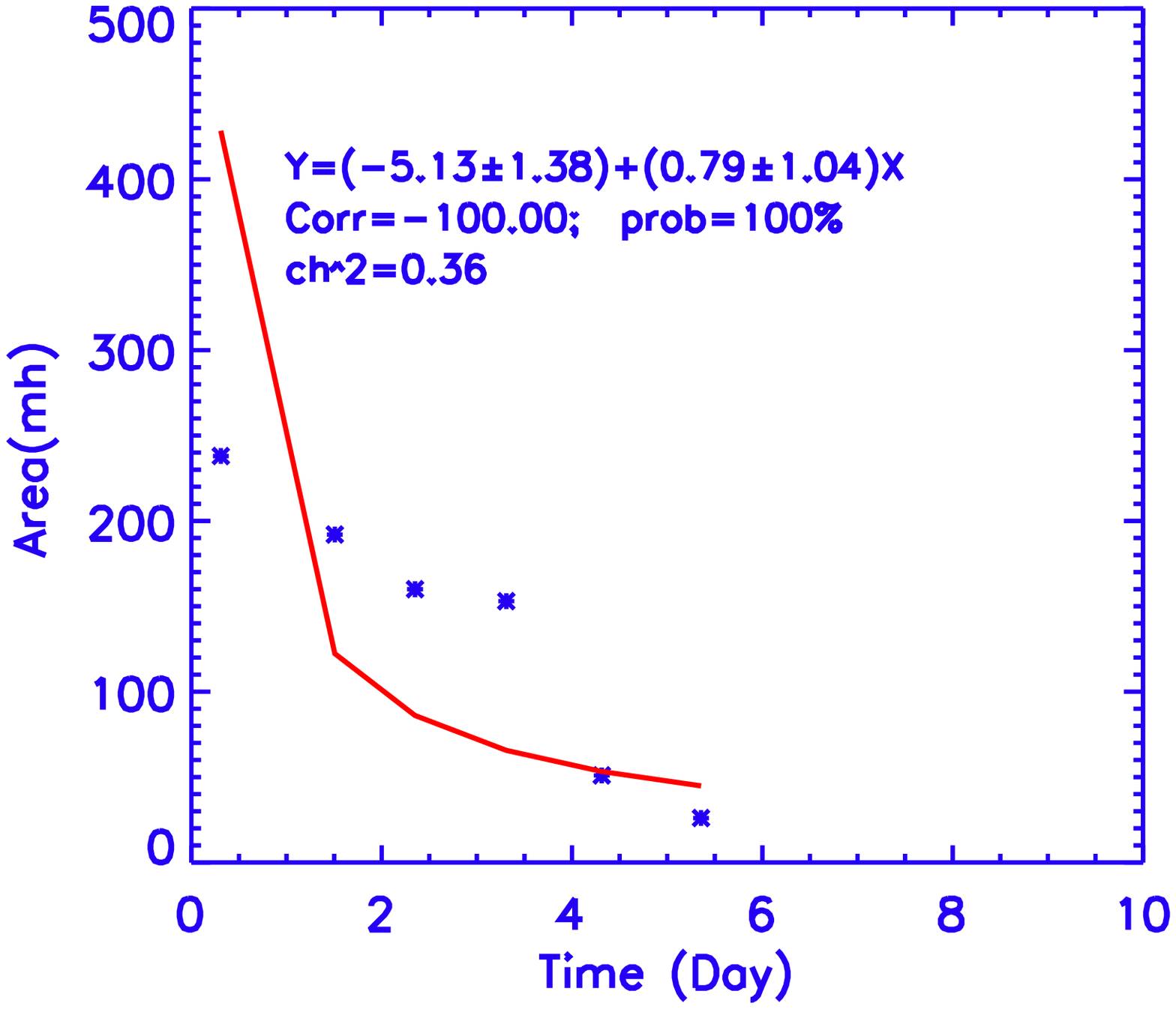}}
{\label{fig:exponential fit}\includegraphics[width=7.5cm,height=7.5cm]
{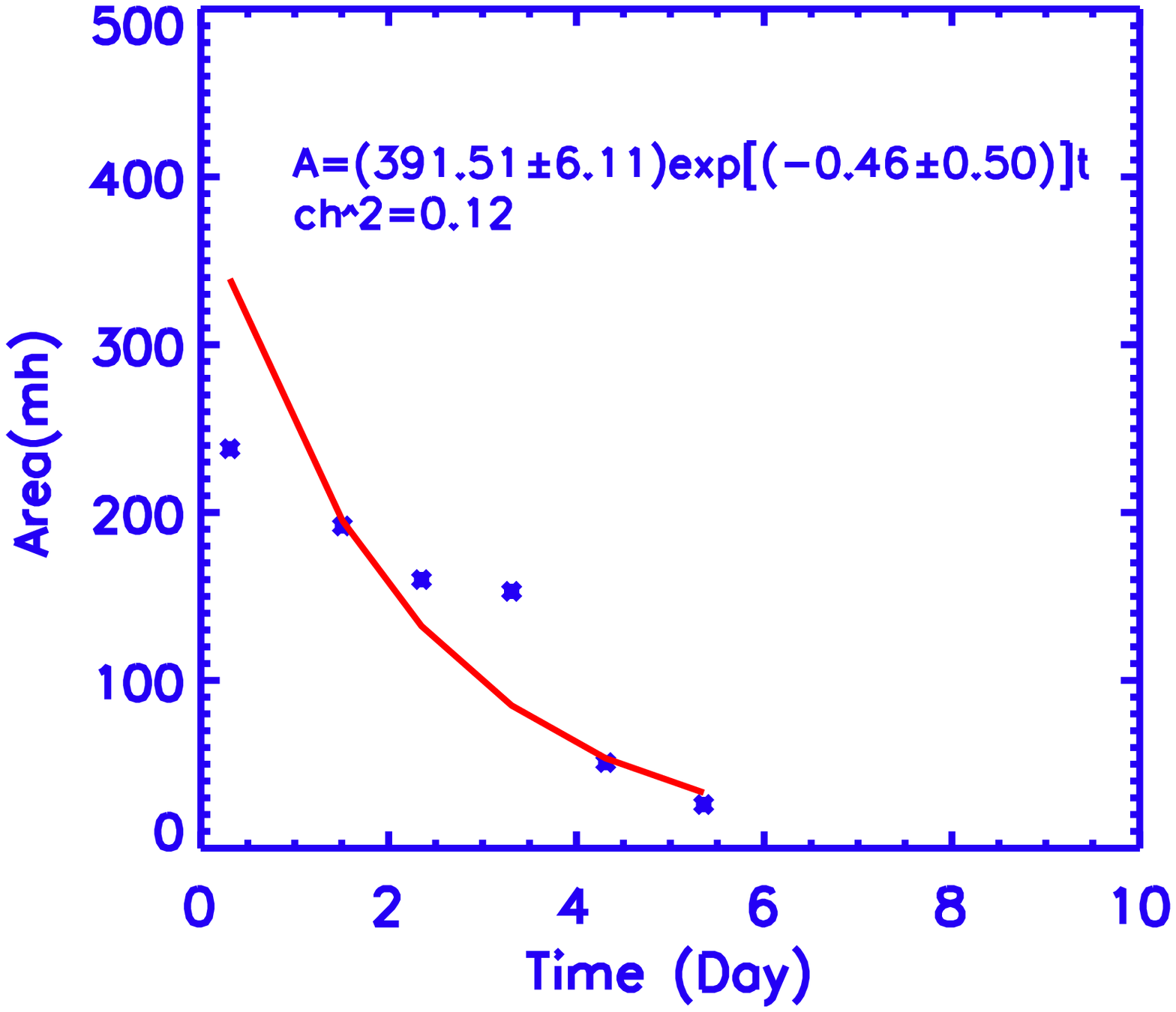}}
\caption{Evolution of decay of area A of non-recurrent sunspot group at a latitude region of 30 -40$^\circ$ that has lifespan of 8 days. In Fig (c), X=ln(Time) and Y=-ln(A). Red line is theoretical area decay curve over plotted on the observed area decay curve (blue cross points).}
\end{figure}
where
\begin{equation}
A_{1} = 1 + {\Omega^{'} sin^{2}\theta \over \eta {U_{0}}^{2}}\ ,
\end{equation}
\begin{equation}
A_{2} = r^{2}sin^{2}\theta {\partial \Omega^{'} \over \partial t}
+{\Omega_{0}^{2} r^{2} sin^{2} \theta \over {\eta}}
+2{\Omega_{0}\Omega^{'} r^{2} sin^{2} \theta \over {\eta}}
+{\Omega^{'} sin^{2}\theta \over \eta {U_{0}}^{3}} {\partial {U_{0}} \over \partial t}
-{\Omega^{'} sin^{2}\theta cot \theta \over \eta U_{0}} \ ,
\end{equation}
and 
\begin{equation}
A_{3} = -\Omega_{0} .
\end{equation}
On both sides of the equation (20), multiply a term  $A = \pi S^{2}$ of the flux tube
(where $S$ is radius of the tube). As amplitude $\Omega^{'}$ of fluctuations in angular velocity
assumed to be negligible compared to steady part of the angular velocity,
then we have the analytical solution for variation of the sunspot area
with respect to time in case of perturbations along the direction of angular velocity
\begin{equation}
A(t)=c_{1}e^{D_{1}t}+c_{2}e^{-D_{2}t} \ ,
\end{equation} 
where $c_{1}$ and $c_{2}$ are integrational constants and,
\begin{equation}
D_{1}={1\over 2}[{\Omega_{0}^{2} r^{2} sin^{2} \theta \over {\eta}}+\sqrt {\Large ({\Omega_{0}^{2} r^{2} sin^{2} \theta \over {\eta}} \Large )^{2}+4\Omega_{0}}] \ ,
\end{equation}
and
\
\begin{equation}
D_{2}={1\over 2}[{\Omega_{0}^{2} r^{2} sin^{2} \theta \over {\eta}}-\sqrt {\Large ({\Omega_{0}^{2} r^{2} sin^{2} \theta \over {\eta}} \Large )^{2}+4\Omega_{0}}] \ ,
\end{equation}
As second term in the square root is negligible compared
to the first term, solution for decay part of the area of the 
sunspot with respect to time is given as
\begin{equation}
A(t)=c_{1}e^{({\Omega_{0}^{2} {R_{\odot}}^{2} x^{2} sin^{2} \theta \over \eta})t}+c_{2} \ ,
\end{equation}
where $x={r\over{R_{\odot}}}$ and ${R_{\odot}}$ is the radius of the sun.
Although, mathematically, solution has an exponential
growth, physically, in the region of negative rotational
gradient as flux tube lifts its anchored feet (due to
buoyancy) toward surface, the distance (difference between
sunspot's anchored depth and the surface)
$r$ decreases, angular velocity $\Omega_{0}$ decreases
and magnetic diffusivity $\eta$ (~ $T^{-3/2}$, where $T$
is ambient temperature) increases and hence resulting
area decreases.

If one keeps the ratio $\Omega_{0}^{2}/ \eta$
constant at a particular depth (say near the surface),
exponent of the decay is directly proportional
to $sin^{2} \theta$. That means spots at the lower co-latitudes $\theta$
 (or higher heliographic latitudes) decay very slow compared to the spots
that decay at the higher co-latitudes (or lower  heliographic latitudes,
i,e., near the equator). This important decay property
will be tested in the following sections.

As for the Alfven wave perturbations opposite to the direction of angular
velocity in the solar interior, solution yields
\begin{equation}
A(t)=c_{1}e^{({-\Omega_{0}^{2} {R_{\odot}}^{2} x^{2} sin^{2} \theta \over \eta})}t+c_{2}
\end{equation}

Hence, in the region of negative rotational gradient, at a  particular latitude
and depth, summation of these two solutions effectively
constitutes the decay of area of the sunspots.
\begin{table}
\caption{$ \chi^2 $ fit for the laws of linear, quadratic and exponential growth of the sunspot.}
\label{Gsun$ch^2$}
\centering
\begin{tabular}{p{2cm}p{2.3cm}p{2.3cm}p{2.3cm}p{2.3cm}}
\hline
Latitude &  Life span & Linear & Quadratic &  Exponential\\
 &  (Days)& &   &\\
\hline
0 - 10$^\circ$& 9 & 6.65 & 4.15 &    0.03\\
0 - 10$^\circ$& 9& 26.35 & 1.64 &    0.02\\
10 - 20$^\circ$& 9 & 4.39 & 2.54 &   0.29\\
10 - 20$^\circ$& 10 & 0.19 & 0.18&   0.02\\
20 - 30$^\circ$& 10& 257.98 & 69.37&    0.11\\
20 - 30$^\circ$& 10 & 487.63 & 9.67&   0.13\\
30 - 40$^\circ$& 8 & 2.60 & 2.60&    0.02\\
30 - 40$^\circ$& 9 & 46.68 & 16.75&   0.39\\
\hline
\end{tabular}
\end{table}
\section{Result and Conclusions}

In order to test results of the physical ideas on the growth
and decay of the sunspots that are presented in the
previous sections, data of time evolution of corrected
areas of non-recurrent sunspot groups from Greenwich Photoheliographic
Results (GPR) are used. For the four latitude zones of $0-10$, $10-20$, $20-30$
and $30-40$ degrees two spot groups that lie between $\pm$ 70 degree from the
central meridian and life spans in the range of 8-10 days are
considered. 

In Figures 2-9, time evolution of growth of
area of the non-recurrent sunspot groups are presented.
It is assumed that sunspot area grows linearly,
quadratically and exponentially and relevant laws
are fitted with the observed growth of area of the
sunspot groups. As measured uncertainties in the areas of  sunspot
groups are not available in GPR, it is assumed that
growth and decay of area curves follow the
Poisson distribution and hence uncertainty in
each of measured area $A(t)$ (where $t$ day of observation) is taken as $[A(t)]^{1/2}$.
By knowing area $A(t)$ values and their uncertainties,
all the three laws are fitted to the observed sunspots'
area growth curves and are over plotted on top
of the each plot. In all the Figures 2-9,
the plots in the top are for the linear and 
quadratic fits and the
the plot at the bottom is fit for the exponential
growth law.  

Similarly, in Figures 10-17, observed decay
of area of the sunspot groups for all
the four latitude zones are presented.
In addition to three (viz., linear,
quadratic, exponential) decay laws,
a law of log-normal distribution is also 
considered for fitting the observed
decay curves. In all the Figures 10-17,
first and second plots in the top row are for linear 
and quadratic decay fits respectively.
In the second row of Figures 10-17,
log-normal and exponential decay fits
are presented.

As for growth of the sunspots, it is interesting
to note that among all the Figures 2-9, exponential
fit is best one. This is also clearly evident
from the $ \chi^2 $ values presented in Table 1. In the Table 1, first column
represents latitude of occurrence of the sunspot, second column 
represents lifespan and, columns 3-5 represent $ \chi^2 $ values for 
linear, quadratic and exponential fits. It is to be noted that low value of $ \chi^2 $ 
means, observed and expected curves are almost similar. 
In Table 2, constants $C_{1}$ and $C_{2}$ of exponential growth and decay parts of 
the area curve are presented. First column represents latitude of occurrence 
of the spot group, second and third columns represent the constants $C_{1}$ and $C_{2}$ 
that are determined from the exponential growth and, fourth and fifth columns 
represent the constants that are determined from fits of exponential decay 
of the sunspot area curve respectively. Another
important property, according to theoretical expectations
presented in section 3 regarding growth of the sunspot, 
as is evident from Table 2 (see the third column) that 
the exponent of the exponential fit for the
high heliographic latitude is high compared to
the exponential fits for the low heliographic
latitudes. That means the spots that formed 
at the high latitudes grow fast (with exponential growth)
and the spots that are formed near the low
heliographic latitudes grow slowly.
\begin{table}
\caption{Values of constants obtained from growth and decay of the exponential fits.}
\label{congd}
\centering
\begin{tabular}{p{2.3cm}p{2.3cm}p{2.3cm}p{2.3cm}p{2.3cm}}
\hline
   & Growth &   & Decay& \\
\hline
 Latitude&  $C_{1}$&   $C_{2}$&    $C_{1}$&  $C_{2}$\\
\hline
0 - 10$^\circ$ & 39.65$\pm$15.93& 0.26$\pm$0.66& 395.44$\pm$6.17 &0.93$\pm$0.56\\
0 - 10$^\circ$ & 4.57$\pm$3.22&0.59$\pm$0.41 & 208.51$\pm$4.71 & 0.98$\pm$0.28\\  
10 - 20$^\circ$ &12.31$\pm$4.18 & 0.72$\pm$0.58 & 138.38$\pm$1.08& 0.43$\pm$0.35\\ 
10 - 20$^\circ$ & 60.95$\pm$7.10& 0.54$\pm$0.93& 403.434$\pm$6.17 &0.55$\pm$0.48\\
20 - 30$^\circ$ & 36.23$\pm$5.31& 0.63$\pm$0.54& 5.02$\pm$1.45& 0.14$\pm$0.27\\
20 - 30$^\circ$ & 12.18$\pm$3.86 & 0.65$\pm$0.34 & 212.73$\pm$6.23&  0.27$\pm$0.53\\
30 - 40$^\circ$ &29.08$\pm$5.53 & 0.14$\pm$0.62& 92.76$\pm$4.02& 0.38$\pm$0.26\\
30 - 40$^\circ$ &30.27$\pm$5.10 & 0.68$\pm$0.61& 391.51$\pm$6.11& 0.46$\pm$0.50\\
\hline
\end{tabular}
\end{table}
\begin{table}
\caption{$ \chi^2 $ fit for the laws of linear, quadratic, log-normal and exponential decay of the sunspot.}
\label{Dsunch2}
\centering
\begin{tabular}{p{2cm}p{2.cm}p{2.cm}p{2.cm}p{2.cm}p{2cm}}
\hline
Latitude &  Life span & Linear & Quadratic & Log-normal& Exponential\\
 &(Days)& &  &  & \\
\hline
0 - 10$^\circ$& 8 & 19.82& 7.77 & 1.17&   0.29\\
0 - 10$^\circ$& 9 & 9.12 & 3.29 & 0.05&   0.03\\
10 - 20$^\circ$& 10 & 53.27 & 6.37& 0.23&  0.08\\
10 - 20$^\circ$& 8& 101.64& 62.45 & 0.26&  0.13\\
20 - 30$^\circ$& 10& 16.60& 17.28& 0.13&   0.08\\
20 - 30$^\circ$& 10 & 24.56 & 3.29&  0.08& 0.05\\
30 - 40$^\circ$& 8 & 17.38& 17.28& 2.44&   0.91\\
30 - 40$^\circ$& 8 & 36.84& 31.51& 0.36&  0.12\\
\hline
\end{tabular}
\end{table}
As for decay of the sunspots, even though log-normal
fit appears to be a very good fit among all the Figures 10-17, 
from criterion of goodness of fit of $\chi^{2}$ value, exponential decay 
fit is best one. In fact this result is also clearly evident
from the values of $ \chi^2 $ presented in Table 3. Similarly we have
another important property from these results. 
According to theoretical expectations
presented in section 4 regarding decay of the sunspots,
exponent of the exponential fit (see the $5^{th}$ column of Table 2) for the
high heliographic latitude is very low compared to
exponent of the exponential fits for the low heliographic
latitudes. That means the spots that are formed
at the high latitudes decay slowly compared to the
spots that are formed near the low
heliographic latitudes.

Even with approximations
by neglecting fluctuations in  poloidal (meridional) and 
toroidal (angular) components of velocity fields,  
theoretical solutions of growth (equation 16) and decay 
(equation 28) parts of sunspot's area evolutionary
phases match with the observed area evolutionary phases.
In order to understand a unique single solution for
understanding growth and area decay curve, one should
solve consistently full set of MHD equations (as the neglected fluctuations
${\partial \Omega^{'} \over \partial t}$ in turn
depend upon fluctuations in the momentum equations).

From the observed characteristics of growth and
decay of the sunspots at different latitudes
on the surface and from the theoretical
ideas presented in this study, one can safely conclude that
sunspots are formed due to constructive interference
of toroidal Alfven wave perturbations and, after attaining
a critical strength in the convective envelope, due
to buoyancy, sunspots raise along isorotational contours and
reach the respective latitudes. It is understood from this
study that growth and decay phases of the sunspots
 not only depend upon the surface physical characteristics
, as this problem (especially decay part) was treated by the earlier studies, but also evolutionary
history  of internal dynamics and magnetic field structure
of the sunspots while they raise toward the surface. As the sunspot is a three dimensional
structure whose evolutionary history not only
depends upon its internal structure but also on
the ambient dynamic properties of the solar convective envelope 
that ultimately yields a combined solution of growth and decay of the sunspot.


\begin{thebibliography}{180}
\expandafter\ifx\csname natexlab\endcsname\relax\def\natexlab#1{#1}\fi


\bibitem[{{Antia} {et~al.}(1998){Antia}, {Basu}, \& {Chitre}}]{Antia1998}
{Antia}, H.~M., {Basu}, S., \& {Chitre}, S.~M. 1998, \mnras, 298, 543

\bibitem[{{Badruddin} {et~al.}(2006){Badruddin}, {Singh}, \&
  {Singh}}]{Badruddin2006} 
{Badruddin}, {Singh}, Y.~P., \& {Singh}, M. 2006, in Proceedings of the ILWS
  Workshop, ed. {N.~Gopalswamy \& A.~Bhattacharyya}, 444--445

\bibitem[{{Bumba}(1963)}]{Bumba1963}
{Bumba}, V. 1963, Bulletin of the Astronomical Institutes of Czechoslovakia,
  14, 91

\bibitem[{{Cowling}(1946)}]{Cowling1946}
{Cowling}, T.~G. 1946, \mnras, 106, 218

\bibitem[{{Feymann}(2007)}]{Feymann2007}
{Feymann}, ~J. 2007, Advances in Space Science, vol. 40, p. 1173                                                                                            
\bibitem[{{Gokhale} \& {Zwaan}(1972)}]{Gokhale1972}
{Gokhale}, M.~H. \& {Zwaan}, C. 1972, \solphys, 26, 52

\bibitem[{{Hasan}(1985)}]{Hasan1985}
{Hasan}, S.~S. 1985, \aap, 143, 39

\bibitem[{{Hathaway} \& {Choudhary}(2008)}]{Hathaway208}
{Hathaway}, D. ~H. \& {Choudhary}, D. ~P. 2008, Solar Physics, 250, 269

\bibitem[{{Hiremath}(2002)}]{Hiremath2002}
{Hiremath}, K.~M. 2002, \aap, 386, 674

\bibitem[{{Hiremath} \& {Mandi}(2004)}]{Mandi2004}
{Hiremath}, K.~M. \& {Mandi}, P.~I. 2004, New Astronomy, 9, 651

\bibitem[{{Hiremath}(2009{\natexlab{a}})}]{Hiremath2009}
{Hiremath}, K.~M. 2009{\natexlab{a}}, ArXiv e-prints; 0906.3110

\bibitem[{{Hiremath}(2009{\natexlab{b}})}]{Hirem2009}
{Hiremath}, K.~M. 2009{\natexlab{b}}, ArXiv e-prints; 0909.4420

\bibitem[{{Hiremath}(2010)}]{Hiremath2010}
{Hiremath}, K. ~M. 2010, {\em Sun and Geosphere, vol.5, no. 1, p.17-22.}

\bibitem[{{Javaraiah}(1997)}]{Javaraiah1997}
{Javaraiah}, J. and {Gokhale}, M. H. 1997, \aap, 327, 795

\bibitem[{{Javaraiah}(2001)}]{Javaraiah2001}
{Javaraiah}, J. 2001, Ph.D Thesis, Bangalore University, India

\bibitem[{{Komitov}(2009)}]{Komitov2009}
{Komitov}, B., 2009, Bulgarian Astronomical Journal, 11, 139

\bibitem[{{Martinez Pillet} {et~al.}(1993){Martinez Pillet}, {Moreno-Insertis},
  \& {Vazquez}}]{MartinezPillet1993}
{Martinez Pillet}, V., {Moreno-Insertis}, F., \& {Vazquez}, M. 1993, \aap, 274,
  521

\bibitem[{{Meyer} {et~al.}(1974){Meyer}, {Schmidt}, {Wilson}, \&
  {Weiss}}]{Meyer1974}
{Meyer}, F., {Schmidt}, H.~U., {Wilson}, P.~R., \& {Weiss}, N.~O. 1974, \mnras,
  169, 35

\bibitem[{{Moreno-Insertis} \& {Vazquez}(1988)}]{Moreno-Insertis1988}
{Moreno-Insertis}, F. \& {Vazquez}, M. 1988, \aap, 205, 289

\bibitem[{{Parker}(1978)}]{Parker1978}
{Parker}, E.~N. 1978, \apj, 221, 368

\bibitem[{{Parker}(1992)}]{Parker1992}
{Parker}, E.~N. 1992, \apj, 390, 290

\bibitem[{{Perry}(2007)}]{Perry2007}
{Perry}, C.~A. 2007, Advances in Space Res, vol 40, p. 353

\bibitem[{{Petrovay} \& {Moreno-Insertis}(1997)}]{Perovay1997}
{Petrovay}, K. \& {Moreno-Insertis}, F. 1997, \apj, 485, 398

\bibitem[{{Petrovay} \& {van Driel-Gesztelyi}(1997)}]{Petrovay1997}
{Petrovay}, K. \& {van Driel-Gesztelyi}, L. 1997, in Astronomical Society of
  the Pacific Conference Series, Vol. 118, 1st Advances in Solar Physics
  Euroconference. Advances in Physics of Sunspots, ed. {B.~Schmieder, J.~C.~del
  Toro Iniesta, \& M.~Vazquez}, 145 

\bibitem[{{Prabhakaran Nayar} {et~al.}(2002){Prabhakaran Nayar}, {Radhika},
  {Revathy}, \& {Ramadas}}]{Prabhakaran2002} 
{Prabhakaran Nayar}, S.~R., {Radhika}, V.~N., {Revathy}, K., \& {Ramadas}, V.
  2002, \solphys, 208, 359

\bibitem[{{Scafetta} \& {West}(2008)}]{Scafetta2008}
{Scafetta}, NW. \& {West}, B. ~J. 2008, \apj, Physics Today, March issue

\bibitem[{{Schmidt}(1968)}]{Schmidt1968}
{Schmidt}, H.~U. 1968, in IAU Symposium, Vol.~35, Structure and Development of
  Solar Active Regions, ed. {K.~O.~Kiepenheuer}, 95

\bibitem[{{Simon} \& {Leighton}(1964)}]{Simon1964}
{Simon}, G.~W. \& {Leighton}, R.~B. 1964, \apj, 140, 1120

\bibitem[{{Sivaraman}(2003)}]{Sivaraman2003}
{Sivaraman}, K.~R. and {Sivaraman}, H. and {Gupta}, S.~S. and
{Howard}, R.~F. 2003, \solphys, 214, 65

\bibitem[{{Solanki} {et~al.}(1992){Solanki}, {Rueedi}, \&
  {Livingston}}]{Solanki1992}
{Solanki}, S.~K., {Rueedi}, I., \& {Livingston}, W. 1992, \aap, 263, 339

\bibitem[{{Solanki} (2003)}]{Solanki}
{Solanki}, S.~K. 2003, The Astron Astrophys Rev, 11, 153

\bibitem[{{Soon} (2005)}]{Soon}
{Soon}, W. ~H. 2005, Geophysical Research Letters, Volume 32, Issue 16, CiteID L16712

\bibitem[{{Spruit}(1979)}]{Spruit1979}
{Spruit}, H.~C. 1979, \solphys, 61, 363

\bibitem[{{Tiwari}(2007)}]{Tiwari2007}
{Tiwari}, ~M.and Ramesh, R., 2007,  Current Science, vol  93, 477

\bibitem[{{Wilson}(1981)}]{Wilson1981}
{Wilson}, P.~R. 1981, in The Physics of Sunspots, ed. {L.~E.~Cram \&
  J.~H.~Thomas}, 83--97

\bibitem[{{Zirin}(1988)}]{Zirin1988} 
{Zirin}, H. 1988, {Astrophysics of the sun} (Cambridge University Press)

\end{thebibliography}
\end{document}